%% file: main_NPJ-finalsub-wfigures.tex
\documentclass[twocolumn, floatfix, dvipsnames, svgnames, superscriptaddress, reprint, 10pt, linenumbers]{wlscirep}
\usepackage[english]{babel}
\usepackage[utf8]{inputenc}
\usepackage[T1]{fontenc}
\usepackage[utf8]{inputenc}
\usepackage{verbatim}
\usepackage{bm}
\usepackage{amsmath}
\usepackage{amsfonts}
\usepackage{pdfpages}
\usepackage{graphics}
\usepackage{float}
\usepackage{hyperref}
\usepackage[capitalize]{cleveref}
\usepackage{multirow}
\usepackage{booktabs}
\usepackage{makecell}
\usepackage{outlines}
\usepackage{lipsum}
\usepackage[separate-uncertainty=true]{siunitx}
\usepackage[version=4]{mhchem}
\usepackage{caption}
\usepackage{subcaption}
\usepackage{ulem}
\usepackage{mhchem}
\usepackage{xfrac}
\usepackage{orcidlink}
\usepackage[symbol]{footmisc}
\input{defs.tex}

\newcommand{\maxnote}[1]{{\color{violet} MV: #1}}
\newcommand{\lgi}[1]{{\color{teal} #1}}
\newcommand{\mk}[1]{{\color{orange} #1}}
\newcommand{\npjcorr}[1]{{#1}}

\newcommand{\todorev}[1]{{}}
\makeatletter
\AtBeginDocument{\let\LS@rot\@undefined}
\makeatother
\begin{document}

\title{
Thermodynamics and dielectric response of \ce{BaTiO3} by data-driven modeling 
} 

\input{authors-finalsub.tex}

\begin{abstract}
Modeling ferroelectric materials from first principles is one of the successes of density-functional theory, and the driver of much development effort, requiring an accurate description of the electronic processes and the thermodynamic equilibrium that drive the spontaneous symmetry breaking and the emergence of macroscopic polarization.
We demonstrate the development and application of an integrated machine learning model that describes on the same footing structural, energetic and functional properties of barium titanate (\bto), a prototypical ferroelectric.  The model uses \abinitio{} calculations as reference and achieves accurate yet inexpensive predictions of energy and polarization on time and length scales that are not accessible to direct \abinitio{} modeling. These predictions allow us to assess the microscopic mechanism of the ferroelectric transition. The presence of an order-disorder transition for the Ti off-centered states is the main driver of the ferroelectric transition, even though the coupling between symmetry breaking and cell distortions determines the presence of intermediate, partly-ordered phases. Moreover, we thoroughly probe the static and dynamical behavior of~\bto~across its phase diagram, without the need to introduce a coarse-grained description of the ferroelectric transition.  Finally, we apply the polarization model to calculate dielectric response properties of the material in a fully \abinitio{} manner, again reproducing the correct qualitative experimental behaviour.  

\end{abstract}

\maketitle

\todorev{AUTHOR COMMENT KEY

\lgi{Lorenzo Gigli}

\maxnote{Max Veit}

\mc{Michele Ceriotti}

\mk{Michele Kotiuga}
}

\section{Introduction}

Ferroelectric materials possess a spontaneous electric polarization that can be switched with an external electric field. 
The discovery of ferroelectricity in barium titanate (\bto), the prototypical ferroelectric perovskite, changed the general understanding and perception of ferroelectrics due in large part to its relatively simple crystal structure~\cite{Jona}.
At low temperatures~\bto~is rhombohedral with a polarization along the \pTi direction; at higher temperatures, it undergoes three phase transitions, first to an orthorhombic phase with the polarization along the $\langle110\rangle$ direction at \SI{183}{K}, then to a tetragonal phase with the polarization along $\langle100\rangle$ at \SI{278}{K}, and finally, at \SI{393}{K}, to a cubic, paraelectric phase\cite{Merz1949}.
It has been long understood that the spontaneous polarization is a result of the titanium atom off-centering within the enclosing oxygen octahedron, but the detailed microscopic nature of the ferroelectric transition has been the subject of intense, ongoing research with a variety of experimental and theoretical techniques. 
The ferroelectric transitions were first described with a displacive model in which the Ti displacements are driven by a transverse phonon instability\cite{cochranCrystalStability1960}.
Almost concurrently, an order-disorder model was proposed to explain the origin of the Ti displacements along any one of the eight local \pTi directions in the cubic phase, as driven by the pseudo Jahn-Teller effect ~\cite{Bersuker2013}, showing how these displacements order at lower temperatures in different ferroelectric phases~\cite{bersukerOriginFerroelectricity1966,Chaves1976}.
These models capture some of the phenomena experimentally observed in characterizing~\bto, such as phonon softening at the transition temperatures~\cite{Yamada1969,Vogt1982} -- consistent with the displacive model -- and diffuse \mbox{X-ray} scattering in all phases except the rhombohedral one~\cite{Comes1968, Comes1970, pasciakdynamic2018} -- consistent with the order-disorder model -- leading also to approaches combining the two models~\cite{Girshberg1999,pircOffcenterMathrmTi2004, pasciakpolarized2010}. In this context, simulations -- especially from first principles -- can offer a precious microscopic understanding of the nature of the phase transitions.

A computer simulation of the ferroelectric phase transition of any given material requires three key ingredients: first, a model of the potential energy surface (PES) that describes the energetic response to atomic and structural distortions,
 second, the free energy surface (FES) sampled at the relevant, finite-temperature thermodynamic conditions, and third, the polarization of individual configurations that determines, through averaging over samples, the macroscopic polarization. 

Density functional theory (DFT) calculations have long been used to explore the PES of~\bto{} as well as the soft phonons and their strong dependence on pressure~\cite{Cohen1990,cohenOriginFerroelectricity1992,ghosezLatticeDynamics1997,ghos+98fe}. Further DFT investigations have found that Ti displacements along local \pTi{} directions can result in dynamically stable structures~\cite{zhang2006, Kotiuga_2021, zhaoIntrinsicLocal2021}.
The phase transitions and rhombohedral-orthorhombic-tetragonal-cubic (R-O-T-C) phase sequence of~\bto{} has been extensively studied and reproduced using effective Hamiltonians solved using both Monte Carlo~\cite{zhongFirstprinciplesTheory1995,tinteQuantitativeAnalysis2003} and molecular dynamics (MD)~\cite{Tinte1999,ponomarevaTerahertzDielectric2008,qiAtomisticDescription2016}; furthermore, similar studies have been carried out on other perovskite systems~\cite{Krakauer1999}, including solid solutions~\cite{walizerFinitetemperatureProperties2006}.
Despite their successes, effective models rely on \npjcorr{the choice of an} explicit parametrization of the Hamiltonian; therefore,  in order to confidently make first-principles-accurate predictions of the thermodynamics, it is desirable to use an unbiased, agnostic approach without any prior assumption on the form of the PES.

To this aim, we introduce an integrated machine learning (ML) framework allowing us to carry out MD without the need to compromise on simulation size and time scales. 
This framework, based on a combination of an interatomic ML potential and a vector ML model for the polarization, is used to simultaneously predict the total energy, atomic forces and polarization of a ferroelectric material in order to explore its complex, temperature-dependent phase diagram as well as to predict its functional properties.
This approach allows us to compute macroscopic observables -- chemical potentials and dielectric susceptibilities, specifically --  with an accuracy equivalent to that of the level of theory of the underlying DFT calculations, but at a much smaller computational cost. Moreover, it is applicable with only minor changes to any perovskite or even any other type of ferroelectric material, including 2-D ferroelectrics\cite{zhangStructuralPhase2020}. \npjcorr{Although we do not reach quantitative agreement with the experimental R-O-T-C transition temperatures, we demonstrate that this limitation in accuracy stems from the DFT reference itself and not the approximation made in modelling the potential energy surface. Thus, we foresee clear, systematic pathways to improving the model potential, with only slight modifications of the ML methodology. Specifically, the} generality of the framework and the relatively small size of the training dataset makes it possible to improve the model accuracy by computing the reference structures with more advanced functionals such as Hubbard-corrected DFT~\cite{Liechtenstein1995,Dudarev1998}, meta-GGAs,~\cite{Perdew2001} and hybrids~\cite{Becke1993}. 

The key advance underlying this work is an integrated ML framework combining an interatomic potential, based on the SOAP-GAP method\cite{bart+10prl}, and a microscopic polarization model, based on the symmetry-adapted Gaussian process regression (SA-GPR) method\cite{gris+18prl}.
The use of ML for materials modelling has gained considerable momentum in the past decade \cite{behl-parr07prl, bart+10prl, rupp+12prl, mont+13njp, behlerFirstPrinciples2017, deringerMachineLearning2019, noeMachineLearning2020, butlerMachineLearning2018, schu+18jcp, friederichMachinelearnedPotentials2021, lopanitsynaFinitetemperatureMaterials2021, imba+21jcp, chen+19pnas, carleoMachineLearning2019, dragoniAchievingDFT2018, bart+18prx, isayevMaterialsCartography2015, sanchez-lengelingInverseMolecular2018,szla+14prb,deri-csan17prb,mora+16pnas,Caro2018,lopanitsynaFinitetemperatureMaterials2021}. Specifically, the prediction of finite-temperature properties of materials as the ones we focus on this paper relies on the construction of ML potential energy surfaces based on a set of reference structures computed with \abinitio{} methods \cite{soss+12prb, eshe+12prl, szla+14prb}.  Such potentials allow the simulation of molecules and complex solids with almost the same accuracy as the reference method used to generate the dataset.  In this way, it is possible to investigate the meso- and macroscopic properties of materials at a considerably reduced computational effort compared to direct \abinitio{} simulations. Notable successes of the machine learning potentials approach include the study of bulk and interfacial properties of metallic alloys from cryogenic temperatures up to the melting point \cite{lopanitsynaFinitetemperatureMaterials2021}; finite-temperature modeling of binary systems with variable concentration, such as GaAs \cite{imba+21jcp}; accurate calculations on the relative stability of competing phases of various compounds, such as sodium \cite{eshe+12prl}, carbon \cite{khal+10prb}, water \cite{chen+19pnas}, iron \cite{dragoniAchievingDFT2018} and silicon \cite{bart+18prx}; as well as MD studies of polycrystalline phase-change materials \cite{soss+12prb} and hybrid perovskites \cite{jinn+19prl}.

Two \npjcorr{important developments} have enabled the great success of machine learning in condensed matter and chemical physics.  First, appropriate regression schemes -- such as kernel methods, typified by Gaussian approximation potentials (GAP)\cite{bart+10prl}; neural networks (e.g. of the Behler-Parrinello type\cite{behl-parr07prl} or more recent graph convolutional approaches\cite{schu+18jcp,parkAccurateScalable2021}); or non-kernel-based linear fitting schemes (\npjcorr{with appropriate representations\cite{thom+15jcp,shap16mms,vanderoordRegularisedAtomic2020,niga+20jcp}}) -- have been designed and specialized for atomistic systems.  The key to nearly all of these methods is the decomposition of a global (extensive) physical observable of the system into local contributions, each written as a function of the neighbourhood of individual atoms.   Note that this decomposition carries with it an implicit assumption of \emph{locality} of the potential energy surface, thus neglecting the effect of long-range electrostatic and dispersion forces.
Several extensions have previously been proposed to include such forces within existing ML frameworks\cite{artr+11prb,bere+18jcp,veit+19jctc,koFourthgenerationHighdimensional2021}, but for the purpose of this work, we use an explicitly short-range model with an appropriately chosen cutoff.

The second advancement is the construction of suitable, physically motivated, representations to predict the target properties of interest\cite{musi+21jcp, himanenDataDrivenMaterials2019, butlerMachineLearning2018, csan+20book, musi+21cr}. In particular, the representation of an atomic configuration should reflect all the physical symmetries of the target property. The framework built around the Smooth Overlap of Atomic Positions (SOAP) descriptor\cite{bart+13prb} and its \npjcorr{covariant} counterparts\cite{gris+18prl,musi+21cr}, which we call the \textit{atom-centered density correlation} framework, is well suited to the task of integrated machine learning modeling of multiple properties, since it allows us to treat these properties within the same unified mathematical framework.   We provide further details on the mathematical framework, as well as the construction of the unified ML model, including the definition of polarization-derived collective variable, in Sec.~\ref{sec:Methods}.
Looking forward, the flexibility and extensibility of this framework will also allow us in future studies to include long-range interactions in a natural and general way, using a recent approach called LODE\cite{gris-ceri19jcp,gris+21cs}. This will allow us to address some of the observed disagreements with DFT benchmarks in the prediction of the phonon spectra, which likely derive from the neglect of long-range forces (see Sec.~\ref{ssec:phon-disp} for additional details).

The modelling of multiple properties within a single ML framework is gaining importance as a way to \npjcorr{extract} richer information from simulations than the PES alone can provide. Such models combine the extensive, accurate, finite-temperature thermodynamic sampling afforded by a ML potential, as in a series of previous works\cite{lopanitsynaFinitetemperatureMaterials2021, imba-ceri21prm, deri+21nature, eshe+12prl, chen+19pnas, dragoniAchievingDFT2018, bart+18prx, soss+12prb, jinn+19prl}, with the expressiveness and utility of an ML property model. Particularly relevant are the studies using a potential energy surface combined with a dipole-moment model for studying the infrared spectra of isolated molecules\cite{gast+17cs,laurensInfraredSpectra2021}.  To date, such combined models have not yet been applied to ferroelectric materials; one important difficulty for ML modelling is the multi-valued character of the polarization in the condensed phase (although see Kapil \textit{et al.}\cite{kapi+20jcp} and Zhang \textit{et al.}\cite{zhan+20prb} for applications in liquid water, where this difficulty is much less severe).  We describe a method of overcoming this difficulty in a systematic and generalizable way in Section~\ref{sec:Methods} and in Supplementary Note 3.
In the following study, we show how a combined modelling study can advance the field of ferroelectrics by providing a rich array of experimentally relevant properties
from one unified mathematical framework.

\section{Results}

\subsection{Summary}
In this section, we summarize the main results obtained via the integrated ML model described in the Methods section.
In Sec. \ref{sec:BaTiO3_phases} we investigate the structural transitions of~\bto, recovering the well-known sequence of phases R-O-T-C
and highlighting the key role of the ML-predicted polarization vector in distinguishing each of the phases. 
In Sec. \ref{sec:micro-nature} we elucidate the microscopic nature of the phase transitions, finding that the Ti off-centering is the \emph{driving} mechanism of ferroelectricity and not just a result of cell distortions.

In Sec.~\ref{sec:thermo} we provide explicit calculations of the thermodynamics of~\bto{} in order to compute phase transition temperatures.  Finally, in Sec.~\ref{sec:dielec-resp}, we compute its temperature- and frequency-dependent dielectric response properties and compare these with experiment.

\subsection{Structural transitions in $\text{B\lowercase{a}T\lowercase{i}O}_3$}
\label{sec:BaTiO3_phases}

\begin{figure*}[tbhp]
    \centering
    \vspace{-1.5cm}\includegraphics[width=0.985\textwidth]{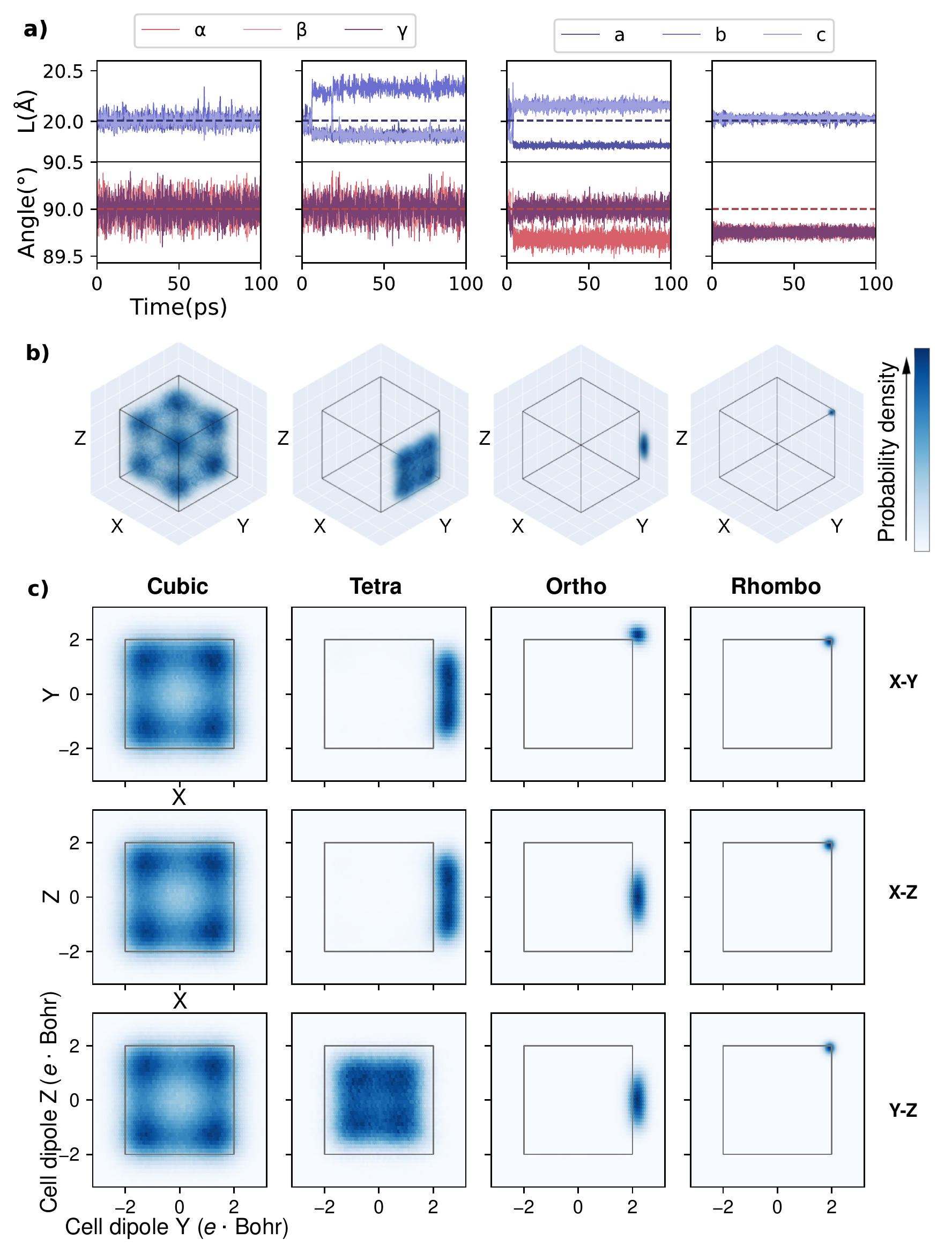}
    \caption{The four phases of \bto{}: (a) time evolution of the lattice vector lengths ($a$, $b$ and $c$) and angles ($\alpha$, $\beta$, $\gamma$) over fully flexible MD simulations at (from left to right) \SIlist{250;100;50;15}{\kelvin}; (b) 3-D histograms of the unit-cell polarization predicted by the SA-GPR polarization model, computed across the same simulations;  (c) projections of those 3-D histograms onto the three principal Cartesian planes (X-Y, X-Z and Y-Z).  The light gray boxes mark the range $\pm\SI{2}{\electron\bohr}$ for easier comparison between the phases.  The structure of the high-frequency (dark blue) regions characterizes the different phases of \batiooo, while the phase transitions are marked by a clear symmetry-breaking pattern that restricts the cell dipole to visit only a subset of the available off-centered sites.}
\label{fig:pol_histograms}
\end{figure*}

The detection of the ferroelectric transitions of \bto{} in MD simulations is challenging due to the small lattice distortions and free-energy differences that differentiate the phases. To overcome these challenges, one has to choose a sufficiently large cell so as to make the transitions clearly visible while allowing for a well-converged statistical sampling of configurations across each of the coexistence regions. As a qualitative indicator of the phase transitions, we track the MD time evolution of the cell parameters $(a, b, c, \alpha, \beta, \gamma)$ and the histograms of the ML-predicted unit-cell polarization components for each phase. The unit-cell polarization correlates strongly with the magnitude and direction of the Ti displacement (see SI).

Figure 1 shows that the high-temperature cubic phase consists of a collection of local minima arranged at the vertices of a cube, as proposed by the eight-site model \cite{bersukerOriginFerroelectricity1966, chavesThermodynamicsEightsite1976, comesChainStructure1968, pircOffcenterMathrmTi2004}. The presence of large thermal fluctuations, as compared to the energy barriers separating the minima, allows for diffusion of the polarization vector across the minima over a timescale of the order of a few ps in the MD trajectories. This makes the eight local Ti minima equally probable, yielding $\langle \mathbf{P} \rangle = 0$. 

A reduction of the temperature results in a structural first-order phase transition, both in agreement with Rappe \textit{et al.}\cite{qiAtomisticDescription2016} and with previous experimental \cite{robertsAdiabatic1952} and theoretical works \cite{zhon+94prl, zhongFirstprinciplesTheory1995} showing a divergence of the latent heat at the Curie point.  Such transition is characterized by a clear breaking of reflection symmetry of the cell dipoles across one Cartesian (X-Y, Y-Z or X-Z) plane.  The polarization vector can only visit four of the eight available cubic sites, marking the onset of the tetragonal phase. Any further decrease of the temperature further reduces the symmetry of the polarization histograms, by successive freezing of the polarization components along a specific axis. At \SI{50}{\kelvin}, the polarization densities show a single and broad minimum corresponding to an orthorhombic state, while at \SI{15}{\kelvin} finally the system is completely frozen in one minimum corresponding to the rhombohedral state. Note that each of the \num{6} tetragonal, \num{12} orthorhombic, and \num{8} rhombohedral minima that are trivially equivalent by symmetry can be reached depending on initial configuration. These states are associated with different distortions of the lattice vectors (that are not symmetry invariants) and one can observe occasional transitions between them.
Thus we can infer from Figure~1 that the phenomenology of the ferroelectric-paraelectric transition agrees well with the eight-site model.  Under this model, the breakdown of ferroelectricity is characterized by thermal fluctuations across cubic off-centered sites that restore, on average, the centrosymmetry of the Ti-displacements.

Furthermore, we see that the action of a large isotropic pressure of \SI{30}{\giga\pascal} in the cubic state fully restores the isotropy of the polarization densities (see Supplementary Figure~4) and generates a paraelectric \bto{} phase down to \SI{0}{\kelvin}. This is consistent with the experimentally observed loss of ferroelectricity in~\bto{} at high pressures~\cite{Decker1989}, as well as with the flattening of the calculated PES ~\cite{Cohen1990,cohenOriginFerroelectricity1992}, the disappearance of all unstable phonon modes of cubic~\bto{},  and the isotropic Ti-displacement distribution observed in Car-Parrinello MD simulations of cubic~\bto{} under pressure\cite{Kotiuga_2021}.
This evidence strengthens the hypothesis that the fluctuations of the unit cell polarization between preferential orientations act as a microscopic precursor of the macroscopic ferroelectricity of the material.

\begin{figure}[tbp]
    \centering
    \includegraphics[width=\columnwidth]{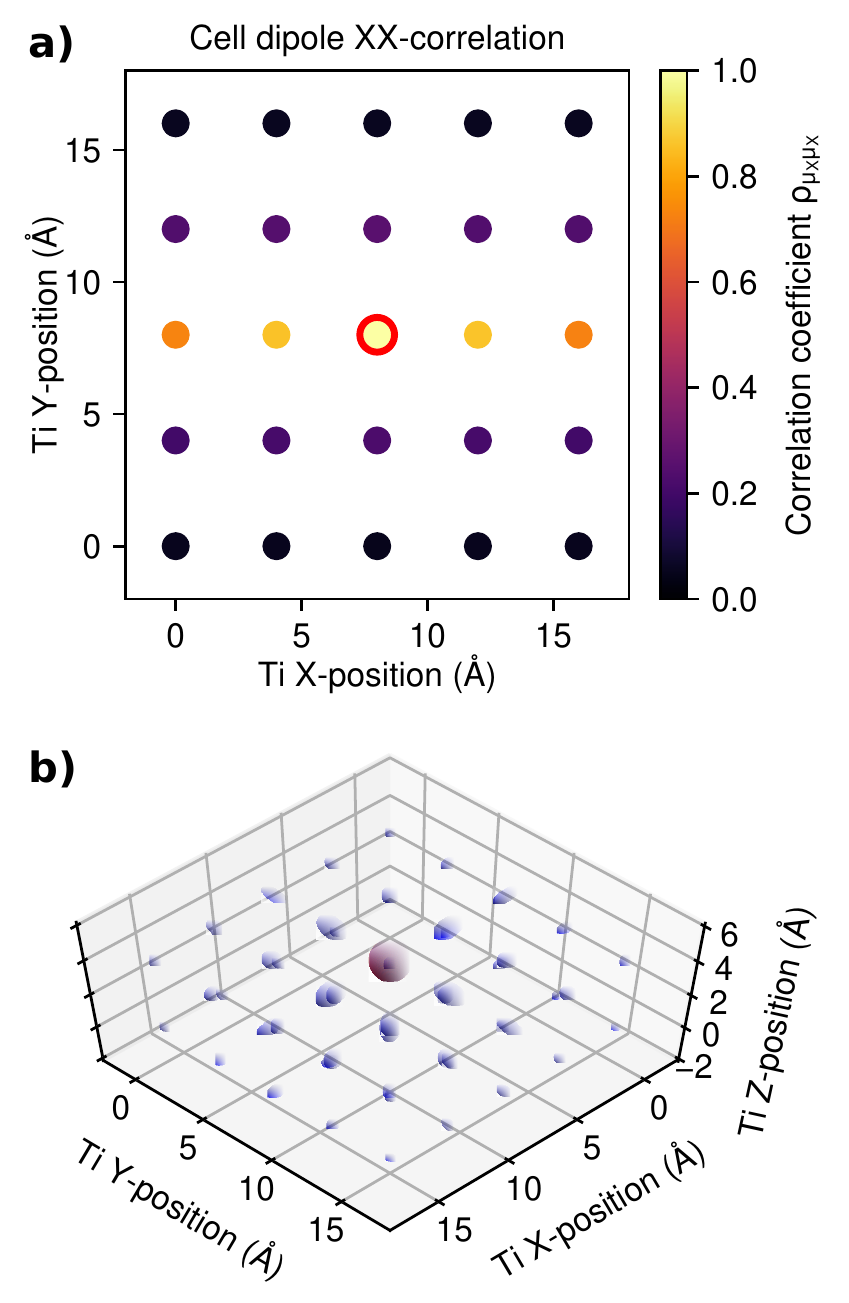}
    \caption{Spatial correlations of the unit-cell dipoles computed on a \sfive{} supercell simulated at \SI{250}{\kelvin}.  (a) Slice at constant Z of the correlation of the X-component of the dipole of the unit cell centered on one Ti-atom (circled in red) with the X-component of the dipole of every other Ti-centered cell.  (b) 3-D view of the dipole correlations for the lower half of the same \sfive{} supercell; the full 3$\times$3 correlation tensor of each dipole with the central Ti-atom (red sphere) is shown as an ellipsoid, where the elongation along the Cartesian axes shows the presence of highly directional, needle-like correlations~\cite{akbarzadehAtomisticSimulations2004}.}
    \label{fig:spatial-correlations}
\end{figure}

Moreover, the emergence of a ferroelectric state is facilitated by the presence of spatial and directional correlations across the structures, which have been proposed to explain X-ray diffraction results~\cite{Comes1970,sennEmergenceLongRange2016} and directly observed in Bencan \textit{et al.} \cite{bencanAtomicScale2021}, with further evidence from first-principles calculations~\cite{vanderbiltFirstprinciplesTheory1998,akbarzadehAtomisticSimulations2004,pasciakpolarized2010}.

Figure~2 shows the extent and directionality of the spatial correlations (specifically, component-wise Pearson correlation coefficients; see Supplementary Note 5 for the exact expression) of the unit cell dipoles, correlated against a central reference cell, in the cubic phase at \SI{250}{\kelvin}.
The correlations are not only large and slowly decaying -- they extend well up to the edges of the \sfive{} supercell -- but they are highly directional, with the slow decay taking place along the direction of the unit-cell dipole.

It has long been assumed that these correlations arise from a combination of an Ising-like nearest-neighbour interaction with a long-range dipole-dipole interaction, as typified e.g. in the model Hamiltonian of Zhong \textit{et al.}\cite{zhongFirstprinciplesTheory1995}.  Indeed, the authors of that study observed that the Coulomb interaction was critical for reproducing the ferroelectric ground state in their model -- when it was turned off, the ground state became antiferroelectric.
However, our simulations show long-range correlations and a ferroelectric ground state even though the energy model itself is \emph{explicitly short-ranged} -- that is, the energy of an atom $i$ is only sensitive to changes within a short-ranged, local environment of \SI{5.5}{\angstrom} of the atom (see Sec. \ref{ssec:training}) -- making the correlations observed in our simulations an emergent phenomenon, not relying on the existence of any explicit long-ranged interaction. This range is sufficient to capture short-range correlations between two neighboring Ti atoms, whose average distance in a typical MD run fluctuates about \SI{\approx{}4.0}{\angstrom}. Furthermore, we observe these correlations even in the disordered, cubic phase, in contrast e.g. to Akbarzadeh \textit{et al.}\cite{akbarzadehAtomisticSimulations2004}, where the strongest correlations were observed only in the ordered phase (albeit in a different material, and where the phase transition was triggered not by temperature, but by including quantum nuclear effects).

Thus, our understanding of the nature of ferroelectricity in \bto{} must take into account these emergent, long-range correlations.  We will see, for instance, how they give rise to spontaneous ferroelectric states -- even in the absence of lattice distortions -- in Sec.~\ref{sec:micro-nature}. On the other hand, these correlations also hamper the statistical and simulation-cell size convergence of various quantities computed from statistical averages of the total dipole moment, as will be discussed in Section~\ref{sec:dielec-resp}.

\subsection{The microscopic mechanism of the ferroelectric transition}
\label{sec:micro-nature}

\begin{figure}[tbhp]
    \centering
    \includegraphics[width=\linewidth]{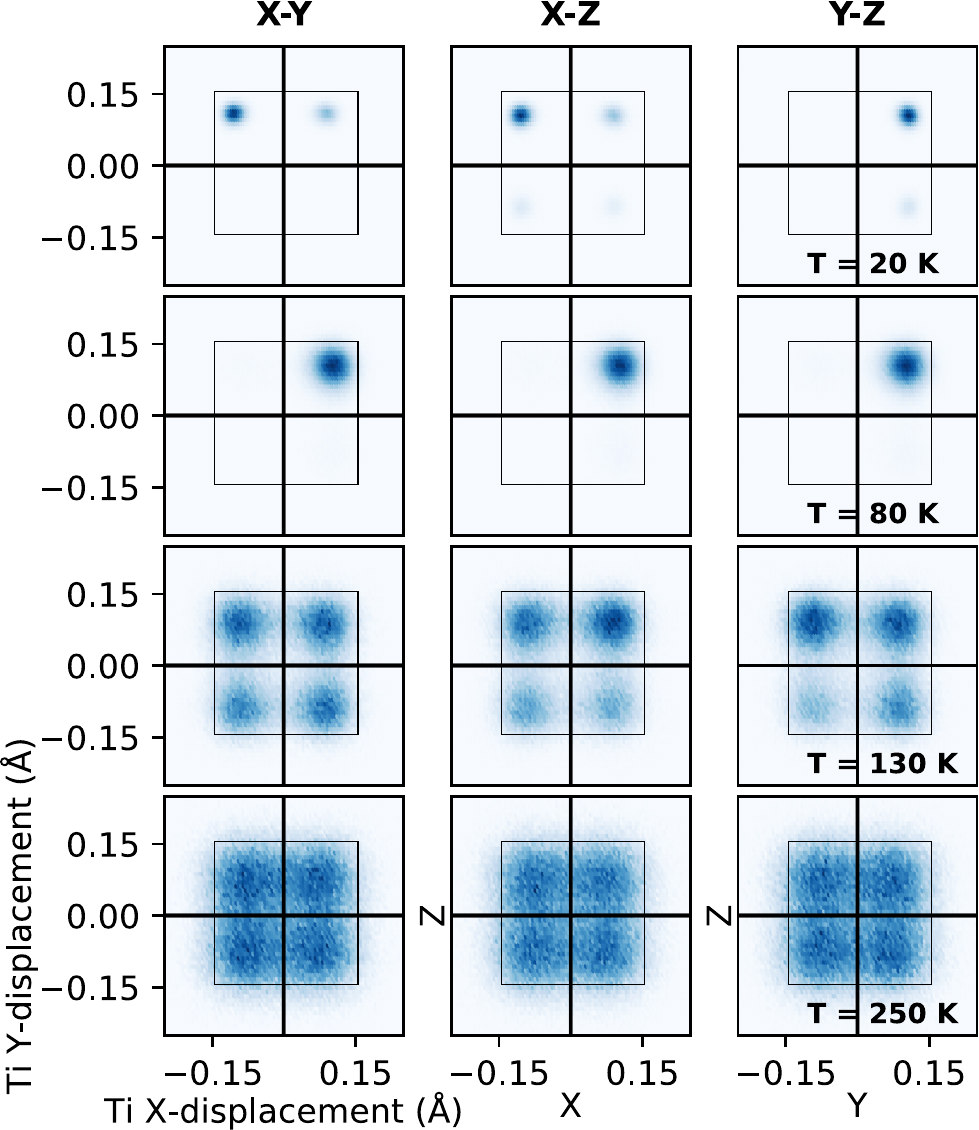}
    \caption{2-D displacement maps, mapped onto the three principal Cartesian planes (X-Y, X-Z and Y-Z), in a series of NpT \sfour{} simulations with restricted cubic geometry. The \SI{20}{\kelvin} trajectory shows fast fast-freezing of the Ti degrees of freedom to the nearest cubic sites (depending on the initial configuration) in the first few \si{\pico\second} of trajectory. In this specific case only states corresponding to positive displacements along the $y-$axis are sampled across the \sfour{} supercell. At \SI{80}{\kelvin} the presence of one single maximum in the 3D density marks the onset of a clear `rhombohedral-like' ferroelectric state (with a polarization parallel to the [-1, +1, +1] axis) showing how the GAP favours aligned displacements even in the presence of geometric constraints on the supercell. Simulations at higher temperature (\SI{250}{\kelvin} and beyond) restore instead the \npjcorr{eight-site} structure of the displacement density.}
  \label{fig:disps-NpT}
\end{figure}

\begin{figure*}[tbhp]
    \centering
    \includegraphics[width=\linewidth]{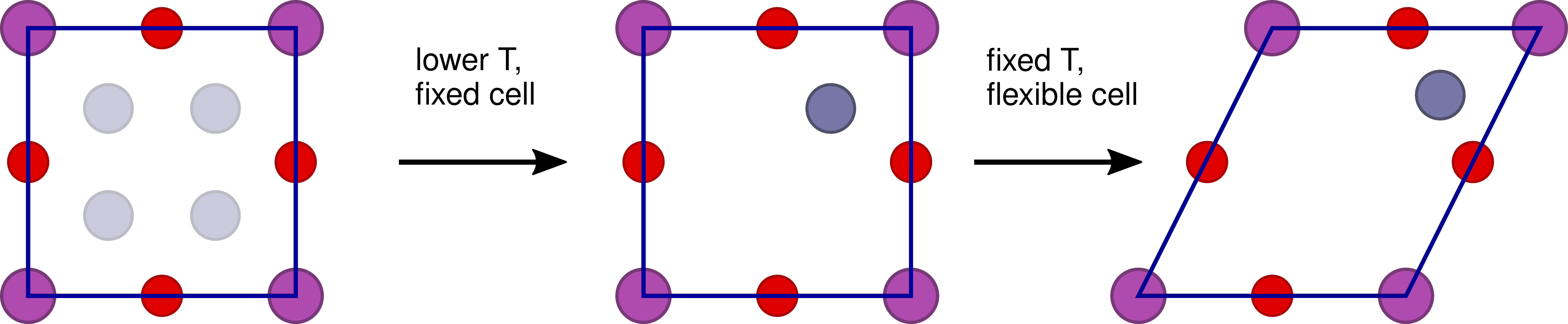}
    \caption{2-D sketch of the sequence of events marking the onset of a ferroelectric state starting from a perfect cubic configuration.}
  \label{fig:sketch-seq}
\end{figure*}

A fundamental question that arises in relation to the structural transitions observed in perovskite ferroelectrics concerns the driving mechanism of the transitions. We have seen in Sec. \ref{sec:BaTiO3_phases} that the presence of the ferroelectric behavior is accompanied by the onset of a macroscopic polarization, mostly driven by ordered displacements of the transition metal atoms, and a cell deformation. This indicates that reducing the temperature makes it energetically favourable to develop polarized states, even at the expense of an internal strain introduced by the subsequent cell deformation. 

One might, however, question whether it is the Ti off-centering or the cell deformation that drives this sequence of transitions, or whether these distinct mechanisms are equally present and competing.
To this end, ambient-pressure simulations of a \sfour{} cell over a wide range of temperatures (between \SIrange[range-phrase={ and }]{20}{250}{\kelvin}) were carried out in a restricted cubic geometry. The geometric constraint inhibits the structural distortions, making it possible to investigate whether the displacements and ferroelectric states are still observable.

Two-dimensional histograms of the Ti displacements for a series of representative trajectories at \SIlist{25;80;130;250}{\kelvin} are provided in Figure~3. The lowest-temperature  trajectory is equivalent to a fast-freezing experiment, where the Ti atoms relax to the closest potential energy minimum, as seen in the top row of Fig~3. We still observe off-centered states according to the eight-site model picture, but no transitions between neighboring cubic sites take place, due to the negligible thermal fluctuations. 

By slightly increasing the temperature, the thermal fluctuations are still smaller than the energy barrier between neighboring cubic sites, but are sufficient to induce rare jumps between them. The Ti atoms consequently freeze in a local minimum, but notably they all collectively jump to a single off-centered state, after a small transient of the order of a few \si{\pico\second}. The system stays trapped in this state for the whole simulation time. 

This gives us evidence that the GAP \textit{inherently} favors correlated ferroelectric displacements, despite the short ranged description of the interactions (in an Ising-like fashion)
and indicates that a low-temperature ferroelectric state arises in fully flexible simulations and in experiments as a consequence of a dipolar ordering, which is suppressed at high-temperatures by thermal fluctuations. 

At higher temperatures, as seen in the lower rows of Figure~3, thermal fluctuations enable transitions between cubic sites with a rate that increases with the temperature. Due to the restored cubic centrosymmetry of the displacements, states with a net polarization are no longer observed, provided sufficiently long MD runs are performed. We find this same sequence of states in NVT cubic simulations (see Supplementary Note 7), showing that constraining the volume does not affect the qualitative picture of Figure~3. Additionally, we note that the intermediate tetragonal and orthorhombic states do not occur in these simulations, as opposed to the fully flexible ones.

In conclusion, the presence of a dipolar ordering is responsible for the emergence of a low-temperature ferroelectric state, as shown in Figure~4 and in agreement with Senn \textit{et al.}\cite{sennEmergenceLongRange2016}. At the same time, the absence of cell distortions considerably affects the shape of the FES, as the intermediate tetragonal and orthorhombic states do not occur with a fixed cell. This effect has also been reported in Zhong \textit{et al.}\cite{zhongFirstprinciplesTheory1995}, where Monte-Carlo simulation with no homogeneous strain showed the disappearance of such phases. 

\subsection{Thermodynamics of \btotitle{}}
\label{sec:thermo}


A challenge in modelling phase transitions such as the ones we focus on in this paper is that they are associated with small structural distortions that are comparable with the thermal fluctuations of individual atoms. 
A commonly-used strategy to improve the signal-to-noise ratio is to use collective variables (CVs), such as the lattice parameters\cite{jinn+19prl}, which are naturally averaged over multiple atomic environments and directly reflect the macroscopic observables associated with the transition. 
Cell vectors, however, are not symmetry-adapted, so that multiple equivalent states are mapped to different values of the CVs. What is more, as we have seen in Sec.~\ref{sec:micro-nature}, cell distortions alone do not drive the different phase transitions of~\bto, making them poor order parameters to distinguish these phases (see Supplementary Note 4for a discussion of our metadynamics simulations that use a symmetrized combination of lattice parameters). 

A more effective characterization of the ferroelectric ordering can be obtained by explicitly using the predicted polarization $\mathbf{P}$ as an order parameter, and, in particular, by building descriptors that show the orientation of the cell polarization relative to the atomic distortions. In Sec. \ref{ssec:ord-param} we provide the construction of a two-component CV, namely $\mathbf{s} = (s_1, s_2)$, that gives us an effective low-dimensional description of the phases of~\bto. 

In Figure~5 we show 2-D contour lines of $\mathbf{s}$ across fully flexible MD runs of a~\bto~\sfour{} supercell between \SIrange[range-phrase={ and }]{10}{250}{\kelvin}. These represent molecular dynamics runs where the GAP predicts coexistence of R-O, O-T and T-C states respectively, with comparable probability.  Four distinct phases are clearly visible, showing how a polarization-derived two-component CV can easily identify the subtle differences between the four phases. The relative positions of the clusters give additional \npjcorr{(but only qualitative)} physical insights: as the C-T-O clusters are maximally distinguishable by $s_1$ and the C center corresponds to the one with lowest $s_1$ value, the first CV is clearly related to the average polarization magnitude. This is predicted to be exactly zero for paraelectric cubic \bto{} in the thermodynamic limit, while positive and increasingly large for the ferroelectric tetragonal and orthorhombic phases. The CV $s_1$ can then be used to discriminate ferroelectric and paraelectric~\bto~states. \npjcorr{A further evidence in this respect is provided in Supplementary Figure~12, where we show how it is possible to reconstruct free energy profiles as a function of $s_1$ across the T-C transition and thus capture their finite-temperature stability.}

On the other hand, $s_2$ maximizes  the difference between the R-O-T ferroelectric states; thus, we can relate it physically to the polarization orientation, in agreement with our observations of Sec. \ref{sec:BaTiO3_phases}. Additional evidence for this interpretation is provided in Supplementary Figure~8, where we analyze the correlation between the CVs, constructed here, with a meaningful physical observable, namely the polarization magnitude, in a series of \sfive{} fully flexible trajectories.

\begin{figure}[tbhp]
    \centering
    \includegraphics[width=\linewidth]{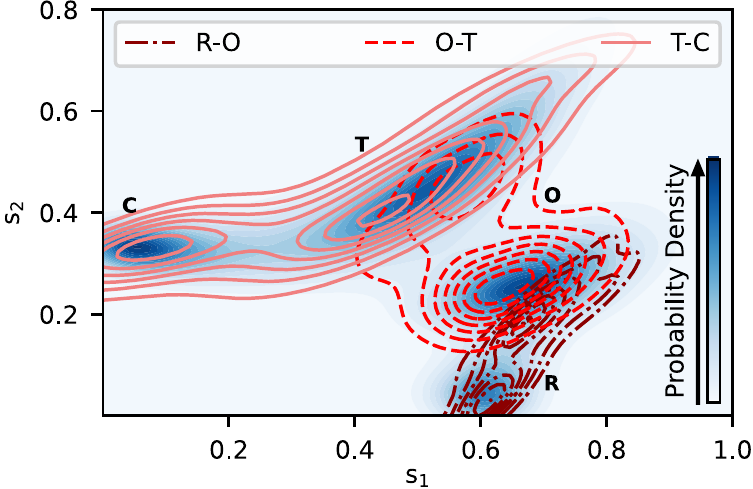}
    \caption{2D contour plots of the two-component CV $\mathbf{s} = (s_1, s_2)$ plane across fully flexible MD trajectories, highlighting the presence of the R-O-T-C clusters.}
  \label{fig:ROTC-clusters}
\end{figure}

\begin{figure*}[t]
    \centering
    \includegraphics[width=\linewidth]{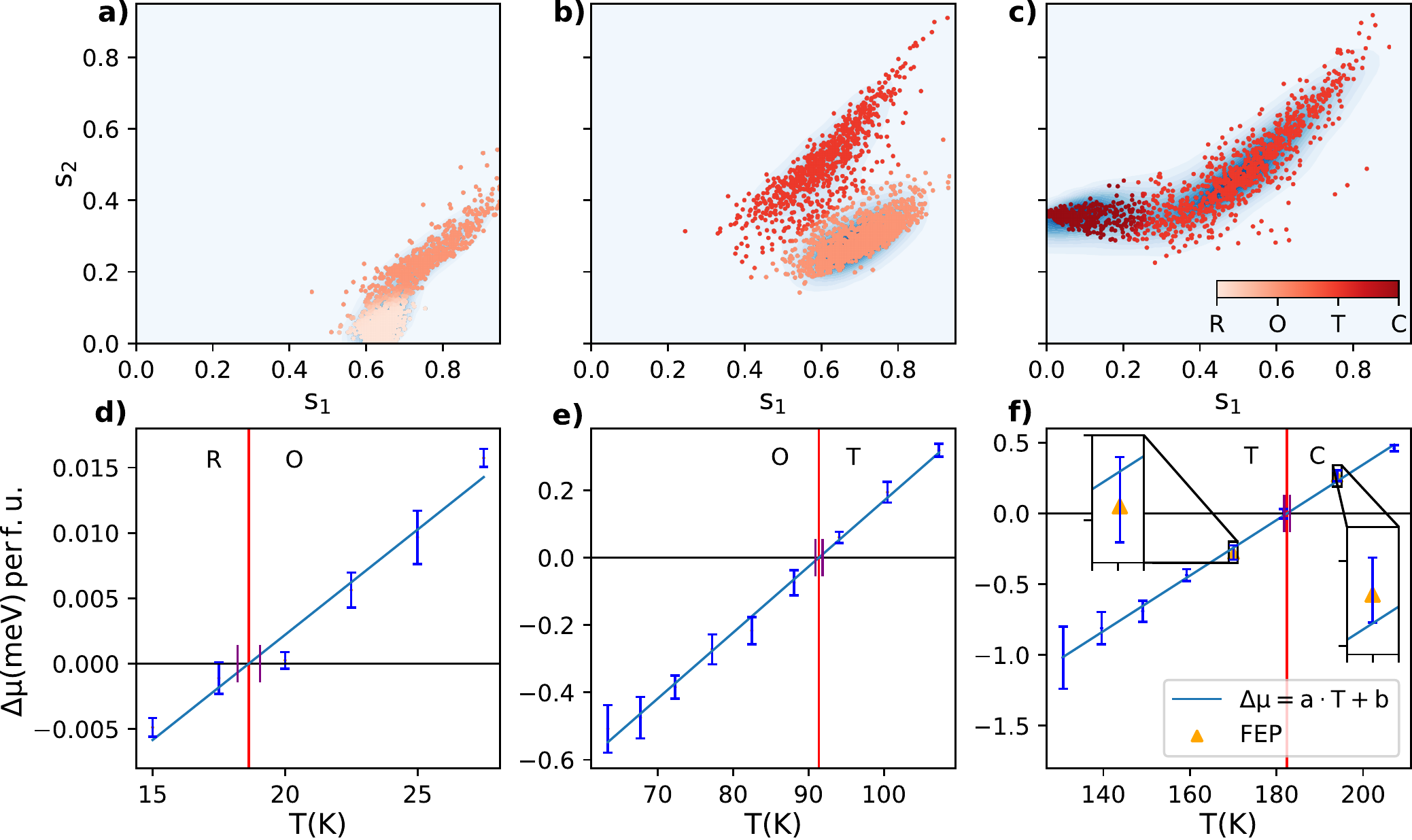} 
        \caption{Phase diagram of~\bto~at ambient pressure. Panels a, b and c show the R-O, O-T and T-C clusters in the $(s_1, s_2)$ plane across the coexistence regions. Each configuration generated in the MD is a point in this plane, coloured according to its probability $P$. The latter is computed via the Probabilistic Analysis of Molecular Motifs (PAMM) \cite{gasp+18jctc} algorithm (see Sec. \ref{ssec:ord-param} for additional details).  $P$ is smoothly increasing in the $[0, 1]$ interval while going from R to O (panel a), then from O to T (panel b) and final from T to C (panel c). Panels d, e and f show the temperature-dependent chemical potential differences across the R-O (d), O-T (e) and T-C (f) transitions. Blue error bars represent standard deviations computed across multiple MD runs, the cyan line the best linear fit across the coexistence regions and the purple error bars the propagated errors on the critical temperatures. In addition, the orange triangles in panel f show the free-energy perturbed chemical potentials, using \num{50} reference tetragonal and cubic structures just below (\SI{170}{\kelvin}) and above (\SI{194}{\kelvin}) the Curie point. \npjcorr{$6 \times$ magnified insets corresponding to these two temperatures show how the} FEP-corrected chemical potentials consistently fall within the error bars due to the MD sampling, confirming the DFT-accuracy of the GAP (see Sec. \ref{sec:thermo}).}
        \label{fig:phasediag}
\end{figure*}

Based on 2-D maps as the one shown in Figure 5, it is possible to cluster the MD trajectories and compute directly the temperature-dependent FES, by calculating the relative concentration of the R-O-T-C phases in sufficiently long MD runs, so that many reversible transitions can be sampled. In practice we never find more than two phases explored at each temperature. 
Fully flexible MD runs of a~\bto{} \sfour{} cell, each with a total simulation time up to \SI{1.6}{\nano\second}, between \SIrange[range-phrase={ and }]{10}{250}{\kelvin} allow us to compute the relative Gibbs free energy $\Delta G^{k, k'}(T)$ of the phase pair $(k, k')$ at temperature $T$ and the corresponding chemical potential difference $\Delta \mu^{k, k'}(T)$ using the following equations: 

\begin{equation}
 \begin{split}
 \label{eq:Gibbs-fe}
 \Delta G^{k, k'}(T) &= - N k_{\mathrm{B}} T \ln \frac{\sum_t P_k(t) w(t)}{\sum_t P_{k'}(t) w(t)} \\
 \Delta \mu^{k, k'}(T) &= \frac{\Delta G^{k, k'}(T)}{N}
 \end{split}
\end{equation}
where $N = \num{320}$ is the number of atoms, $w(t)$ is the weight of the $t$-th structure, $P_k(t)$ is the probability that the $t$-th structure belongs to the phase $k$, and $k_{\mathrm{B}}$ is the Boltzmann constant. For the O-T and T-C transition, unbiased MD simulations are used, i.e. with $w(t) = 1$, for every $t$. In the case of the R-O transition, $w(t)$ represents the weight of the $t$-th structure computed via the iterative trajectory
reweighting (ITRE) technique \cite{gibe+20jctc}, which is used to remove the time-dependent bias in the distribution of the microstates introduced by metadynamics.  

The estimates of the critical temperatures at ambient pressure are computed by linear fits of the relative chemical potentials $\Delta \mu$ profiles, see Equation~\eqref{eq:Gibbs-fe}, and are reported for each pair of phases in Table 1. We note that our computed temperatures differ significantly from the experimentally observed transition temperatures,
\SI{393}{\kelvin} (T-C), \SI{278}{K} (O-T), and \SI{183}{K} (R-O)\cite{Merz1949}. 
This underestimation of the critical temperatures, stemming from an underestimation of the free-energy barriers between the phases, could come, in principle, from a variety of mechanisms, particularly the neglect of long-range electrostatics (and consequently of the LO-TO splitting in the training set structures), as well as the presence of finite-size effects that could stabilize the high-temperature disordered phases in the MD. In fact, a significant size dependence of the finite temperature properties of another perovskite, $\text{PbTiO}_3$, has been reported in a recent ML-driven study by Xie \textit{et al.}\cite{xie_ab_2022}.

\begin{center}
\begin{table}[htp]
    \caption{Critical temperatures of the T-C, O-T and R-O transition at ambient pressure. $R^2$ represents the coefficient of determination of best linear fits of $\Delta \mu (T)$ across the coexistence regions, as shown in Figure 6.}
    \label{tab:phasetemps}
\begin{tabular}{ ccc}
 \hline\hline
  phase transition & $T_{\mathrm{c}} (K)$ & $R^2$ \\ 
 \hline
 T-C & $182.4 \pm 0.7$ & $0.99$\\ 
 O-T & $91.4 \pm 0.5$ & $0.99$ \\ 
 R-O & $18.6 \pm 0.4$ & $0.97$ \\ 
 \hline\hline
\end{tabular}
\label{table:tempcrit}
\end{table}
\end{center}

However, previous work based on effective Hamiltonian models already pointed out this same underestimation of the critical temperatures and connected them to a shortcoming of the underlying exchange-correlation functionals \cite{qiAtomisticDescription2016,zhongFirstprinciplesTheory1995}, which can be compensated by rescaling the potential energy surface or introducing an artificial negative pressure.

\begin{figure*}
    \centering
    \includegraphics[width=\linewidth]{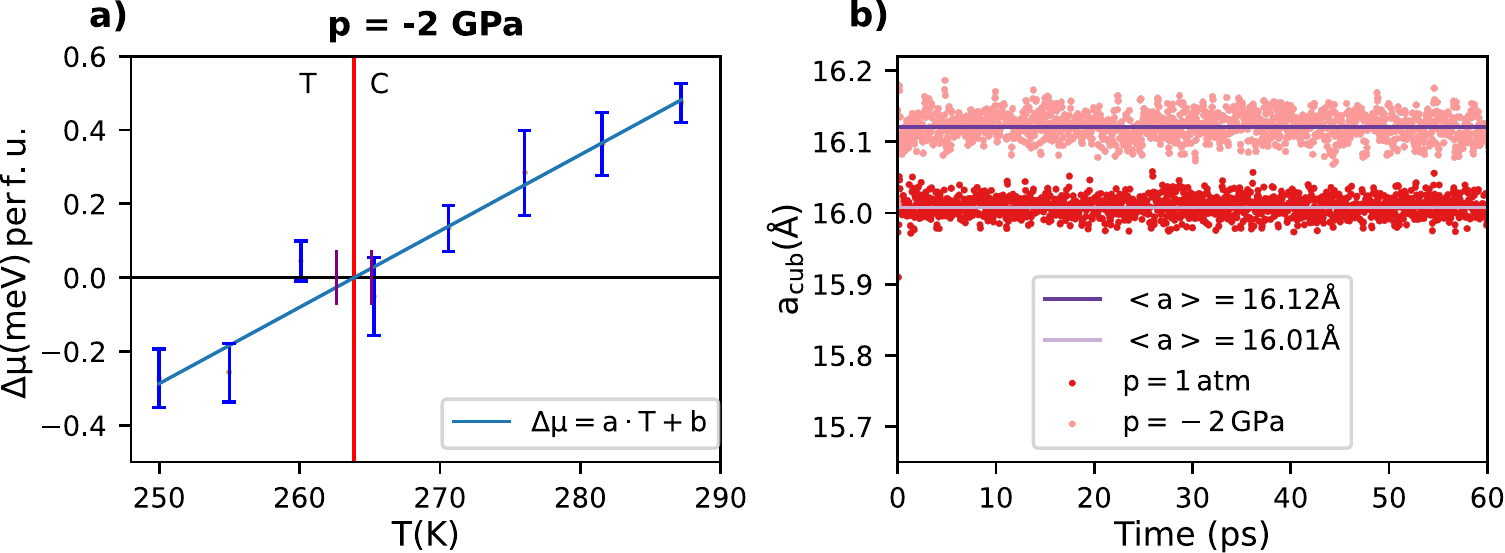}
    \caption{Results of the MD simulations with an externally applied negative pressure $p = -2\,$GPa. Panel a) shows the chemical potential differences across the T-C transition, obtained via NST fully flexible simulations and with an estimated Curie temperature of $T_{\mathrm{c}} = \SI{264 \pm 1}{\kelvin}$, as opposed to \SI{182.4 \pm 0.7}{\kelvin} calculated at ambient pressure (see sec. \ref{sec:thermo}). Panel b) shows instead a comparison between the time evolution of the lattice constant in NpT simulations \npjcorr{at 250 K} with negative pressure and ambient pressure. These show how the effect of a negative pressure slightly increases the average \npjcorr{unit-cell lattice parameter} (by only \SI{0.7}{\percent}), due to the large bulk modulus of \batiooo. This has however important consequences on the relative stability of T and C phases, with a shift of the Curie point of \SI{82}{\kelvin} with respect to the ambient pressure estimate.  Error bars have the same meaning as in Figure~6.}
  \label{fig:negpress}
\end{figure*}

To confirm that pressure can significantly affect the transition temperature, in Figure~7 we investigate the sensitivity of the Curie temperature to negative pressure ($p = \SI{-2}{\giga\pascal}$) within our ML framework. 
We observe a shift of the Curie temperature that is increased by a significant \SI{82}{\kelvin} with respect to the ambient pressure estimate via NST simulations, while a very small variation in the \npjcorr{lattice constant} of the MD supercell (\SI{0.7}{\percent} at \SI{250}{\kelvin}) is seen in NpT. We note that \npjcorr{the corresponding} change in \npjcorr{the volume (\SI{2.1}{\percent})} is within the variation of calculated volumes of cubic~\bto{} with different DFT exchange-correlation functionals\cite{Kotiuga_2021}.
Furthermore, experimental data\cite{fischerElasticityCaTiO31993} on the elastic properties of~\bto{} show that the bulk modulus of~\bto{} is in the range of $\approx \SI{200}{\giga\pascal}$, implying that the action of a relatively high pressure would result in a small change in the cell volumes while completely modifying the free-energy landscape. This effect induces the Curie temperature shift. \npjcorr{We note in this respect that the PBEsol functional, that we used to compute energies and atomic forces of the training set structures, predicts a slightly underestimated lattice constant - $\SI{4.0}{\angstrom}$ at $400\,$K from Car-Parrinello MD \cite{Kotiuga_2021} as compared to the experimental value of $\SI{4.012}{\angstrom}$ of cubic \batiooo \cite{kay1949}. This is a minuscule underestimation, that is however not negligible in the calculation of these tiny free-energy barriers. Moreover, we also rule out that the main source of discrepancy with the experimental transition temperatures might arise from an incorrect prediction of thermal expansion (see Supplementary Figure~11), as previously shown in Tinte \textit{et al.}\cite{tinteQuantitativeAnalysis2003}, where the underestimation of the critical temperatures by an effective Hamiltonian model was related to the approximations made in the construction of the PES. }

This sensitivity of the relative free energies on the equilibrium volume shows how it is possible to tune the applied pressure to obtain a better agreement with experiment. Moreover, while the use of an external pressure is a common strategy to improve the accuracy of \abinitio{} MD -- including in the recent ML-driven study by Xie \textit{et al.}\cite{xie_ab_2022}, to correct the so-called supertetragonality problem -- our strategy opens up other avenues for improvement.  For instance, since our ML potential is trained on a relatively small set of \num{1458} self-consistent energy calculations, one could systematically test more accurate and demanding electronic structure approximations \cite{Liechtenstein1995,Dudarev1998, Perdew2001, Becke1993}, by running them on the existing dataset, to improve the quantitative agreement between simulations and experiment.
In the following, we show that the error committed by the GAP in energy predictions translates into very small free-energy errors, that provide no shift in the estimated critical temperatures.


So far, we have shown the capability of the GAP of both qualitatively describing the emergence of ferroelectric states in \bto{} and reproducing the correct phase sequence. Seeing however the substantial disagreement of the critical temperature predictions of the ML model with the experiments, we shall now assess its accuracy as compared to the underlying DFT method. 
A \npjcorr{compelling} test in this direction is provided by the free-energy perturbation (FEP) method. From the collected MD trajectories we extract a validation set of 50 tetragonal and cubic structures, just below (\SI{170}{\kelvin}) and above (\SI{194}{\kelvin}) the Curie point and recompute their energies with self-consistent DFT calculations. 
This allows us to compute how the error of the
GAP-predicted energies on the test set propagates to the error of the chemical potential estimate at a given temperature.

The FEP on the chemical potentials is first computed as a correction on the Gibbs free energy $G^k$ of phase $k = \mathrm{T, C}$: 
\begin{equation}
\label{eq:FEP-Gibbs}
    \Delta G^k_{\mathrm{FEP}} (T) = - k_{\mathrm{B}} T \ln \, \biggl< \, \exp \left(- \frac{E^k_{\mathrm{GAP}} - E^k_{\mathrm{DFT}}}
 {k_{\mathrm{B}} T} \right) \, \biggr>
\end{equation}
where $\langle \cdot \rangle$ represents the average over the test set structures, $E^k_{\mathrm{GAP}} - E^k_{\mathrm{DFT}}$ the deviation between GAP and DFT total energies in phase $k$, and $T$ the temperature. 
The FEP on the Gibbs free energies can then be translated into a correction on the chemical potential differences as follows: 

\begin{equation}
\label{eq:FEP-chempot}
    \mu^k_{\mathrm{FEP}} = \frac{G^k_{\mathrm{GAP}} + \Delta G^k_{\mathrm{FEP}}}{N},
\end{equation}
where $N$ is the number of atoms. 
Equation~\eqref{eq:FEP-Gibbs} represents an average of Boltzmann factors: if the energy deviations between the DFT and GAP estimates are small compared to the thermal fluctuations at temperature T for both the tetragonal and cubic phases, the correction on the corresponding chemical potential is negligible, due to the exponential factors. This propagation of errors can however become significant or even dominant if the energy deviations are of the same order of magnitude or larger than the thermal fluctuations. 

Panel (f) of Figure~6 shows the effect of the FEP correction on the estimate of the chemical potentials for the two selected temperatures. The GAP shows good performance in the prediction of both tetragonal and cubic structures (compared to $k_{\mathrm{B}} T$) and the FEP correction is one order of magnitude smaller than the actual prediction of $\Delta \mu$ and is still well within the error bars computed with the MD runs. The correction is hence negligible and no shift in the Curie point is observed, providing \npjcorr{strong numerical evidence} of the DFT accuracy of the GAP in free-energy predictions.

\subsection{Dielectric response of~\btotitle{}}
\label{sec:dielec-resp}

Let us now turn our attention to using the polarization model developed and described in Section~\ref{ssec:ml-pol-model} to compute experimentally measurable quantities.  As previously mentioned in Sharma \textit{et al.}\cite{sharmaDipolarCorrelations2007} and elsewhere in the literature on the modern theory of polarization~\cite{restaTheoryPolarization2007,Spaldin2012}, the polarization of a condensed-phase system is well defined only modulo the quantum of polarization; however, we can still  
compute experimentally observable quantities as changes and fluctuations in its value.


\begin{figure}
    \centering
    \includegraphics[width=\columnwidth]{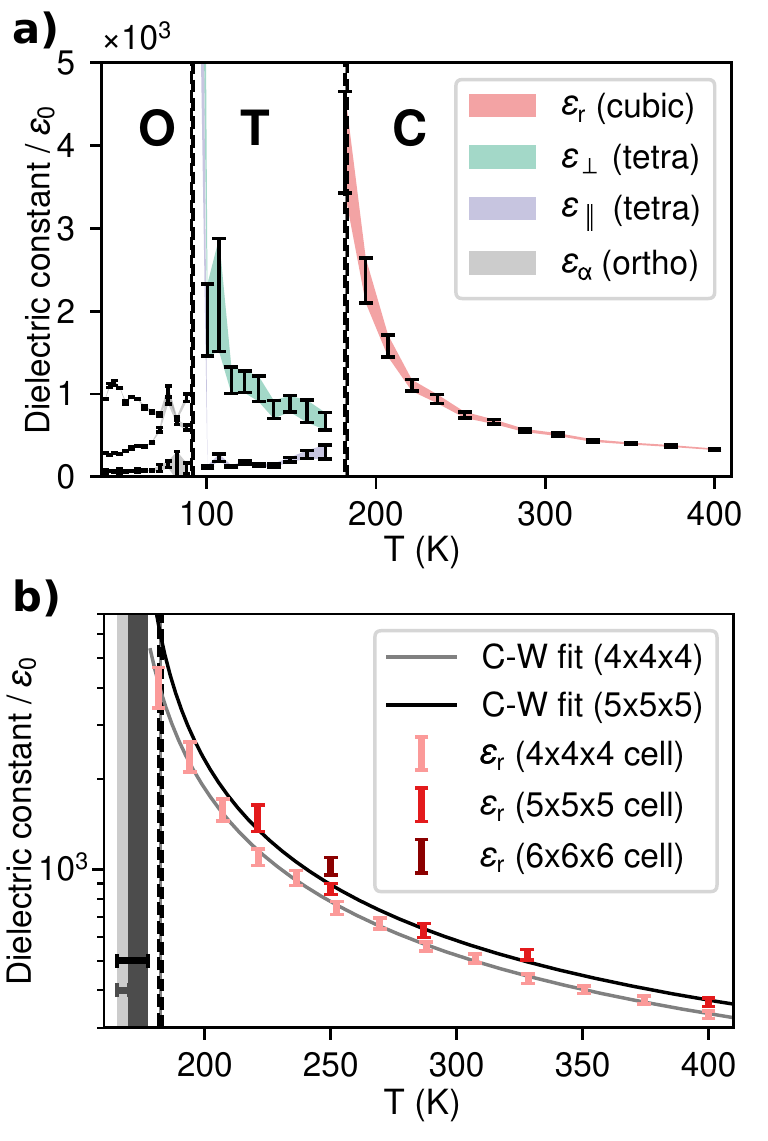}
    \caption{(a) Temperature dependence of the static dielectric constant computed for a \sfour{} supercell \npjcorr{across the orthorhombic, tetragonal, and cubic phases.  The phase transition temperatures computed from the chemical potential (see Table~\ref{tab:phasetemps}) are shown as vertical dashed lines.  (b) Supercell-size comparison of dielectric constants computed in the cubic phase, including fits} to the Curie-Weiss law \npjcorr{Equation~\eqref{eq:curie-weiss-const-offset} -- note the log scale on the $y$-axis}.  The effective Curie temperature $T_{\mathrm{c},\varepsilon}$ predicted by each fit is shown by \npjcorr{the shaded vertical spans and the error bars towards the bottom left of the figure.  Error bars show one standard deviation and are computed as described in Supplementary Note 6.}}
    \label{fig:dielectric-temp-dep}
\end{figure}

The first of these experimentally relevant quantities is the static dielectric constant, which can be computed directly from the fluctuations of the system's total dipole~\cite{sharmaDipolarCorrelations2007}\npjcorr{. In the cubic phase:}
\begin{equation}
    \varepsilon_r = 1 + \frac{\left<M^2\right>}{3\varepsilon_0 V k_\mathrm{B} T} \npjcorr{+ \varepsilon_\infty}
    \label{eq:static-dielectric-const}
\end{equation}
where both the total dipole magnitude $M$ and the vacuum permittivity $\varepsilon_0$ are expressed in SI units \npjcorr{and} the average value of the cell dipole $\left<\mathbf{M}\right>=0$ by symmetry.  \npjcorr{The optical (electronic) dielectric constant $\varepsilon_\infty$ from both measurements and calculations\cite{zhon+95prb,ghosezLatticeDynamics1997,hashimotoDielectricProperties2015} is in the range of \SIrange{5}{6}{}, which is much smaller than the typical range of $\varepsilon_r$ we calculate for this material, so this term will be neglected in the following analysis.  In any case, the analyses below are nearly or completely insensitive to such a small constant shift.}
\npjcorr{For non-cubic phases, we must modify the equation to subtract off the (now non-zero) average polarization, replacing $\left<M^2\right>$ with $\left<M_\alpha M_\beta\right> - \left<M_\alpha\right>\left<M_\beta\right>$, where $\alpha$ and $\beta$ are Cartesian components of the total dipole vector\cite{macdowellDielectricConstant2010}.  Notably, since the non-cubic phases have anisotropic structure, the dielectric \emph{tensor} $\varepsilon_{r,\alpha\beta}$ will also generally be anisotropic.  Indeed, experimental measurements on single-domain crystals of \bto{} have shown a pronounced dielectric anisotropy especially in the tetragonal phase\cite{Merz1949,liDielectricElastic1996}.}

\npjcorr{Comparison of our results with experiments is complicated by the dramatic variation in the} measured value with temperature, composition, and grain size \cite{ostapchukGrainsizeEffect2006,davisDielectricProperties1953,chuTemperatureDependence1992}.  We \npjcorr{therefore study} the temperature dependence explicitly, as shown in Figure~8.  \npjcorr{The calculated values for the orthorhombic, tetragonal, and cubic phases agree qualitatively with the calculations of Hashimoto and Moriwake\cite{hashimotoDielectricProperties2015}, which used a similar computational methodology but with a shell-model potential, as well as with measurements on single-domain crystals\cite{Merz1949,liDielectricElastic1996}.  In the tetragonal phase, we see the expected strong anisotropy between the components parallel ($\varepsilon_\parallel$) and perpendicular ($\varepsilon_\perp$) to the polarization axis -- as we can already see from Figure~1, the polarization fluctuations in the tetragonal phase are strongly suppressed along the polarization direction, which matches the much smaller value of $\varepsilon_\parallel$ seen here.  In the orthorhombic phase, the experimental measurements are averages over different domains and thus do not show the same pattern of anisotropy -- namely, the splitting into three separate principal components -- seen here, but this splitting is present in Hashimoto and Moriwake\cite{hashimotoDielectricProperties2015}.}

\npjcorr{In the cubic phase, }the expected temperature dependence follows a version of the Curie-Weiss law~\cite{ostapchukGrainsizeEffect2006}:
\begin{equation}
    \varepsilon_r = \frac{C}{T-T_{\mathrm{c},\varepsilon}} + \varepsilon_{T\rightarrow\infty}
    \label{eq:curie-weiss-const-offset}
\end{equation}
where $T_{\mathrm{c},\varepsilon}$ is the (dielectric) Curie temperature, which should -- in the limit of infinite system size and statistical sampling -- agree with the tetragonal-cubic phase transition temperature computed above, $T_{\mathrm{c,C\!-T}}=\SI{182.4}{\kelvin}$.

From the temperature dependence data in Figure~8, we determine the best fit parameters for the \sfour{} cell data to be $\varepsilon_{T\rightarrow\infty}=\npjcorr{\num{90\pm18}}$, $T_{\mathrm{c},\varepsilon} = \npjcorr{\SI{167.4\pm2.3}{\kelvin}}$, and $C=\SI{57100\pm3600}{\kelvin}$.  The most important discrepancy to note here is that the Curie point predicted by this fit is still about \npjcorr{\SI{15}{\kelvin}} lower than the thermodynamic phase transition temperature predicted for ambient pressure in Sec.~\ref{sec:thermo}.  This discrepancy is \npjcorr{likely a result of finite-size effects} due to the small supercell\npjcorr{, which are known to broaden and shift critical points\cite{binderFiniteSize1987}.}  The \sfive{} \npjcorr{fit, on the other hand,} yields $\varepsilon_{T\rightarrow\infty}=\npjcorr{\num{99\pm18}}$, $T_{\mathrm{c},\varepsilon} = \npjcorr{\SI{171\pm6}{\kelvin}}$, and $C=\npjcorr{\SI{62200\pm2500}{\kelvin}}$: the Curie temperature $T_{\mathrm{c},\varepsilon}$ is \npjcorr{slightly} closer to the predicted phase transition temperature $T_\mathrm{c,C\!-T}$\npjcorr{, which is now within the \SI{95}{\percent} confidence interval of the fit parameters}.  However, even with the larger supercell, we still note a discrepancy from the parameters determined by fits to experimental data\cite{rupprechtDielectricConstant1964,ostapchukGrainsizeEffect2006} -- namely, the Curie-Weiss constant $C$ is under-predicted by a factor of about 2 with respect to experiment. This difference could be due to approximations inherent in the underlying DFT functional, either directly or indirectly due to the underestimation of the phase transition temperatures.  We test this hypothesis in more detail in the following section by investigating the negative-pressure simulations.


The equation for the static dielectric constant, Equation~\eqref{eq:static-dielectric-const}, is in fact only the zero-frequency limit of the whole frequency-dependent response function.  We can compute the frequency-dependent susceptibility (and thus the relative dielectric constant) via linear response theory, from the one-sided Fourier transform of the dipole-dipole autocorrelation function~\cite{lofflerFrequencydependentConductivity1997,frenkelUnderstandingMolecular2002} \npjcorr{(again for the cubic phase)}:
\begin{align}
    \chi(\omega) &= \frac{1}{3\varepsilon_0 V k_B T}\Big(\left<M^2\right>\nonumber\\
      &\qquad{} -\mathrm{i}\omega\int_0^\infty \left<\mathbf{M}(\tau)\cdot \mathbf{M}(t + \tau)\right>e^{-\mathrm{i}\omega t}\mathrm{d}\;t\Big)\nonumber\\
      &= (\varepsilon_r - 1)\left(1 - \mathrm{i}\omega\int_0^\infty\tilde{C}_{MM}(t)e^{-\mathrm{i}\omega t} \mathrm{d}\;t\right)
    \label{eq:dielectric-freq-dep}
\end{align}
where $\tilde{C}_{MM}(t) = \frac{1}{\left<M^2\right>} \left<\mathbf{M}(0)\cdot\mathbf{M}(t)\right>$ is the normalized dipole-dipole autocorrelation function and $\varepsilon_r$ is the \emph{static} dielectric constant computed from Equation~\eqref{eq:static-dielectric-const}.

\begin{figure}
    \centering
    \includegraphics[width=\columnwidth]{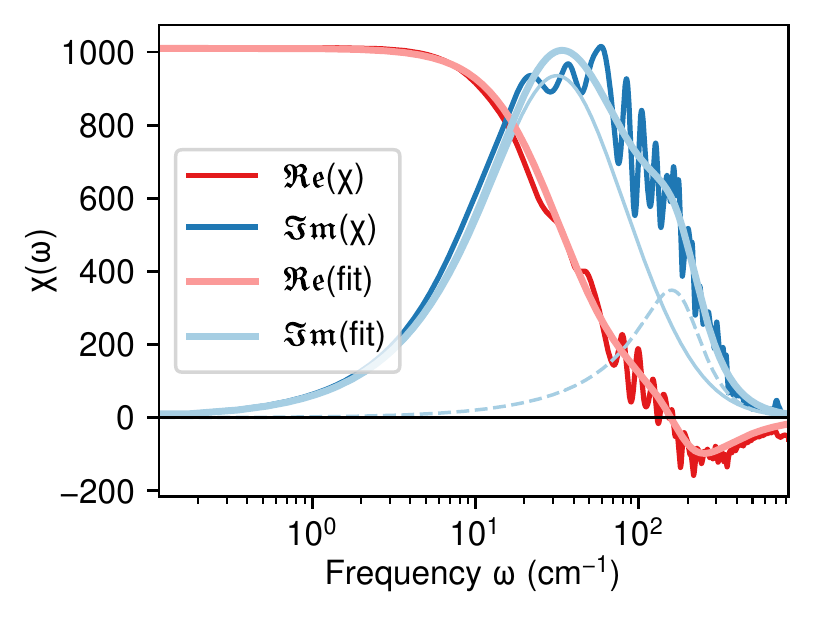}
    \caption[stupid latex]{Frequency-dependent dielectric response of cubic \batiooo{} at \SI{250}{\kelvin} computed with a \ssix{} supercell.  The imaginary part of the spectrum is fitted to a sum of two simple harmonic oscillator response functions, as in Ponomareva \textit{et al.}\cite{ponomarevaTerahertzDielectric2008}; the two separate oscillator responses are shown in thin light blue lines, while the fitted sum is shown as a thick light blue line. 
    The real part of the response spectrum is also shown together with the corresponding real part of the fitted function.}
    \label{fig:pol_freq_dependence}
\end{figure}

We show the frequency-dependent susceptibility for a \ssix{} supercell trajectory at \SI{250}{\kelvin}, computed using Equation~\eqref{eq:dielectric-freq-dep}, in Figure~9.  In general, we see the same structure as predicted for the high-temperature cubic phase by both theoretical effective-Hamiltonian MD calculations\cite{ponomarevaTerahertzDielectric2008} and observed experimentally\cite{ostapchukGrainsizeEffect2006}, namely, that of a large absorption peak corresponding to the soft-mode (TO$_1$) phonon frequency. \npjcorr{Note the slight negative dip in the real dielectric constant is expected and seen in many previous observations\cite{ponomarevaTerahertzDielectric2008,macdowellDielectricConstant2010,liDielectricElastic1996}.  This does not imply that the real or imaginary part of the \emph{refractive index} $n=\sqrt{\epsilon_r}$ is anywhere negative.} It was previously proposed\cite{luspinSoftMode1980,vogtSoftmodeSpectroscopy1982} that the 'soft-mode' part of the absorption spectrum of \batiooo{} could be described with a single, strongly damped harmonic oscillator of the form
\begin{equation}
    \chi(\omega) = \frac{A\omega_0^2}{\omega_0^2 - \omega^2 + \mathrm{i}\gamma\omega}
    \label{eq:sho-response}
\end{equation}
with amplitude $A$, damping constant $\gamma$, and \emph{resonant frequency} $\omega_0$ (which is always larger than the actual apparent \emph{peak} frequency).  However, a later study~\cite{prestingModeSoftening1983} uncovered possible inadequacies of this single-oscillator model especially in the high-frequency range ($\omega_0 \approx \SI{100}{\per\centi\meter}$), and suggested a two-oscillator model as a possible replacement, though it was not yet justified by the available experimental data.

More recently, Ponomareva \textit{et al.}\cite{ponomarevaTerahertzDielectric2008} both measured high-accuracy infrared spectra and computed theoretical spectra from MD simulations of the effective Hamiltonian model of Walizer \textit{et al.}\cite{walizerFinitetemperatureProperties2006}, and they found strong evidence that the spectrum indeed is best modeled by \emph{two} harmonic oscillators.  The computed spectrum from our model at \SI{250}{\kelvin}, shown in Figure~9, further supports this picture:  we also find that the imaginary part of the spectrum could only be satisfactorily described with two oscillators, although with different parameters from those calculated in Ponomareva \textit{et al.}\cite{ponomarevaTerahertzDielectric2008}: we find one oscillator with fundamental frequency $\omega_1 = \SI{86}{\per\centi\meter}$ and damping ratio $\sfrac{\gamma_1}{\omega_1} = \num{3.0}$, and another with fundamental frequency $\omega_2 = \SI{187}{\per\centi\meter}$ and damping ratio $\sfrac{\gamma_2}{\omega_2} = \num{1.1}$.  Comparing these parameters with those calculated in\cite{ponomarevaTerahertzDielectric2008}, we find both frequencies to be rather high\npjcorr{, so our agreement with their results remains mostly qualitative for now.}

On the one hand, \npjcorr{the differences we observe} could be due to the large oscillations and lack of resolution at high frequencies due to the limited sampling time imposed by the relatively large computational cost of our model.  However, it is more likely that both these discrepancies have the same origin as the underestimation of the phase transition temperatures discussed above -- either inaccuracies in the underlying DFT model or some other effect not yet accounted for.  As noted in Sec.~\ref{sec:thermo}, the phase transition temperatures can be compensated by applying a negative pressure.  Indeed, Ponomareva \textit{et al.}\cite{ponomarevaTerahertzDielectric2008} associate the higher-frequency mode with short-range correlations between (mostly) neighbouring unit cell dipoles, so it is likely that this frequency shift has the same origin as the pressure effect.

To investigate this discrepancy further, and to assess the effect of negative pressure on the dielectric response, we compute frequency-dependent susceptibility spectra for all the negative-pressure simulations previously run for Sec.~\ref{sec:thermo} (specifically, Figure~7), where the material remained in the cubic phase.  The spectra are also compared to those derived from ambient-pressure simulations, specifically those used to compute the temperature dependence of the dielectric constant in Figure~8.  The comparison is shown in Figure~10.  On the one hand, we see the main peak shifting towards higher frequencies as the temperature increases, as expected from previous theoretical and experimental studies\cite{vogtSoftmodeSpectroscopy1982,ponomarevaTerahertzDielectric2008}.  On the other hand, we also see the peak shifting towards \emph{lower} frequencies when a negative pressure is applied at any given temperature.  While the peak frequencies for the negative-pressure simulations still do not match experimental data for the same temperatures, the shifts are in the right direction.

Furthermore, all simulations show a small narrow peak or edge at around \SI{340}{\per\centi\meter}, independent of the temperature. The frequency of the mode does depend on pressure, but due to the large bulk modulus of \bto{} the mode shifts very little: only about \SI{7}{\per\centi\meter} under \SI{-2}{\giga\pascal} of pressure.

\npjcorr{Although this peak likely represents a feature of our model and not just a simulation artefact, we do not yet have enough information to} confidently identify \npjcorr{this peak with known vibrational modes of \bto{}\cite{luspinSoftMode1980,ostapchukGrainsizeEffect2006,hlinkaInfraredDielectric2006}.}

In fact, the difficulties we encounter here in reproducing the results of simpler, experimentally accurate -- but empirically adjusted -- models are reminiscent of the difficulties encountered previously, e.g. in Veit \textit{et al.}\cite{veit+19jctc}, in applying more accurate (in the sense of reproducing the quantum PES) ML potentials that must in turn account for more accurate physics, such as many-body dispersion and quantum nuclear effects, in order to arrive at the right predictions for the right reasons.  Rather than being a deficiency in the machine learning simulation approach, we see this as an opportunity to discover interesting physical behaviours and mechanisms that were overlooked before.

The calculations presented here are a promising first step towards using the ML PES and polarization framework as a generally applicable tool to predict experimentally relevant \emph{response} properties.  \npjcorr{This tool will be a valuable future asset for} investigating new candidate ferroelectric materials or gaining more insight into the underlying behaviour of existing ones.

\begin{figure}
    \centering
    \includegraphics[width=\columnwidth]{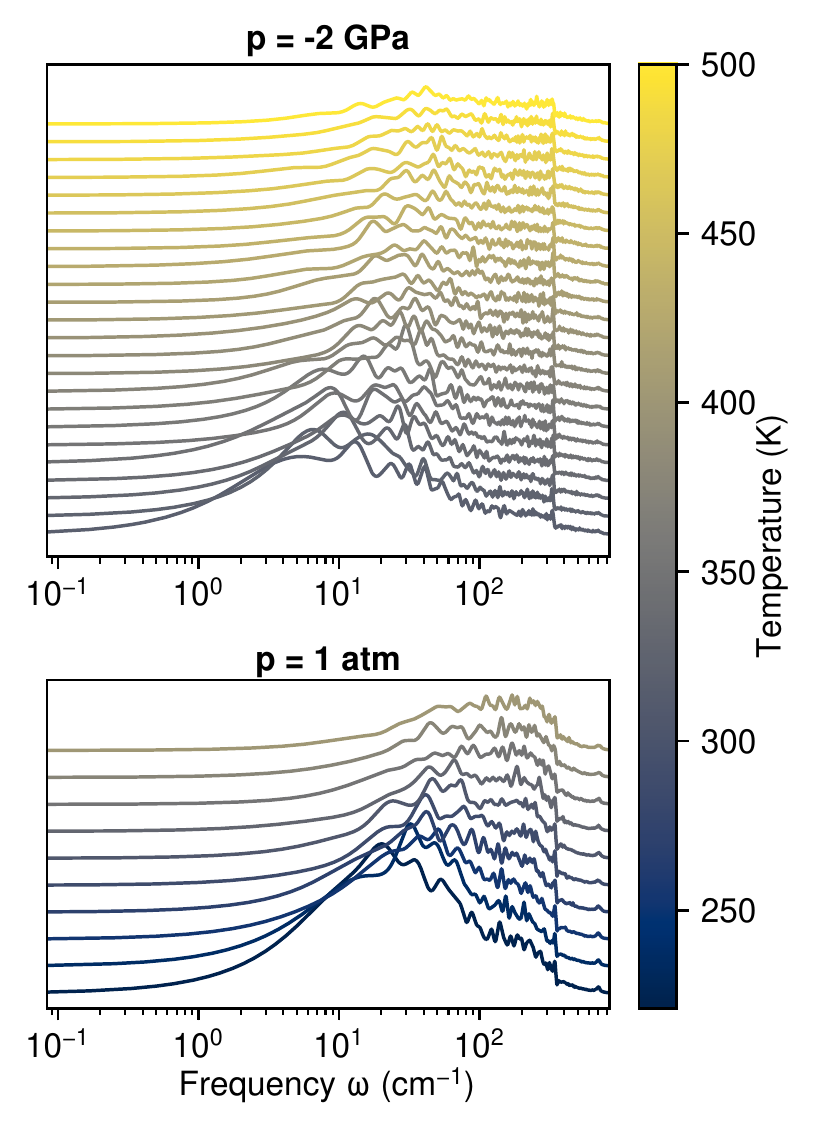}
    \caption{Temperature and pressure dependence of the imaginary part of the dielectric response spectrum, all computed in the cubic phase on a \sfour{} supercell.  For each of two pressures, spectra computed at different temperatures are displayed on the same graph, vertically offset from one another for clarity.  The units of the y-axis are arbitrary, though all spectra on the same plot have the same scale.  While the increase in temperature broadens the main peak and shifts it to higher frequencies, the negative pressure instead shifts the main peak to lower frequencies for a given temperature (i.e. closer to experiment, as well as the theoretical calculations of Ponomareva \textit{et al.}\cite{ponomarevaTerahertzDielectric2008}).}
    \label{fig:spectra_under_pressure}
\end{figure}

\section{Discussion}
\label{sec:conclusions}

In this work, we introduce a modern, general ML framework to describe at once the finite-temperature and functional properties (dielectric response) of perovskite ferroelectrics, and apply it specifically to model barium titanate (\batiooo{}). This framework \npjcorr{matches} the accuracy of the underlying DFT method and does not require to preselect a given effective Hamiltonian model\cite{zhongFirstprinciplesTheory1995, garciaTemperaturedependentDielectric1998}.
The simulations made possible by this framework recover the correct R-O-T-C phase ordering in fully flexible simulations and allow to investigate the emergence of Ti off-centerings. In particular, we highlight the driving mechanism of the ferroelectric transition, showing how the presence of these off-centered displacements gives rise to a low-temperature ferroelectric phase, due to a long-range dipolar ordering. Moreover, the interplay between the displacements and the cell deformations leads to the emergence of intermediate tetragonal and orthorhombic phases.
We further proceed to reconstruct the thermodynamics of \bto{} (see Sec. \ref{sec:thermo}), by means of a two-component, polarization-derived, CV. 

Finally, we apply the ML polarization model to calculate dielectric response properties of experimental interest, including the static and frequency-dependent dielectric constants, and investigate their dependence on temperature.
While we do not reach quantitative agreement with experimental measurements for many of the properties computed here, we see several clear, systematic pathways to improving the model potential and its predictions, such as including long-range electrostatic effects, simulating larger system sizes, as well as addressing the
possible deficiencies in the underlying DFT model for both energies and polarizations. We expect that such improvements will allow us to reach quantitative agreement with the experiments. Our results obtained with negative pressure calculations and the FEP show how this discrepancy can in fact be traced back in part to the sensitivity of the transition temperatures to cell volume combined with the deviation of the DFT cell volume from the experimental one. This effect suggests that a more in-depth investigation of the effects of pressure -- which is well known to influence the onset of ferroelectricity -- could provide further insights into the deviation from experiments.
A closer agreement could also be obtained by combining, as recently proposed, different DFT schemes to describe simultaneously energy, structure, and electronic density of perovskites oxides \cite{williamsCombiningDensity2020}. 

We also plan to make improvements on the model performance by means of the feature sparsification technique, as detailed in Musil \textit{et al.}\cite{musi+21jcp}. The latter has proven to reduce the computational cost (in energies and force predictions) by a factor of 3 or 4 for realistic systems and, in combination with larger-scale parallelization techniques, will allow us to treat larger, more complex systems.

Importantly, this ML framework automates the construction of a model of the PES and the polarization and can then be used to investigate finite-temperature properties in detail and with first-principles accuracy. Since the ML-PES was made with no explicit assumption on the functional form of the underlying PES and no prior definition of the relevant degrees of freedom of the system, this strategy is generalizable to other materials, to study, e.g., 2-D ferroelectrics \cite{zhangStructuralPhase2020} and solid solutions with variable stoichiometries, that are known to possess different and more complex ferroelectric states. For instance,  $\mathrm{Ba}_x \mathrm{Sr}_{1-x} \mathrm{Ti}_y \mathrm{Zr}_{1-y} \mathrm{O}_3$ is known to display a rich phase diagram, depending on composition, and shows both ferroelectric and relaxor ferroelectric phases\cite{maitiStructurePropertyPhase2008}.  Furthermore, the framework developed is easily applicable to study the role of nuclear quantum effects, for instance in incipient ferroelectrics such as \ce{SrTiO3} and  \ce{KTaO3} where quantum fluctuations appear to suppress the ferroelectric state\cite{zhon-vand96prb,akbarzadehAtomisticSimulations2004}. Further extensions of this framework include the investigation of the role of a finite electric field in the MD and its effect on the polarization. This will allow us to simulate, for instance, hysteresis loops, which are key to measure the energy storage of ferroelectric devices.

In conclusion, we have shown how a comprehensive, data-driven modeling framework for a perovskite ferroelectric material, based on DFT reference data, can capture the mechanisms of the ferroelectric transition, as well as make predictions of thermodynamic and functional properties with first-principles accuracy.  The work opens the door for a new avenue of fruitful research into the understanding and characterization of known ferroelectric materials, as well as the discovery and design of new candidate compounds with improved industrially relevant properties.

\section{Methods}
\label{sec:Methods}

In Sec.~\ref{ssec:symm-adapted} we summarize the construction and properties of the symmetry-adapted features; a more thorough discussion of this family of features and an introduction to the notation we use here is given in Section 3 of Ref.~\citenum{musi+21cr}.
With these features defined, we detail how the potential energy surface and the polarization models are constructed in Secs.~\ref{ssec:ml-pes} and~\ref{ssec:ml-pol-model}, respectively.
Turning our attention to the specifics of modeling~\bto, we report the training and validation of our ML model for~\bto{} in Sec.~\ref{ssec:training}, and in Sec.~\ref{ssec:ord-param} we develop physically-inspired order parameters which we use to characterize and interpret our results (see Sec. \ref{sec:thermo}). In Sec. \ref{sec:comp-details} we report the computational details on the ML-MD simulations.

\subsection{Symmetry-adapted features}
\label{ssec:symm-adapted}
\npjcorr{To construct the family of features that are relevant for this paper, we make use of the atom-centered density correlation framework \cite{will+19jcp}. The starting point is the definition of a set of features, namely $\rep<anlm||A;\rho_i>$, from an expansion of the atomic density for an environment $i$ of structure $A$, as in Equation~(31) of Willatt \textit{et al.} \cite{will+19jcp}.  The different indices in the bra identify the chemical species ($a$), radial function ($n$) and angular momentum $(l,m)$, the latter being especially important to track the symmetry of the features.}

Symmetry-adapted descriptors can be obtained as a symmetrized average (referred to by an overline decoration) of the  tensor product of $\nu$ sets of expansion coefficients, resulting in density-correlation features $\rep<q||\frho[\lambda\mu]_i^{\nu}>$.
While the generic index $q$ only enumerates the features, the other indices encode the physical meaning of these descriptors.  There are two fundamental parameters: (a) the body-order exponent $\nu$, which indicates that the features describe the relative position of $\nu$ neighbors of the central atom, and (b) the $\lambda, \mu$ coefficients, which
determine how the descriptor transforms under rotations -- namely as spherical harmonics $Y^\mu_\lambda$.  
This framework allows us to build features that are not only \emph{invariant} to rotations, but also explicitly \emph{covariant} (more generally called \emph{equivariant}) features of any tensor order.  
Such equivariant features were first introduced by Glielmo \textit{et al.}\cite{glie+17prb}, for vector features, and in Grisafi \textit{et al.}\cite{gris+18prl} for tensors of arbitrary order.  Equivariant features are now gaining considerable popularity, especially for graph convolutional neural networks to predict scalar and tensor properties~\cite{andersonCormorantCovariant2019,batznerSEEquivariant2021,schuttEquivariantMessage2021,qiaoUNiTEUnitary2021,parkAccurateScalable2021}.
In this work we only deal with spherical invariants or SOAP descriptors\cite{bart+13prb} \npjcorr{- corresponding to $\lambda = 0, \mu = 0$ -} and $\lambda = 1, \mu = (-1, 0, +1)$ \npjcorr{features}, representing spherical equivariants of order $1$.
For instance, SOAP power spectrum features, which are invariant under rotations, are obtained from the contraction of two sets of coefficients ($\nu = 2)$:
\begin{equation}
    \npjcorr{\rep<an; a'n'; l||A;\frho[00]_i^2> \equiv} \\
    \npjcorr{\sum_m \rep<anlm||A;\rho_i> 
     \rep<A;\rho_i||a'n'lm>.}
\end{equation}     
These features can thus be written as $\rep<q||A;\frho_i^2>$.

\npjcorr{Similarly, the simplest example of equivariant features only encodes information on the radial distribution of neighbors. They are equivalent to the density coefficients themselves:}
\begin{equation}
 \rep<q\!=\!(an) ||A; \frho[\lambda\mu]_{{i}}^1> =
 \rep<an\lambda \mu||A; \rho_i >,
\end{equation}
An extension of this construction allows one to build symmetry-adapted tensors of arbitrary rank and body-order\cite{niga+20jcp}.

Given that, in order to learn dipole moments and polarizations, we only need the special case of vector-valued features, we find it convenient to exploit the relationship between real-valued spherical harmonics of order $\lambda=1$ and the Cartesian coordinates $\alpha=(x,y,z)$ to define Cartesian equivariants
\begin{equation}
\label{eq:cartesian-equi}
\rep<q ||A; \frho[\alpha\!=\!(x,y,z)]_{i}^{\nu}> \equiv
\rep<q ||A; \frho[\lambda\!=\!1\, \mu\!=\!(1,-1,0)]_{i}^{\nu}>,
\end{equation}
The Cartesian equivariants of Equation~\eqref{eq:cartesian-equi} now explicitly transform as a 3-vector under rotations: 

\begin{equation}
\label{eq:cartesian-equi-transform}
\rep<q ||\hat{R}A; \frho[\alpha]_{i}^{\nu}> = \sum_{\alpha'}
R_{\alpha\alpha'} \rep<q ||A; \frho[\alpha']_{i}^{\nu}>.
\end{equation}

$\hat{R}A$ indicates an arbitrary rotation of a structure A, while $R_{\alpha\alpha'}$ is its representation as a 3$\times$3 Cartesian matrix. We use this family of features to model the polarization of a $\mathrm{BaTiO}_3$ structure and to build an order parameter to distinguish the R-O-T-C phases (see Sec. \ref{sec:thermo}).
We refer the reader to Refs.~\cite{musi+21jcp,musi+21cr} and the documentation of librascal\cite{rascalgithub} for implementation details.

\subsection{Potential energy surface}
\label{ssec:ml-pes}

A Gaussian approximation potential (GAP) is constructed by linear regression of energies $E$ and atomic force components $\left \{\mathbf{f}_r \right \}_{r = 1}^N$, where $N$ is the number of atoms, in the space of the kernels of these descriptors, representing the degree of correlation between the structures.

In order to control the computational cost of the calculation of energies and forces, we also construct a sparse set of representative atomic environments $J$ that are used to define a basis of kernels $k(\cdot, J_j)$ in order to approximate the structure-energy relation. This is discussed further in Sec. \ref{ssec:training}.

We write the target properties as a sum of kernel contributions:

\begin{equation}
\begin{split}
     E & =\sum_{i \in A} \sum_{j \in J} b_j k(A_i, J_j)  \\ 
     \mathbf{f}_r & = -\grad_r E, 
 \end{split}
 \label{eq:ml-pes}
\end{equation}

where the kernel is built as a function of a set of atom-centered invariant features $\rep<q||A_i>$, the index $j$ runs over all environments $J_j$ in the sparse set $J$ and $b_j$ are the weights on each sparse environment to be determined via ridge regression.
Here we use SOAP powerspectrum features, $\rep<q||A; \frho_i^2>$, and 
we compute the kernel between atomic environments as a scalar product raised to an integer power
$k(A_i, A'_{i'}) = (\sum_q \rep<A_i||q>\rep<q||A'_{i'}>)^\zeta$, using $\zeta=4$ here, to introduce non-linear behavior.

\subsection{Polarization model}
\label{ssec:ml-pol-model}

Besides this potential energy model, we construct a fully flexible, conformationally sensitive \emph{dipole moment} surface for the material by employing the symmetry-adapted Gaussian process regression (SA-GPR) framework~\cite{gris+18prl}, previously benchmarked in the context of molecules in Veit \textit{et al.}\cite{veit+20jcp} and proven to extend to the condensed phase in Kapil \textit{et al.}\cite{kapi+20jcp}.  Even though the cell polarization (or dipole) is not uniquely defined in periodic boundary conditions\cite{Spaldin2012,restaTheoryPolarization2007}, we can still make a model for only a single branch of this polarization manifold with suitable pre-processing of the training data, detailed \npjcorr{in Supplementary Note 3}.  This branch choice is essentially equivalent to fixing the polarization to be a single continuous function whose linearization about $\mathbf{P}=0$ is the product of Born effective charges and displacement from some non-polar reference structure, in the spirit of Zhong \textit{et al.}\cite{zhongGiantLOTO1994}.

As with existing SA-GPR approaches, the total dipole of the cell is decomposed into vector-valued atomic contributions.  In analogy to Equation~\eqref{eq:ml-pes}, we express the total dipole $\mathbf{M}$ and polarization $\mathbf{P}$ of a structure $A$ as:

\begin{eqnarray}
    \mathbf{M}(A) &=& \sum_{i \in A} \sum_{j\in J'} \mbf{b}_j \mathbf{k}(A_i, J_j),\\
    \label{eq:dipole-model}
    \mathbf{P}(A) &=& \frac{\mathbf{M}(A)}{V_A}.
    \label{eq:pol_dipole_pervol}
\end{eqnarray}

Our model works with total dipoles rather than polarizations as only the former are size extensive.  A key advantage of this model is that we represent the dipole moment as a sum of atom-centred contributions (effectively, `partial dipoles', in analogy to partial charges), giving us a spatially resolved, atomistic picture of how the different parts of the system contribute to the total polarization. Note that in contrast to the model described in Veit \textit{et al.}\cite{veit+20jcp}, we do not define an additional partial-charge model, since such a model would depend on the choice of the unit cell and be incompatible with the modern theory of polarization.  The only situation in which we use nonzero partial charges is in the linearized effective-charge model used to shift the training-set polarizations to the same branch; these effective charges are not used in the production model.  As already remarked in Sharma \textit{et al.}\cite{sharmaDipolarCorrelations2007} and later in Veit \textit{et al.}\cite{veit+20jcp}, this information can give us a much deeper insight into the physics of the system than predicting the total dipole alone.  
\npjcorr{In this study, we use this information to define \ce{Ti}-centered \emph{unit-cell dipoles} by an appropriate sum of atomic partial dipoles. The dipole of the \ce{Ti} atom is added to the dipoles from neighbouring \ce{O} and \ce{Ba} atoms, with the neighbouring contributions weighted (by $\sfrac{1}{2}$ for \ce{O} and $\sfrac{1}{8}$ for \ce{Ba}) so that the sum of the unit-cell dipoles is still equal to the total cell dipole. These unit-cell dipoles were used to make Figures~1 and~2.}

The model is trained on the same set of structures as the potential energy surface from Section~\ref{ssec:ml-pes}, but uses different training data and, generally, a different sparse set $J'$ each with a different set of weights $\{\mathbf{b}\}$.  These weights take the form of 3-component vectors, corresponding to the kernel $\mbf{k}(A_i, A'_{i'})$, which is now a rank-2 Cartesian tensor (i.e. a 3x3 matrix) for any pair of environments.
This kernel is computed, as in the scalar case, as an inner product of symmetry-adapted features $\rep<q ||A; \frho[\alpha]_{{i}}^2>$
\begin{equation}
\label{eq:symm-adapt-feat}
k_{\alpha\alpha'} (A_i, A'_{i'}) = \sum_q \rep<A; \frho[\alpha]_i^2|| q> \rep<q ||A'; \frho[\alpha']_{{i'}}^2>.
\end{equation}

\subsection{Training the ML model for \ce{BaTiO3}}
\label{ssec:training}

As pointed out in Sec. \ref{ssec:ml-pes}, constructing a GAP model requires defining a representative set of environments, to control the computational cost in evaluating energies and atomic forces of structures. The representative environments should be ideally as diverse as possible so as to provide a good extrapolation across all the phases of interest. Specifically, for our case study of \ce{BaTiO3}, we use Farthest-Point Sampling (FPS) to select a total of $250$ environments centred around barium and titanium atoms, and $500$ around oxygen atoms from the initial training dataset obtained via DFT optimizations (additional details are given below).  

A second crucial parameter is the radial cutoff in the neighbor density $\rho_i(\mathbf{x})$, defined in Sec. \ref{ssec:symm-adapted}. This defines the size of the atomic environment, centered around atom $i$. Choosing large cutoff radii means including more neighbors in the density expansion and allows, in general, a more accurate representation of the environment. This happens however at the expense of increasing the computational complexity. For the purpose of constructing a GAP for \ce{BaTiO3}, we choose a radial cutoff of $\SI{5.5}{\angstrom}$ around each center which is larger than the average separation of first nearest Ti neighbours ($\approx \SI{4.0}{\angstrom}$).  This cutoff allows us to capture the short-ranged Ti--Ti interactions that ultimately result in long-range emergent dipole correlations, a distinctive feature of polarized states in $\mathrm{BaTiO}_3$, as seen in Sec. \ref{sec:BaTiO3_phases}. 

The training dataset is constructed in an iterative fashion, which also means it can be systematically extended. Energies and forces are calculated using DFT as implemented in Quantum ESPRESSO~\cite{Giannozzi2009,Giannozzi2017} with the PBEsol~\cite{Perdew2008} functional, and managed with AiiDA~\cite{Pizzi2016,Huber2020,Uhrin2021}; further details can be found in Supplementary Note 2. An initial training set of $N_0 = \num{518}$ cubic structures (obtained from DFT optimizations with the PBEsol functional) is used to train a preliminary GAP. 
Molecular dynamics simulations with \ipi{}\cite{ipicode} are then performed in all the R-O-T-C geometries and for a total simulation time up to \SI{500}{\pico\second}.
Among all uncorrelated structures thus generated with MD - the correlation being computed via the time-dependent autocorrelation function of the total energy - only the most diverse according to their SOAP descriptors are then selected via FPS and recomputed with DFT self-consistent calculations. 
These are then used to extend the training dataset and refit the GAP, thus restarting the loop and obtaining an increasingly accurate description of the PES.
The final dataset built with this procedure has a total of \num{1458} structures, with an adequate sampling of all the phases of~\bto. Specifically, on top of the initial training set of \num{518} structures, we added \num{100} structures coming from a first round of replica-exchange molecular dynamics (REMD) simulations in the NVT ensemble, and additional \num{840} structures coming from sampling of each of the R-O-T-C phases (\num{210} per phase) in a second round of NpT REMD calculations.

\begin{figure}
    \centering
    \includegraphics[width=\columnwidth]{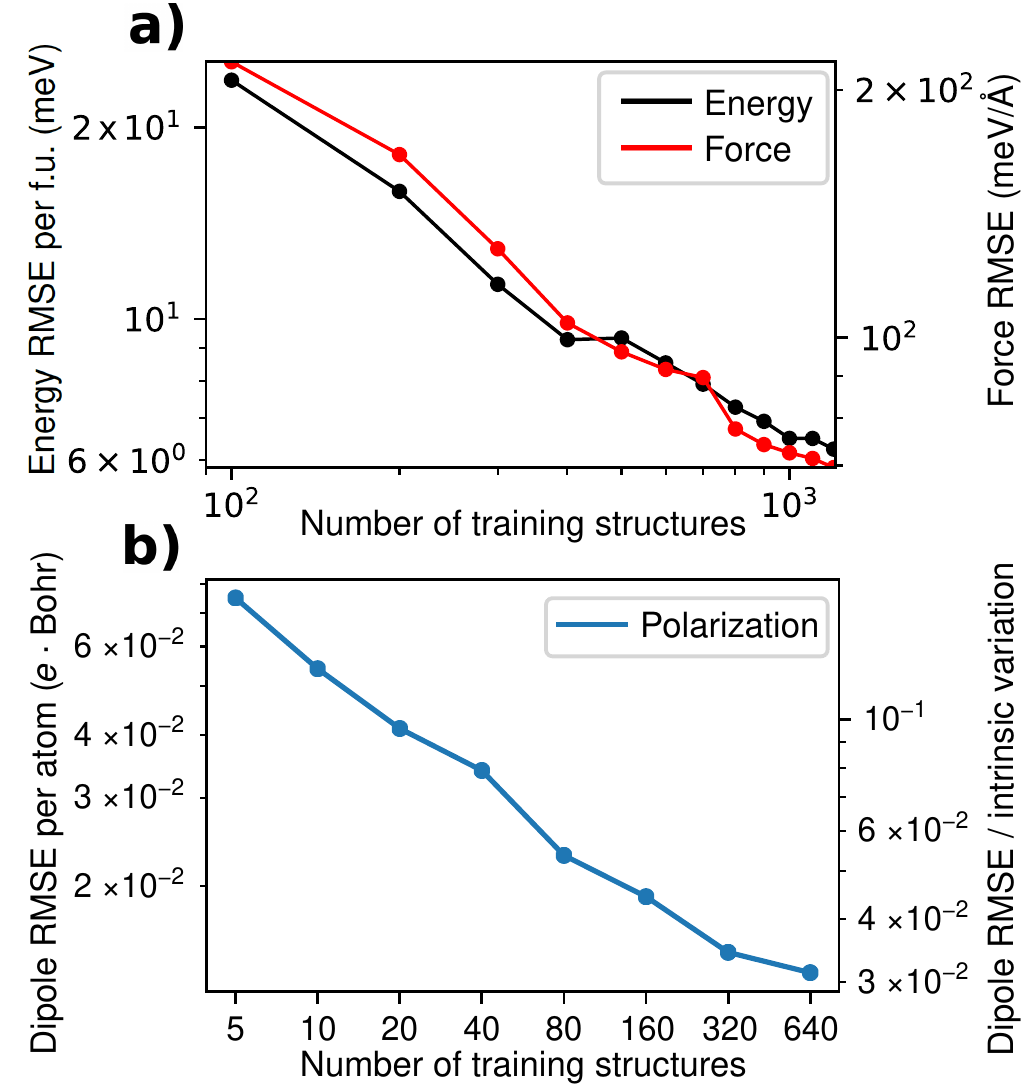}
    \caption{Learning curves of: (a) the GAP model for energies and atomic force components, trained on a total of $1200$ training structures with $258$ randomly selected structures used as validation set, and (b) the SA-GPR polarization model, trained on a total of \num{640} structures with \num{200} validation structures.}
    \label{fig:lc_gap}
    \label{fig:pol_lc}
\end{figure}

The learning curve of the GAP, trained on a total of \num{1200} training structures, is shown in Figure~11a, with \num{258} randomly selected structures used as validation set. The Root Mean Square Error (RMSE) decreases significantly with an increasing number of training points and the final accuracy of the potential in energy estimations is about \SI{6}{\milli\electronvolt} per formula unit (f.u.). This level of accuracy is sufficient to capture several interesting features of the physics of $\text{BaTiO}_3$, including the structural R-O-T-C phases, the presence of needle-like correlations even in the high-temperature paraelectric phase, and to enable predictions of the free-energy surface, that have the same degree of accuracy as the underlying DFT method (see Sec.\ref{sec:thermo}).

The polarization model, in contrast to the GAP, is trained only on the set of $N_\text{T,pol}=\num{840}$ structures sampled from the NpT REMD calculations described above, with \num{210} structures coming from each of the four phases. A total of \num{200} randomly selected structures are withheld for testing; the largest model has therefore been trained with $N_\text{max,pol} = \num{640}$ structures.  
The learning curve of the polarization model, shown in Figure~11b, shows good performance; the largest model ($N=640$ structures) achieves an accuracy \SI{3}{\percent} of the intrinsic variation of the total dipoles in the training set, corresponding to an RMSE of \SI{0.013}{\electron \bohr} per atom, or \SI{0.07}{\electron \bohr} per unit cell -- which is still small compared to the scale of unit-cell polarizations shown, for example,  in Figure~1.

\subsection{Phonon dispersions}
\label{ssec:phon-disp}

\begin{figure}
    \centering
    \includegraphics[width=\columnwidth]{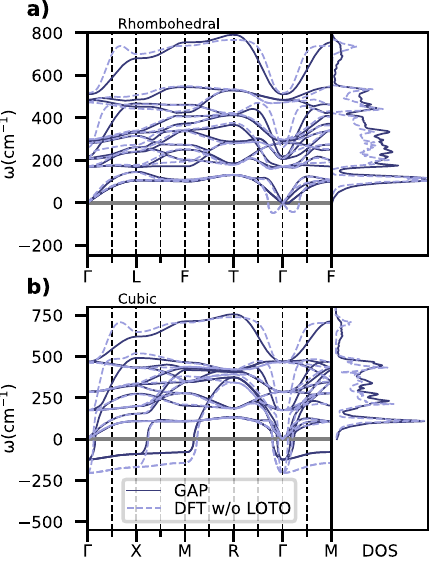}
    \caption{Phonon dispersion and density of states (DOS) of the stable ($\omega>0$) phonons for \bto{} calculated using finite differences with a \sfour{} $q$-mesh for (a) the 5-atom rhombohedral ground state and (b) the 5-atom cubic structure (high-symmetry $k$-point labels from Ref.~\citenum{Hinuma2017}). We compare the GAP predictions with the DFT calculations without long-range electrostatic contributions, i.e., without LO-TO splitting. The vertical dashed lines indicate the explicitly calculated $q$-points.} 
    \label{fig:phonons}
\end{figure}

A crucial test to evaluate the performance of the GAP is to compute phonon spectra and the corresponding density of states (DOS) and compare them with the DFT phonon spectra. In Figure 12, we directly compare the outcome on a \sfour{} $q$-mesh, taking two representative structures as reference: the $5$-atom cubic structure and the rhombohedral ground state, optimized via variable-cell DFT calculations.
The calculations were carried out via the finite difference method using the atomic simulation environment~\cite{larsenAtomic2017} (ASE) for the GAP calculations and phonopy~\cite{phonopy} in conjunction with Quantum ESPRESSO for the DFT calculations. Since no explicit correction for the long-range electrostatics was explicitly taken into account in constructing the ML model, we compare the GAP predictions with the DFT calculations without such contributions. We stress, however, that this contribution due to long-range electrostatic interactions should be included to recover, e.g., the correct LO and TO mode splitting at $\Gamma$ and to stabilize the TA mode of the rhombohedral structure along the  T-$\Gamma$ and $\Gamma$-F paths (see the Supplementary Figure~10 for the DFT dispersion with LO-TO splitting). It has been shown in the work of Libbi \textit{et al.}\cite{Libbi2020} that short-ranged potentials in polar materials can capture the correct phonon dispersions if the appropriate long-range dielectric model is subtracted before fitting the short-ranged potential, and then added back analytically - in analogy to what is done to Fourier interpolate phonon dispersions~\cite{Giannozzi1991}. We also show, in  Supplementary Figure~10, the full phonon spectra once these dielectric contributions are considered.
The spectra show an overall good agreement, especially for the low-frequency acoustic modes, with the most apparent discrepancies occurring for the highest LO mode.
These discrepancies are likely to be caused by two main effects: (a) the training set construction and (b) the locality of the GAP. First, we recall that the interatomic potential is only trained on \stwo{} structures, so that long-wavelength modes that correspond to the periodicity of a \sfour{} cell lie in the extrapolative regime of the potential. Second, the GAP is only sensitive to atomic displacements within the chosen radial cutoff, so phonon modes with a small momentum $q$, and thus involving long-wavelength excitations outside this radial cutoff, are not guaranteed to be well reproduced. 
These effects are likely the root of
disagreement between modes that lie along  the $\Gamma$-X and $\Gamma$-L paths, like $(0,0, \frac{1}{4})$. Additional studies in this direction to investigate the role of the long-range electrostatic contribution on top of the GAP will shed light on this discrepancy and likely offer a better agreement with the reference DFT calculations. Furthemore, the inclusion of the LO-TO splitting will allow us to perform a finite-temperature study of the phonon dispersion across the T-C transition, to be compared with a recent study by Zhang \textit{et al.}\cite{ZhangFiniteTemperature2022}. As we have seen, however, long-range electrostatic contributions are not essential to model the thermodynamics and phase transitions of \bto{}.

\subsection{Validation with local dipole rotations}
\label{ssec:loc-dipole}

\begin{figure*}[tbhp]
    \centering
    \includegraphics[width=\linewidth]{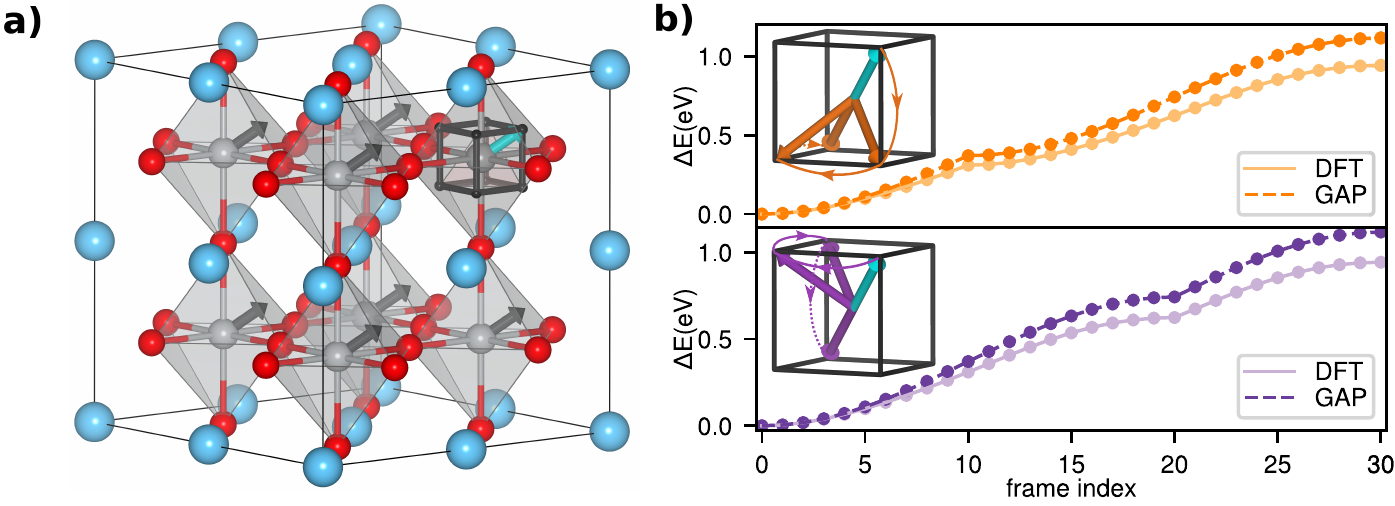}
    \caption{Validation of the GAP. Panel (a) shows the starting configuration of the two selected paths: the Ti dipoles are aligned along the \pTi state and marked by black arrows. They are kept frozen along the two paths shown as insets in panel (b). The dipole that is rotated is instead marked by a cyan arrow. Panel (b) shows a comparison between GAP and DFT predictions for the two paths that connect the \pTi and \npjcorr{\mTi} states.}
  \label{fig:benchmarks}
\end{figure*}

As a further test, we evaluate the accuracy of the GAP by modeling some of the distortions associated with the ferroelectric transition. In particular, the states associated with the presence of off-centered Ti atoms relative to the O cage
and the energy barrier separating them is key.  As we will discuss in Sec.~\ref{sec:BaTiO3_phases}, the long-range ordering of these displacements is the fundamental driver of ferroelectricity in \bto{}.

To test the performance of the GAP in reproducing these states we construct two paths, representing a local dipole rotation, across the phase space of a \stwo{} cubic supercell with a lattice parameter of \SI{8}{\angstrom}.
We start with the DFT-optimized structure with all Ti displaced by \SI{0.082}{\angstrom} along the \pTi direction resulting in local dipoles, as depicted by the arrows in panel a of  Figure 13. This is a rhombohedral structure -- spacegroup R3m (160) -- with Ba and Ti occupying the 1a position ($z_\textrm{Ba} = \num{-0.0004}$ and $z_\textrm{Ti} = \num{0.51116}$), and O occupying the 3b position ($x_\textrm{O} = \num{0.48823}$, $z_\textrm{O} = \num{-0.01872}$). For reference, the cubic structure with no dipole moment would have  $z_\textrm{Ba} = 0$, $z_\textrm{Ti} = \num{0.5}$, $x_\textrm{O} = \num{0.5}$, and $z_\textrm{O} = 0$, resulting in a cubic structure with spacegroup Pm$\bar 3$m (221).
One dipole, depicted in cyan, is then rotated about the barycenter of the enclosing oxygen octahedron to align with \npjcorr{\mTi} while keeping the magnitude of the Ti displacement constant and all other atoms fixed. 
The two paths, shown as the insets in panel b, have the same endpoints but visit different vertices of the cube centered at barycenter of the octahedron with the \npjcorr{\mTi} and \pTi displacements defining a diagonal. 

Physically, these paths represent the energy cost due to a relative rotation of one local dipole starting from a perfect ferroelectric state. A comparison between the GAP and the DFT energy variations across these paths (see panel b of Figure 13) shows that the GAP correctly reproduces the energy profile and favours states that correspond to aligned Ti-displacements, a feature that we have also seen in low-temperature MD simulations (see Sec. \ref{sec:micro-nature}). From a quantitative perspective, the GAP overestimates the energy barriers by some non negligible, but still reasonable, \SI{18}{\percent} for both paths. We stress, however, that these paths lie within the extrapolative regime of the potential, as they are constructed artificially and no MD simulation visits configurations that are close to them, except for the starting, completely ordered, structure that is visited at low temperature (see sec. \ref{sec:BaTiO3_phases}).

\subsection{Physically-inspired order parameters}
\label{ssec:ord-param}

\begin{figure}
    \centering
    \includegraphics[width=0.9\columnwidth]{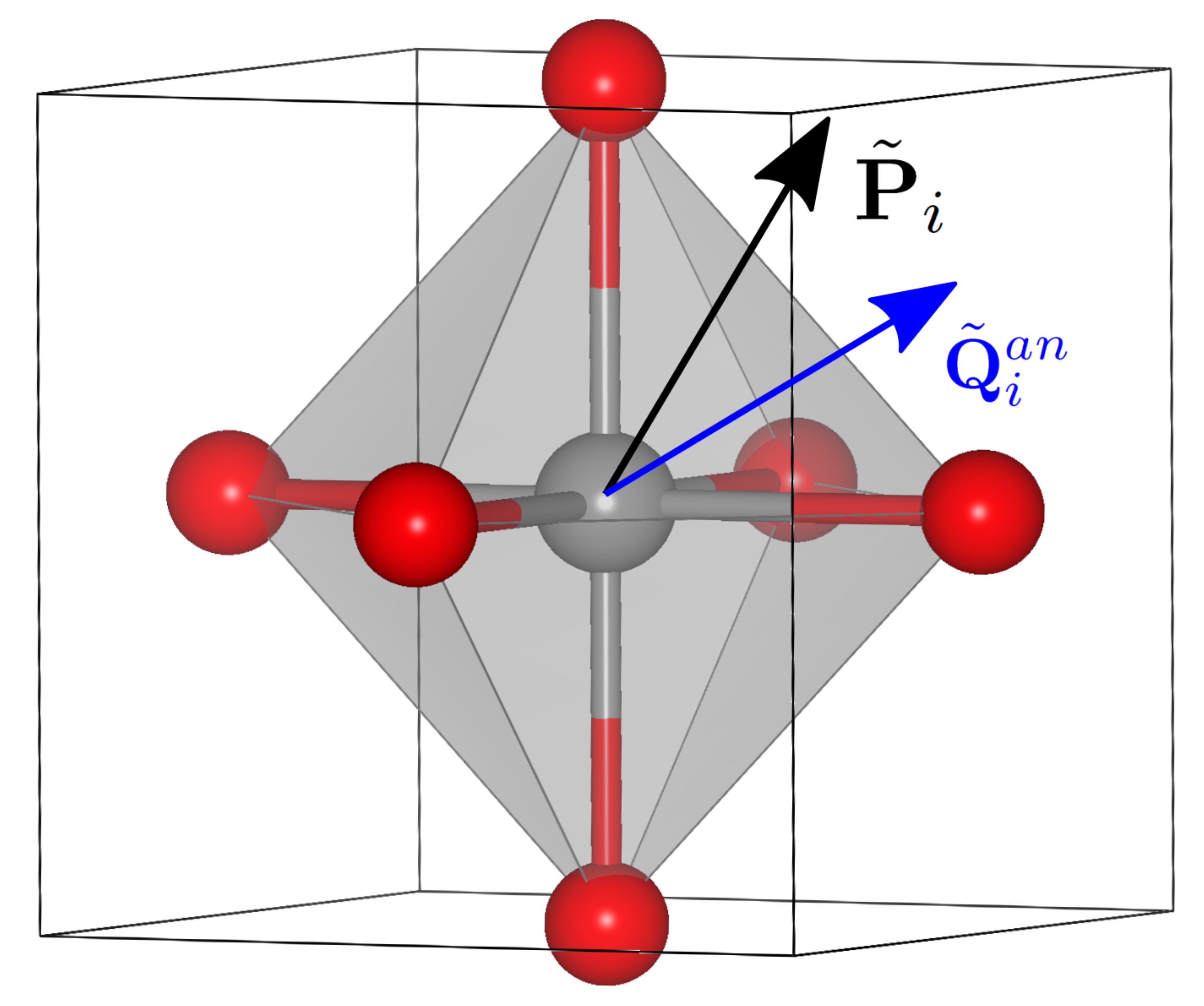}
    \caption{Sketch of the $i$-th Ti-center of structure A, along with its enclosing oxygen cage and the two vectors $\tilde{\mbf{P}}_i(A)$ and $\tilde{\mbf{Q}}^{an}_i(A)$ associated with it. The final CV is computed as a scalar product of these quantities, after first summing over all Ti-centers.} 
    \label{fig:CV-constr}
\end{figure}

As mentioned in Sec. \ref{sec:thermo}, the construction of a CV that can effectively distinguish the structural phases of \bto{} is key for the prediction of its phase diagram. In this section, we provide the construction of a two-component CV, namely $\mathbf{s} = (s_1, s_2)$, by explicitly using the predicted polarization $\mathbf{P}$ as an order parameter. As we shall see, we will build a set of invariant descriptors that correspond, for each structure, to a scalar product of vectors. These are constructed using the equivariant features $\rep<an ||A; \frho[\alpha]_{i}^{\nu}>$ defined in Equation \ref{eq:cartesian-equi}, averaged over Ti-centred environments. Physically, they will carry information about the orientation of $\mathbf{P}$ relative to the 'mean' atomic distortion, which we call $\mathbf{Q}$ (see also Figure 14).

Firstly, in order to compute the CV efficiently for long MD runs, we need to define an easy-to-compute proxy for $\mathbf{P}$, which we will denote as $\tilde{\mbf{P}}$. In practice, we
find that some of the neighbor density coefficients \npjcorr{$\rep<anlm||A;\rho_i>$ introduced in Sec. \ref{ssec:symm-adapted}} correlate strongly with $\mbf{P}$ (see the correlation plots in Supplementary Figure~7). We can then define $\tilde{\mbf{P}}$ by restricting ourselves to \ce{Ti}-centered environments, as follows: 

\begin{equation}
    \tilde{P}_{\alpha}(A) = \sum_{i\in A,\mathrm{Ti}} \tilde{P}_{\alpha, i}(A) =  \sum_{i\in A,\mathrm{Ti}} \rep<a\!=\!\mathrm{O}\,n\!=\!6||\frho[\alpha]_i^1> 
\end{equation}
where $a\!=\!\mathrm{O}$ represent the atomic species (the oxygen) onto which we project the Ti-centred density.
Note that here we use the expression for the Cartesian equivariants defined in Sec. \ref{ssec:symm-adapted}, so that $\alpha = (x, y, z)$ and $\tilde{\mbf{P}}$ transforms like a vector under rotations. It represents in fact a sum of vectors $\tilde{\mbf{P}_i}$, each assigned to one Ti-center, as shown in Figure 14. Similarly, we average the full neighbor density coefficient $\rep<an||\frho[\alpha]_i^1>$ over all \ce{Ti} centers to obtain a measure of the mean structural deformations: 

\begin{equation}
    \color{blue}
    \tilde{Q}^{an}_{\alpha}(A) =
    \sum_{i\in A,\mathrm{Ti}} \tilde{Q}^{an}_{\alpha, i}(A)  = \sum_{i\in A,\mathrm{Ti}} \rep<an||\frho[\alpha]_i^1>.
\end{equation}

Finally, we compute the scalar product of $\tilde{\mbf{P}}$ and $ {\color{blue}\tilde{\mbf{Q}}}$ to construct a set of invariants:

\begin{equation}
O^{an}(A)=\tilde{\mbf{P}}(A)\cdot {\color{blue} \tilde{\mbf{Q}}^{an}(A)}
\end{equation}

and perform a principal component analysis (PCA) on the scalar descriptors $O^{an}(A)$ to obtain two physically-motivated and symmetry-invariant order parameters. This step allows us to obtain the scalar components that mostly contribute to the observed variance of the $O^{an}$ invariants across a dataset of structures.  
In particular, by performing a PCA analysis over the entirety of the MD trajectories as a function of all simulated temperatures, we find that the first two PCs, corresponding to $s_1$ and $s_2$ can neatly separate all four phases (see Sec.~\ref{sec:thermo}). 

At each temperature, we then perform a separate clustering using the Probabilistic Analysis of Molecular Motifs (PAMM) \cite{gasp+18jctc} algorithm, that determines a Gaussian mixture model in which each cluster corresponds to a different phase.
Using the posterior probabilities associated with the mixture model (named probabilistic motif identifiers in Ref.~\citenum{gasp+18jctc}) we can associate with each MD frame a smooth probability $P_k(t)$, based on the corresponding values of the CVs $(s_1(t), s_2(t))$, that represents the probability that the corresponding structure at time $t$ belongs to the cluster $k=$(R, O, T, C). These probabilities are then used to determine the relative stability of the different phases.  The advantage of this technique, as compared to perhaps simpler methodologies, such as tracking the temperature evolution of the lattice parameters, is the fact that it is fully automatized, rotationally-invariant and makes direct use of the polarization vector - the key ingredient to physically describe the onset of ferroelectricity.

\subsection{ML-MD Computational details}
\label{sec:comp-details}

All the machine learning data that we have generated to investigate the physics of~\bto~combines the use of molecular dynamics simulations performed with \ipi{}\cite{kapi+19cpc} -- the MD integrator -- and librascal\cite{musi+21jcp,rascalgithub} -- the engine to compute the total energy, atomic force components and stress tensor of a~\bto~structure. In all cases, we choose the smallest simulation cell size that provides converged results; this is to optimize the tradeoff between adequate sampling in time and adequate sampling in system size that is possible under a given computational budget.

In particular, the results of Sec.~\ref{sec:BaTiO3_phases} correspond to NST fully flexible simulations of a \sfive{} cell, i.e. with an external constant stress tensor $\sigma = \text{diag}(p, p, p)$ with $p = 1$\,atm. The full flexibility of the cell allows the system to relax the off-diagonal components of the MD computed stress tensor, as the system undergoes the structural R-O-T-C phase transitions as a function of the temperature. In this case, we choose the simulation size so as to show well separated structural minima as a function of the temperature, while maintaining the simulations computationally inexpensive.

The results of Sec.~\ref{sec:micro-nature} correspond instead to isotropic NpT simulations of a \sfour{} cell over a wide range of temperatures (between \SIrange[range-phrase={ and }]{20}{250}{\kelvin}) with a restricted cubic geometry. This supercell size is sufficient to identify the Ti off-centering as the physical mechanism governing the emergence of ferroelectricity.

Fully flexible MD runs of a~\bto~\sfour{} cell with a total simulation time up to \SI{1.6}{\nano\second} between \SIrange[range-phrase={ and }]{10}{250}{\kelvin}
are performed for quantitative estimation of the temperature-dependent free energies  (see Sec.~\ref{sec:thermo}). In particular, unbiased MD is used to generate trajectories across the coexistence regions of the O-T and T-C transitions (between \SIrange[range-phrase={ and }]{40}{250}{\kelvin}), while well-tempered metadynamics \cite{barducciWellTemperedMetadynamics2008} runs across the R-O transition are needed to enable collective jumps between R and O states within times that are affordable by classical MD runs. Additional details on the metadynamics runs are given in Supplementary Note 4. The relatively small supercell size in this case allows both efficient sampling of the structural transitions and simulation times, on the order of nanoseconds, that are required to converge the chemical potential estimates. 

The spatial correlations shown in Sec.~\ref{sec:BaTiO3_phases} are calculated on a \sfive{} supercell trajectory of length \SI{400}{\pico\second}, while the static and frequency-dependent dielectric constant in Section~\ref{sec:dielec-resp} were calculated on a \ssix{} supercell trajectory of length \SI{200}{\pico\second} in order to ensure supercell-size convergence of the static value. The temperature dependence of the dielectric constant, being a more expensive calculation requiring multiple trajectories, instead used both a \sfour{} and a \sfive{} supercell, simulated for \SI{250}{\pico\second} each, to explicitly assess the rate of supercell-size convergence.

All the NpT/NST simulations were carried out with an isotropic/anisotropic barostat, leaving the cell volume/vectors free to equilibrate at finite temperature. Thermalization of the cell degrees of freedom is achieved by means of a generalized Langevin equation (GLE) thermostat\cite{ceri+11jcp}, while thermalization of the atomic velocity distribution is realized via stochastic velocity rescaling (SVR) \cite{buss+07jcp}. This combination of thermostats allows for an optimal equilibration of the system's relevant degrees of freedom on a timescale of the order of picoseconds, without significantly interfering with the dynamical properties of the system, especially the polarization vectors.
The characteristic times of the barostat, the SVR thermostat and the MD timestep are \SI{1}{\pico\second}, \SI{10}{\femto\second}, and \SI{2}{\femto\second} respectively.

\section{Data Availability}
All numerical data supporting the results of this paper and allowing to reproduce the results are openly available on the Materials Cloud Archive\cite{MatCloud}.

\section{Code Availability}
In order to generate the data needed for this paper, we made use of the librascal\cite{rascalgithub}, i-PI\cite{ipicode}, and TenSOAP\cite{TENSOAP} codes. These codes are all publicly available on github.  Some additional scripts necessary for data analysis and processing beyond that provided in these codes is provided along with the research data in the Materials Cloud Archive\cite{MatCloud}.

\section{Acknowledgements}

We thank Federico Grasselli for insightful suggestions and a critical reading of the manuscript. 
L.G., M.K. and M.C. were supported by the Samsung Advanced Institute of Technology (SAIT).  M.V., G.P., N.M., and M.C. acknowledge support by the MARVEL National Centre of Competence in Research (NCCR), funded by the Swiss National Science Foundation (grant agreement ID 51NF40-182892). G.P. acknowledges the swissuniversities `Materials Cloud' project (number 201-003). G.P. and N.M. acknowledge support from from  the  European  Centre  of  Excellence  MaX  “Materials  design  at the Exascale” (824143). This work was supported by a grant from the Swiss National Supercomputing Centre (CSCS) under project IDs mr0 and s1073.

\section{Author contributions}
G.P., N.M. and M.C. jointly supervised the project. L.G. and M.V. jointly developed, trained, and benchmarked the ML framework. L.G. ran the MD simulations. L.G. and M.V. analyzed the results of the MD simulations. M.K. performed the DFT calculations. All authors contributed to the discussion and writing of the paper.

\section{Competing Interests}
The authors declare no competing interests.

\bibliography{mk.bib,mc.bib,BaTiO3_modeling.bib}

\clearpage
\newpage

\end{document}




\title{Thermodynamics and dielectric response of~\bto~by data-driven modeling \\
\large{Supplementary Information}}

\input{authors-finalsub.tex}

\maketitle

\section{The ML framework}
\label{sec:ML-framework}
The GAP used to derive all the MD data provided in the main paper and to compute the dielectric properties relies on the construction of an appropriate density-based representation of the atomic environments, namely the Smooth Overlap of Atomic Positions (SOAP) method. 

We construct a feature vector for a given atomic environment $i$ by computing the SOAP power spectrum with the Librascal package (\texttt{https://github.com/cosmo-epfl/librascal}), following the approach of Musil et al. \cite{musilEfficientImplementation2021}.

The power spectrum (with $n_\text{max} = 8$ and $l_\text{max} = 6$) is computed by using Gaussian Type Orbitals (GTO) radial basis functions and gaussian smearing of the atom density ($\sigma_\text{width} = \SI{0.5}{\angstrom}$), with a radial cutoff of $\SI{5.5}{\angstrom}$. The use of a radial cutoff that is larger than the average separation of first nearest Ti neighbors ($\approx \SI{4.0}{\angstrom}$) allows us to capture the short-ranged Ti-Ti interactions that ultimately result in long-range emergent dipole correlations, a distinctive feature of polarized states in $\text{BaTiO}_3$. The atomic environment is defined by a smooth cutoff function with a long-algebraic decay of the form: 

\begin{equation}
u(r) = \frac{c}{c + (r/r_0)^m}
\end{equation}
with $c = 1$, $r_0 = \SI{3.5}{\angstrom}$ and $m = 4$.

The GAP is then constructed using Eq.~(13) of the main paper with a non-linear kernel ($\zeta = 4$), fitting both total energies and atomic forces. The number of sparse environments per species is given by $n_{\text{O}} = 500, n_{\text{Ti}} = n_{\text{Ba}}= 250$, enabling us to achieve DFT-level accuracy while keeping the MD runs computationally inexpensive. \npjcorr{See Sec. \ref{sec:MD-sims} for details on the computational cost.} 

Energy and force regularizers are set to \SI{3.0E-3}{\electronvolt} and \SI[per-mode=symbol]{3.0E-2}{\electronvolt\per\angstrom} respectively. 

The polarization model was constructed with similar (equivariant) SOAP parameters, namely, $n_\text{max}=8$, $l_\text{max}=6$, $\sigma_\text{width} = 0.3$, with radial cutoff $r_c = \SI{5.0}{\angstrom}$ and cutoff function parameters $c = 1$, $r_0 = 3.5$, and $m = 4$.  These parameters were used for both the scalar ($\lambda = 0$) and vector ($\lambda = 1$) power spectra, which were used to compute nonlinear kernels with a SOAP power of $\zeta = 2$ and train a kernel model with the regularizer set to a factor of \num{0.01} times the intrinsic dataset variance.


\section{DFT computational details for training sets and validation}
\label{sub:DFT}

All of our first-principles DFT-based calculations are performed using Quantum ESPRESSO~\cite{Giannozzi2009,Giannozzi2017} with the PBEsol~\cite{Perdew2008} functional, RRKJ pseudopotentials~\cite{Rappe1990} with a wavefunction and charge-density energy cutoffs of \SIlist{60;600}{\rydberg}, respectively.

The initial structures used to train the ML interatomic potential are the initial, final and intermediate structure obtained during the variable-cell structural optimization of the 27 cubic microscopic templates of the cubic perovskite structure. These cubic microscopic templates are \stwo{} supercells of the 5-atom primitive cell which can host atomic displacements while preserving the cubic point symmetry. The methodology to determine all such structures using group-subgroup relations and the symmetries of the structures can be found in the work of Kotiuga \textit{et al.}\cite{Kotiuga_2021}. The total energies are converged self-consistently with a tolerance of $\SI{1E-12}{\rydberg} \approx \SI{1.3E-11}{\electronvolt}$ and the forces with a tolerance of \SI{1E-5}{\rydberg\per\bohr}  using a \ssix{} $k$-mesh in the \stwo{} supercells of the 5-atom primitive cell.

For the structures selected by FPS and to perform the FEP correction, the total energies are converged self-consistently with a tolerance of 1$\times$10$^{-10}$ Ry = 1.3$\times$10$^{-9}$ eV using a \sfour{} $k$-mesh in the \stwo{} supercells of the 5-atom primitive cell (and equivalent meshes in different cell sizes, e.g., \stwo{} in the \sfour{}  supercells).

The polarization calculations are carried out using the Berry phase technique. A calculation is carried out for each of the three components of the polarization with a large number of $k$-points in the direction of the polarization for the Berry phase integration. In our calculations we use 24 $k$-points along the direction of interest. For example for the polarization along the $x$ direction, we use a $24\times4\times4$ $k$-mesh.

The phonon dispersions were calculated using finite differences using phonopy\cite{phonopy} in conjunction with Quantum ESPRESSO. The dielectric constant and Born effective charges needed to apply the non-analytical correction\cite{Gonze1997} were calculated using density functional perturbation theory as implemented in Quantum ESPRESSO. 

\npjcorr{An scf calculation (energy and forces) for a \stwo{} cell is about $2$ core-hours. A calculation of the polarization along the $3$ crystal directions is about $64$ core-hours.}

\section{Polarization training data}
\label{sub:polarization_training_data}

\begin{figure}
    \centering
    \includegraphics[width=0.45\textwidth]{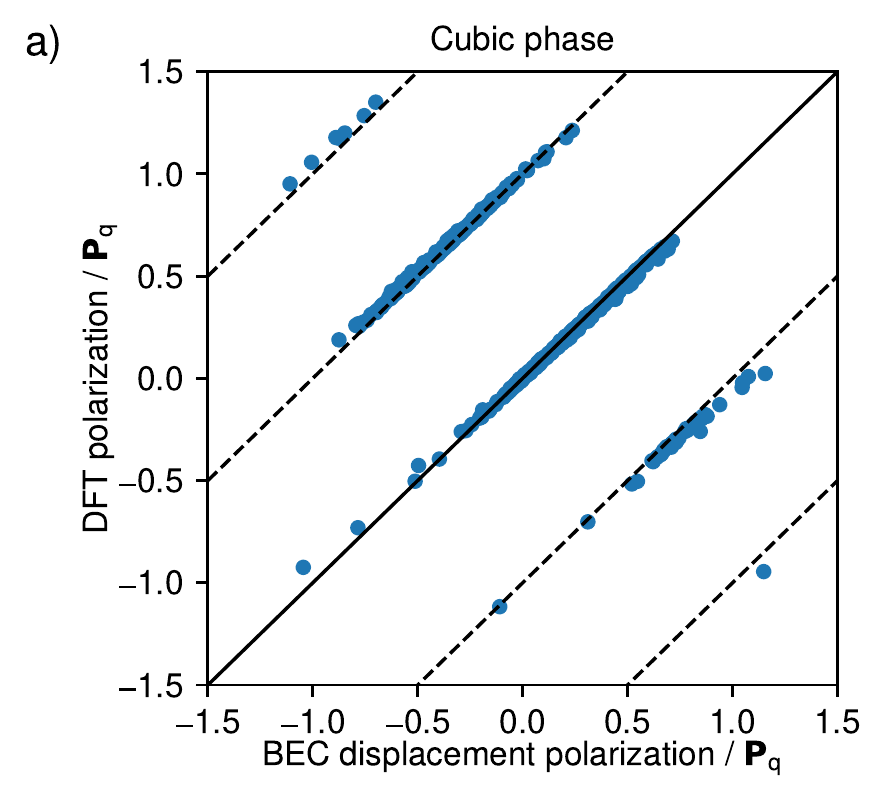}
    \includegraphics[width=0.45\textwidth]{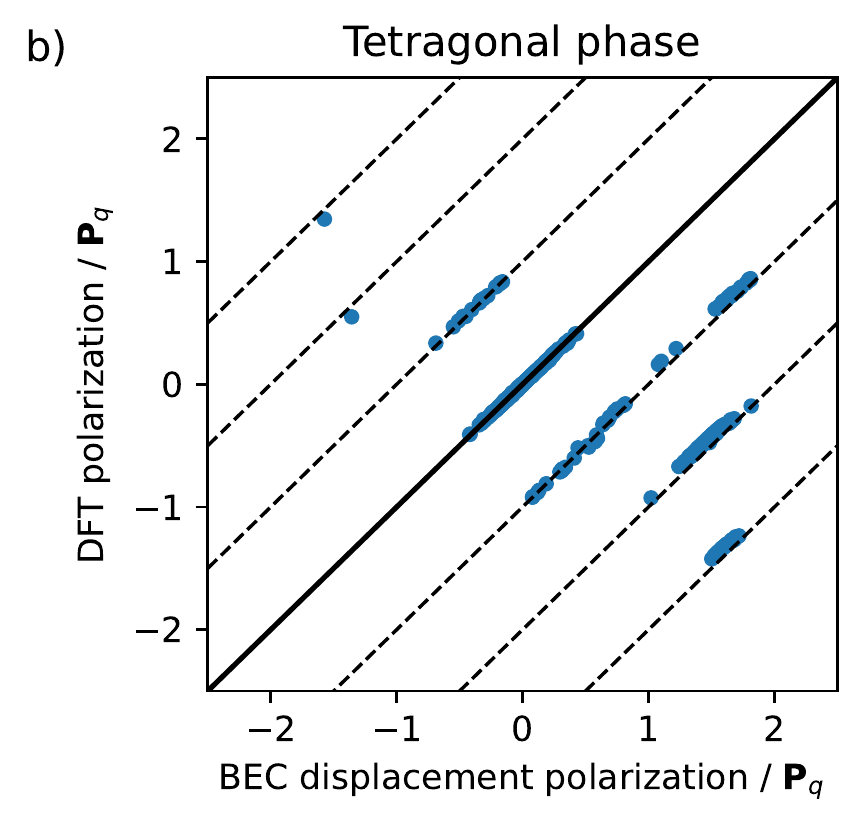}
    \includegraphics[width=0.45\textwidth]{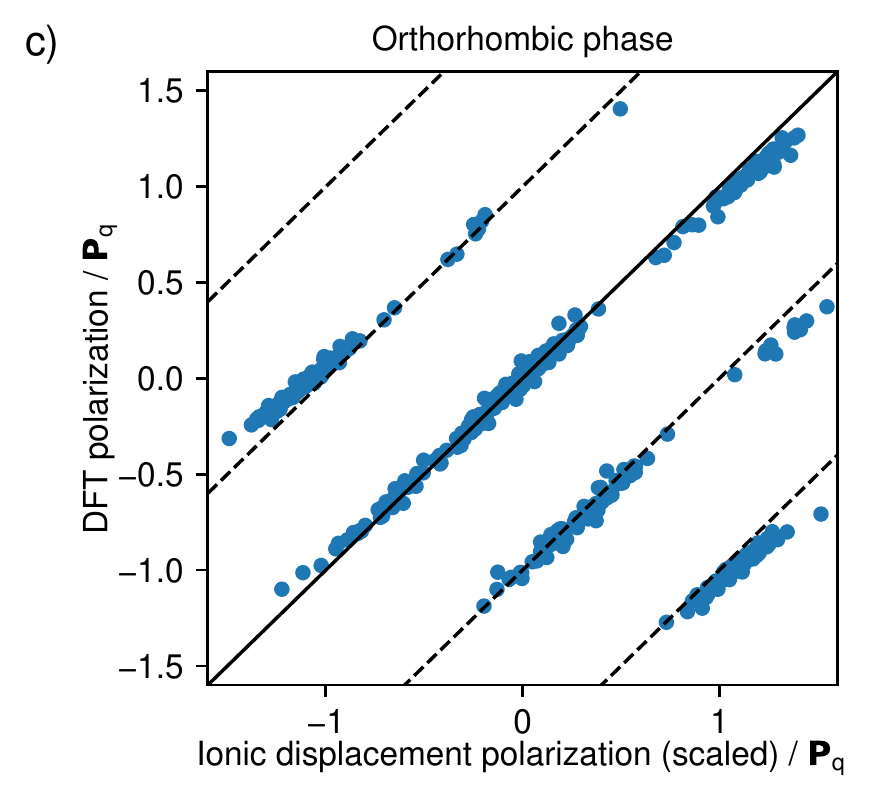}
    \includegraphics[width=0.45\textwidth]{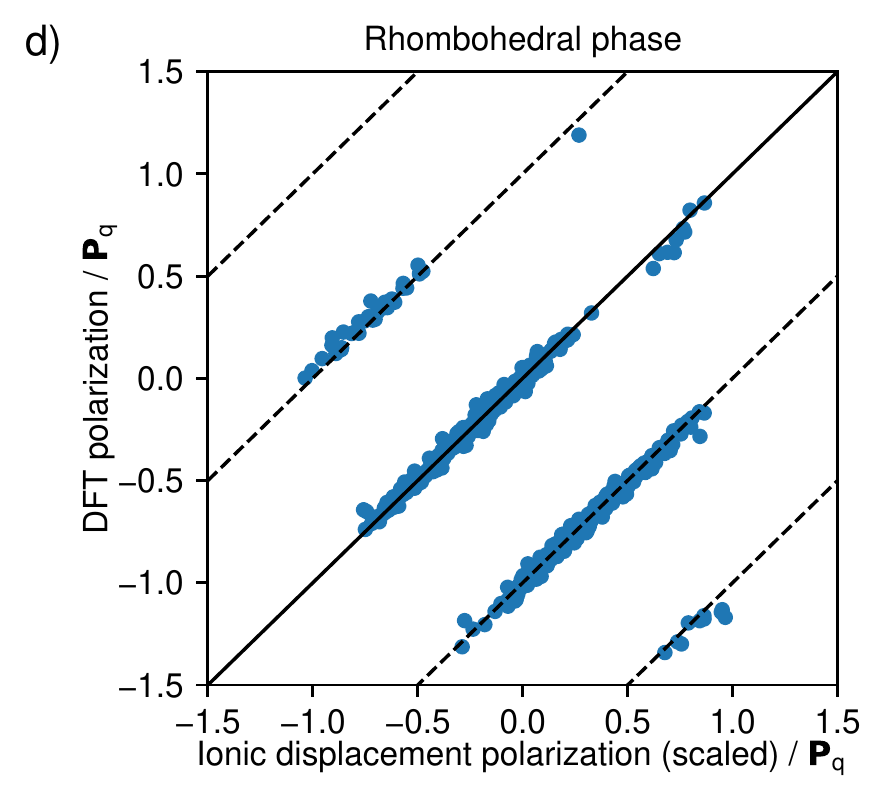}
    \captionsetup{labelformat=npj}
    \caption{Separation of the computed polarization into different branches for each of the four phases of \batiooo{}: (a) cubic, (b) tetragonal, (c) orthorhombic, and (d) rhombohedral, based on the Born effective charge displacement polarization (cubic and tetragonal phases) and a scaled-formal-charge displacement polarization (orthorhombic and rhombohedral phases).  The solid diagonal line is $x=y$ and passes through $(0, 0)$.  The dashed lines are drawn with shifts of $n\mathbf{P}_q, n\in \mathbb{Z}$, i.e. integer multiples of the polarization quantum, in the DFT polarizations; the DFT polarizations can therefore be shifted by the appropriate integer multiple to bring all polarization data points onto one continuous branch.}
    \label{fig:pol_branches_separation}
\end{figure}

\npjcorr{As mentioned in Section IV~C of the main text, we must resolve the ambiguity in the definition of the polarization in periodic boundary conditions in order to train our (single-valued, cell-invariant) polarization model.  Since the dipole model (Eq.~(13) in the main text) is additive over atom-centred contributions, and each dipole contributes a position-independent vector amount to the total dipole, the model is invariant to shifts in the definition of the periodic unit cell.  However, this is not in general true of polarizations defined in periodic systems~\cite{Spaldin2012}.  When calculating the polarizations of the training structures, we run into the well-known ambiguity in defining the condensed-phase polarization, which is only defined modulo the \emph{polarization quantum} (a shift of $\mathbf{P}_q = \frac{e \mathbf{h}_i}{2\pi V_A}$ in the direction of any of the lattice vectors $\mathbf{h}_i$)~\cite{Spaldin2012,restaTheoryPolarization2007}.  Such shifts would appear in the model training data as discontinuities or multi-valuedness making it nearly impossible to obtain a good ML fit.

We therefore fold all the calculated polarizations of the training data onto the same branch of the polarization manifold, defined along the coordinate given by a simple displacive polarization model that is single-valued, like our model.  \sfigref{fig:pol_branches_separation} shows a plot of the calculated polarizations along this coordinate.  Note that this displacive polarization model is only used as a guide to calculate the correct polarization shifts, and is not used in the final model.
The individual data points of the training set are in practice spread sufficiently densely along this coordinate so that we can clearly separate the different branches and calculate the shifts (in integer multiples of the polarization quantum) needed to bring them onto a single, continuous surface as viewed in the coordinates of displacement from the high-symmetry structure.  This procedure also fixes the zero of the training data, and therefore the predictive model, so that the structure's total dipole moment $\mathbf{M}(A)$ is well defined.}


 These branches are plotted for the model training data, separately for each phase, in \sfigref{fig:pol_branches_separation}.

The simple, displacive polarization model is in fact defined as a linearization about the most highly-symmetric structure (i.e. the perfect cubic, tetragonal, orthorhombic or rhombohedral structure, depending on the phase).  This model is defined as
\begin{equation}
    \mathbf{P}^*({\mathbf{R}}) = \frac{\partial \mathbf{P}}{\partial \mathbf{R}} (\mathbf{R} - \mathbf{R}^0)
\end{equation}
where $\mathbf{R}^0$ are the atomic positions in the 'reference', high-symmetry structure, and the difference $(\mathbf{R} - \mathbf{R}^0)$ is taken in the minimum-image convention to avoid any ambiguity in the definition of the periodic unit cell (for sufficiently small displacements).  The partial derivative corresponds to the Born effective charges (see Spaldin\cite{Spaldin2012}, Eq.~(18-19) ), modulo some normalization factors, so we call this model the 'Born effective charge polarization'.  This is similar to the linearized approximation made in Zhong \textit{et al.}\cite{zhongGiantLOTO1994}.  The Born effective charges (BECs) \npjcorr{have been calculated using density functional perturbation theory with the PBEsol functional}.  We can also substitute scaled formal ionic charges for phases where the Born effective charges have not been calculated or made available.

The Born effective charges used for the linearized polarization model were, in units of $e$:

\begin{center}
\begin{tabular}{l c c c c c c c c c}
Phase & Ba$_\perp$ & Ba$_\parallel$ & Ti\npjcorr{$_\perp$} & \npjcorr{Ti$_\parallel$} & O\npjcorr{$_{\parallel;\parallel}$} & \npjcorr{O$_{\perp;\perp}$} & O\npjcorr{$_{x;yy,y;xx}$} & \npjcorr{O$_{\perp;\parallel}$} & \npjcorr{O$_{\parallel;\perp}$} \\
\hline
Cubic       & 2.72 & 2.72 & 7.44 & 7.44 & -5.85 & -5.85 & -2.12 & -2.12 & -2.12 \\
Tetragonal  & \npjcorr{2.73} & \npjcorr{2.79} & \npjcorr{7.08} & \npjcorr{5.75} & \npjcorr{-5.64} &  \npjcorr{-4.66} & \npjcorr{-2.15} & \npjcorr{-1.92} & \npjcorr{-2.0}
\end{tabular}
\end{center}

where, in the tetragonal phase, the $z$ axis is taken along the direction of polarization\npjcorr{, and the notation $O_{\perp;\parallel}$, for example, denotes the effective charge of the oxygen atom \emph{located perpendicularly} ($\perp$) to the polarization axis, relative to the closest \ce{Ti} atom, \emph{displaced} along the axis \emph{parallel} ($\parallel$) to the polarization.  The special case O\npjcorr{$_{x;yy,y;xx}$} denotes displacement and derivatives taken along two perpendicular directions, both perpendicular to the polarization axis.}  All charges are taken to be diagonal tensors; i.e. $\frac{\partial P_\mu}{\partial R_\nu} = q_\mu \delta_{\mu\nu}$.

For the orthorhombic and tetragonal phases, we instead used the following formal charges: Ba:~$2+$, Ti:~$4+$, and O:~$2-$.  The resulting 'ionic' polarizations are scaled by a factor of \num{1.8} to align them with the DFT-computed polarizations.


\section*{Supplementary Note 3 - MD simulations}
\label{sec:MD-sims}
All the MD simulations presented in the main paper have been carried out using the i-Pi package \cite{ipicode}, interfaced with Librascal for the energy and force calculations via a socket interface. 
\npjcorr{The i-Pi/Librascal interface is not yet parallelized, so all the MD simulations were performed on single core. Given the fact that the computational cost of the GAP model scales linearly with the system's size (see Eq. ($12$) of the main paper), we expect a considerable scale-up in the size of these calculations to be possible once a parallel version is available.}

The simulations were performed for increasing $\text{BaTiO}_3$ cell sizes (from \sfour{} to \ssix{}) depending on the specific aim: \sfour cells were used to reconstruct the thermodynamics, as they enable at once phase separation of the R-O-T-C phases across the coexistence regions and a reasonable statistics of transitions over timescales affordable with standard MD. Larger cells were instead used to highlight the microscopic mechanisms of ferroelectricity (see Sec. II~C of the main paper) and to compute dielectric properties. A \ssix{} cell was used instead to converge the dielectric constant calculations.
\npjcorr{The computational cost of a typical \sfour{} MD run of \bto~is about $0.3$ core-hours/ps (about $80$ ps per day on a single core), compared to an estimated $5$k core-hours/ps to perform a Car-Parrinello MD run with the PBEsol functional. So the GAP provides a gain of $4$ orders of magnitude in computational efficiency to run MD. Still, this efficiency is considerably lower than semi-empirical force fields due to the calculation of the SOAP descriptors ($2688$ separate components per each atomic environment due to the choice of $n_{\text{max}}$ and $l_{\text{max}}$) and the kernels with the reference environments in the sparse set. In general, the computational cost of a ML model fitted with the methodology outlined in \ref{sec:ML-framework} grows linearly with SOAP vector size, which itself grows linearly with $l_{\text{max}}$ and quadratically with $n_{\text{max}}$. We refer the reader to Supplementary Reference \cite{musilEfficientImplementation2021} for further details.}

\section{Well-Tempered Metadynamics}

\begin{figure}[tbhp]
    \centering
    \includegraphics[width=\linewidth]{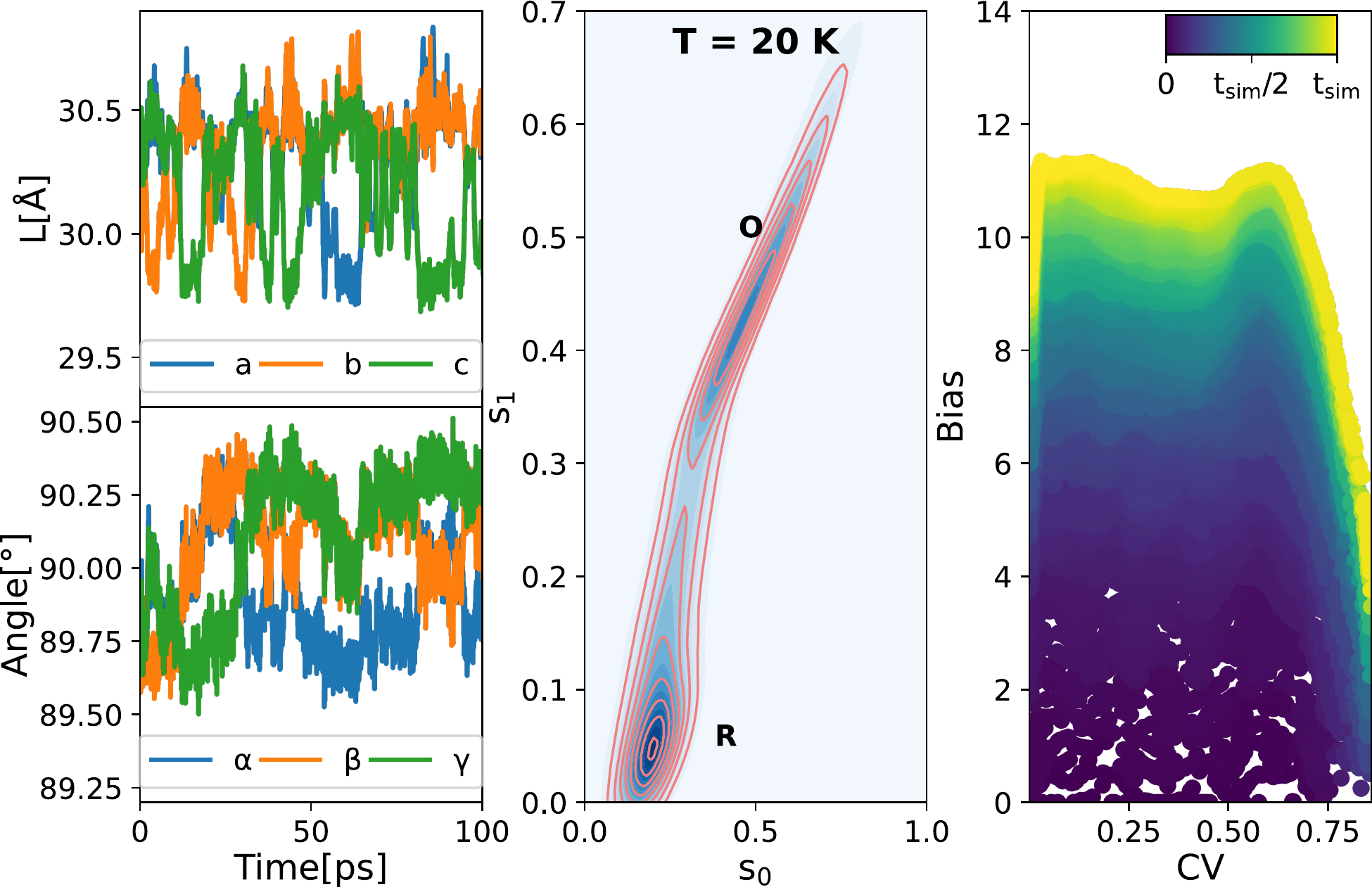}
     \captionsetup{labelformat=npj}
    \caption{Well-Tempered Metadynamics run at $T = 20\,$K. The left set of panels shows the time-evolution of the cell vector lengths (top) and angles (bottom). The central panel shows the R-O clusters in the $(s_0, s_1)$ plane, as sampled along the metadynamics trajectory. The right panel finally shows the accumulated bias as a function of the CV, coloured according to time (in units of the total simulation time $t_{\text{sim}}$).}
   \label{fig:metadynamics}
\end{figure}

As mentioned in Sec. II~D~1 of the main paper, Well-Tempered Metadynamics \cite{barducciWellTemperedMetadynamics2008} has been employed to enhance reversible transitions between R and O states, inhibited in standard MD runs by the tiny thermal fluctuations. 

The metadynamics runs were performed using the PLUMED/i-Pi interface \cite{PLUMED}, by dynamically constructing a bias potential on top of the GAP, in a range of temperatures between \SIlist{15;40}{\kelvin}.

The collective variable (CV) \npjcorr{$\eta$} that has been used to dynamically construct the bias is a symmetrized combination of the lattice parameters. In particular, it is proportional to the variance of the cell vector lengths ($a$, $b$ and $c$): 

\begin{equation}
    \npjcorr{\eta} = 100 \frac{\sqrt{(a^2+b^2+c^2)/3-((a+b+c)/3)^2}}{\sqrt{a^2+b^2+c^2}}
\end{equation}

This CV is seen to be slightly defective, as it visibly separates the R-O states, but does not represent appropriately the transition state ensemble. This results in hysteresis and a residual potential energy barrier between the R-O phases, even after long bias deposition times (of the order of hundreds of ps). We are in fact far from reaching a perfectly diffusive regime between the R and O states (see \sfigref{fig:metadynamics}). The choice of this CV however allows us to sample many reversible transitions over the course of the metadynamics run. An example of a typical metadynamics trajectory is shown in \sfigref{fig:metadynamics}, providing information about the time evolution of the cell lengths and angles, the order parameter and the bias deposition.

\section{Correlations of positions and dipoles}

The spatial correlations shown in Sec.~II~B and Fig.~2 of the main text are calculated using the usual Pearson correlation formula:
\begin{equation}
    \rho^{\mu\nu}_{0j} = \frac{\left<M^\mu_0 M^\nu_j\right>}{\sqrt{\left<\left(M^\mu_0\right)^2\right>\left<\left(M^\nu_j\right)^2\right>}}
\end{equation}
where $\mu,\nu$ are Cartesian components, $0$ represents the central atom with respect to which the correlations are being computed, $j$ represents the test atom whose position is plotted in the figure, and the averages are taken across a trajectory.  This value is strictly bounded in the range $[-1, 1]$.

\section{Static dielectric constant}

It is desirable to have an error estimate for the dielectric constants computed in Sec.~II~E~1 of the main text.  Since the static dielectric constant is computed using the mean squared dipole moment, it is formally related to the \emph{variance} of the dipole moment norm, which we compute using the \emph{estimator}
\begin{equation}
    \mathbb{E}_\text{samp}(M^2) = \left<M^2\right>_\text{samp} = \frac{1}{N}\sum_{i=1}^N M^2
    \label{eq:dipole-var-estimator}
\end{equation}
where $N$ is the number of total samples of the dipole moment distribution.

We are interested in the variance of this variance estimator; the formula can be found in statistics textbooks (e.g. Scott\cite[Ch.~6.2]{ScottDavidW2020SACM}; see also Cho \textit{et al.}\cite{eungchunchoVarianceSample2005}).  Assuming the mean $\mu_X$ of the random variable $X$ is zero, it is:
\begin{equation}
    \variance(\mathbb{E}_\text{samp}(X^2)) = \frac{1}{N}\left((\mathbb{E}_\text{samp}\left(X^4\right) + \frac{3 - N}{N - 1}\mathbb{E}_\text{samp}\left(X^2\right)^2\right)
    \label{eq:variance-of-variance}
\end{equation}

\begin{figure}
    \centering
    \includegraphics{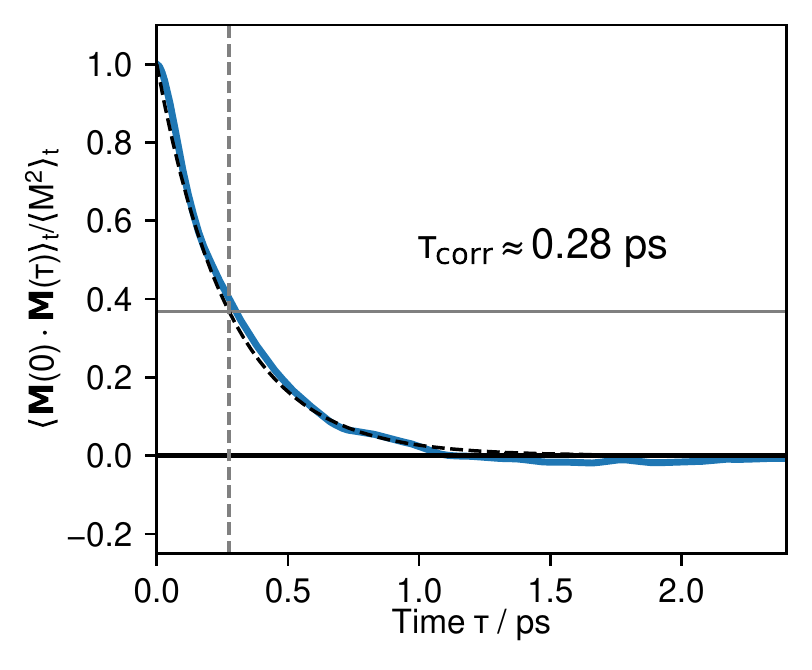}
     \captionsetup{labelformat=npj}
    \caption{Normalized autocorrelation function (thick blue line) of the supercell dipole moments $\mathbf{M}(t)$ for a \sfive{} trajectory at \SI{250}{\kelvin}, along with its exponential approximation (dashed black line).  The autocorrelation time is marked with a dashed vertical line, which is the time needed for the exponential to fall to $\frac{1}{e}$ (horizontal gray line).}
    \label{fig:dip_acorr}
\end{figure}

We can insert $M^2$ in place of $X^2$ in this formula to obtain the variance of the dielectric constant estimator (even though $\mathbf{M}$ is formally a vector quantity, its component-wise mean $\mu_\mathbf{M}=0$ still, and we are interested in the variance of the scalar quantity $M^2$, so this formula is still applicable if we use $M^4 \equiv (M^2)^2$).  The only remaining issue is that we cannot use the total number of samples $N$ in the formula above, since the samples are highly correlated (they come from consecutive samples of a molecular dynamics trajectory) and this formula assumes independent samples.  We can correct for this correlation by estimating the \emph{autocorrelation time} $\tau_\text{corr}$ of the trajectory, as illustrated in \sfigref{fig:dip_acorr}.  Informally, the autocorrelation time can be interpreted as the time needed for a sample to 'decorrelate' from its value at a previous time.  It can be estimated by modelling the (normalized) autocorrelation function as an exponential, e.g. as described in Sokal\cite{Sokal1996} (Sec. 3); the result is also shown in \sfigref{fig:dip_acorr}.  We then obtain an \emph{effective} number of samples
\begin{equation}
    N_\text{eff} = \frac{T_\text{samp}}{\tau_\text{corr}} = N\frac{\Delta t}{\tau_\text{corr}}
\end{equation}
where $T_\text{samp}$ is the total sampling time and $\Delta t$ is the sampling interval in time.  We can then proceed to use Eq.~\eqref{eq:variance-of-variance} for the correlated samples by substituting $N_\text{eff}$ for $N$ wherever it appears explicitly (the expectation values should be calculated as normal, using all the samples available).

The errors on the Curie temperatures shown in Fig.~8b of the main text are computed in a Monte Carlo fashion, by computing multiple fits with the data points each independently offset by an amount sampled from a normal distribution with the standard deviation equal to the error bar for that point.  The error bar represents the standard deviation of the resulting ensemble of fit parameters (with some outliers removed due to the unstable nature of the fit).

\section{Supporting data}

In the following, we show additional plots to further substantiate the results of the main paper. We provide: 

\begin{itemize}
    \item predictions of the ML polarization model on a NST fully flexible simulation of a \sfive{} cell, with a temperature equal to \SI{250}{\kelvin} and a pressure of \SI{30}{\giga\pascal};
    \item a correlation plot between the unit-cell polarization components and the Ti-displacement components in the series of \sfive{} simulations discussed in Sec. II~B of the main paper;
    \item 2D displacement maps in a series of $NVT$ \sfour simulations with restricted cubic geometry (to be compared with Sec. II~C of the main paper);
    \item correlation plots between the total polarization components of ~\bto~and the components of the vector descriptor $\rep<a\!=\!\ce{O}\,n\!=\!6||\frho[\alpha]_i^1>$ which is observed to provide the best separation between the R-O-T-C clusters (in \sfive{} MD trajectories); 
    \item 2D contour plots showing the correlation between the total polarization norm and $s_1$ and $s_2$ in the series of \sfive{} MD trajectories described in Sec. II~B of the main paper;
    \item convergence of the FEP corrections to the chemical potentials across the T-C transition;
    \item the comparison of the phonon dispersion calculated from finite difference with the GAP potential, DFT with and with out LO-TO splitting;
    \item \npjcorr{the behavior of thermal expansion derived from ML-MD at $p = 1\,$atm and $p = -2\,$GPa;}
    \item \npjcorr{free energy profiles as a function of $s_1$ and the polarization $P$ across the T-C transition};
    \item \npjcorr{supercell-size convergence of the frequency-dependent dielectric response spectra at $T=\SI{250}{\kelvin}$};
    \item \npjcorr{ensemble-averaged dielectric response spectrum for the 4x4x4 cell at \SI{250}{\kelvin}}. 
\end{itemize}

\begin{figure}[tbhp]
    \centering
    \includegraphics[width=0.9\linewidth]{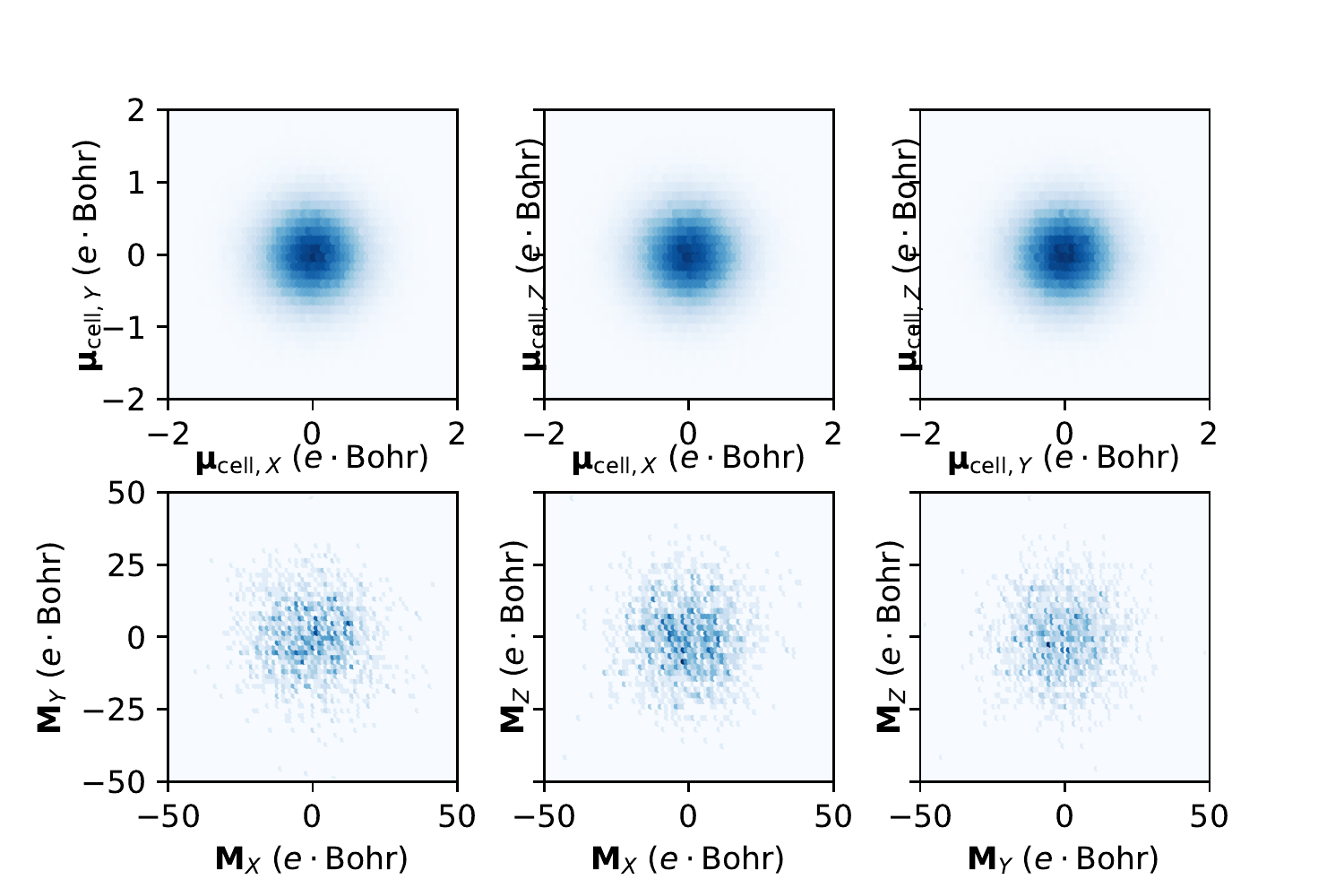}
     \captionsetup{labelformat=npj}
    \caption{2-D histograms of the unit-cell polarization (top row) and supercell polarization (bottom row) predicted by the SA-GPR model in a fully flexible simulation at \SI{250}{\kelvin} with an externally applied diagonal stress tensor $\sigma = \diag(p, p, p)$ and $\mathrm{p} = \SI{30}{\giga\pascal}$.  The action of a large positive pressure restores the isotropy of the polarization densities preventing any ferroelectric state, as stated in Sec.~IIB of the main paper.}
   \label{fig:pol-30GPa}
\end{figure}

\begin{figure}[tbhp]
    \centering
    \includegraphics[width=0.8\linewidth]{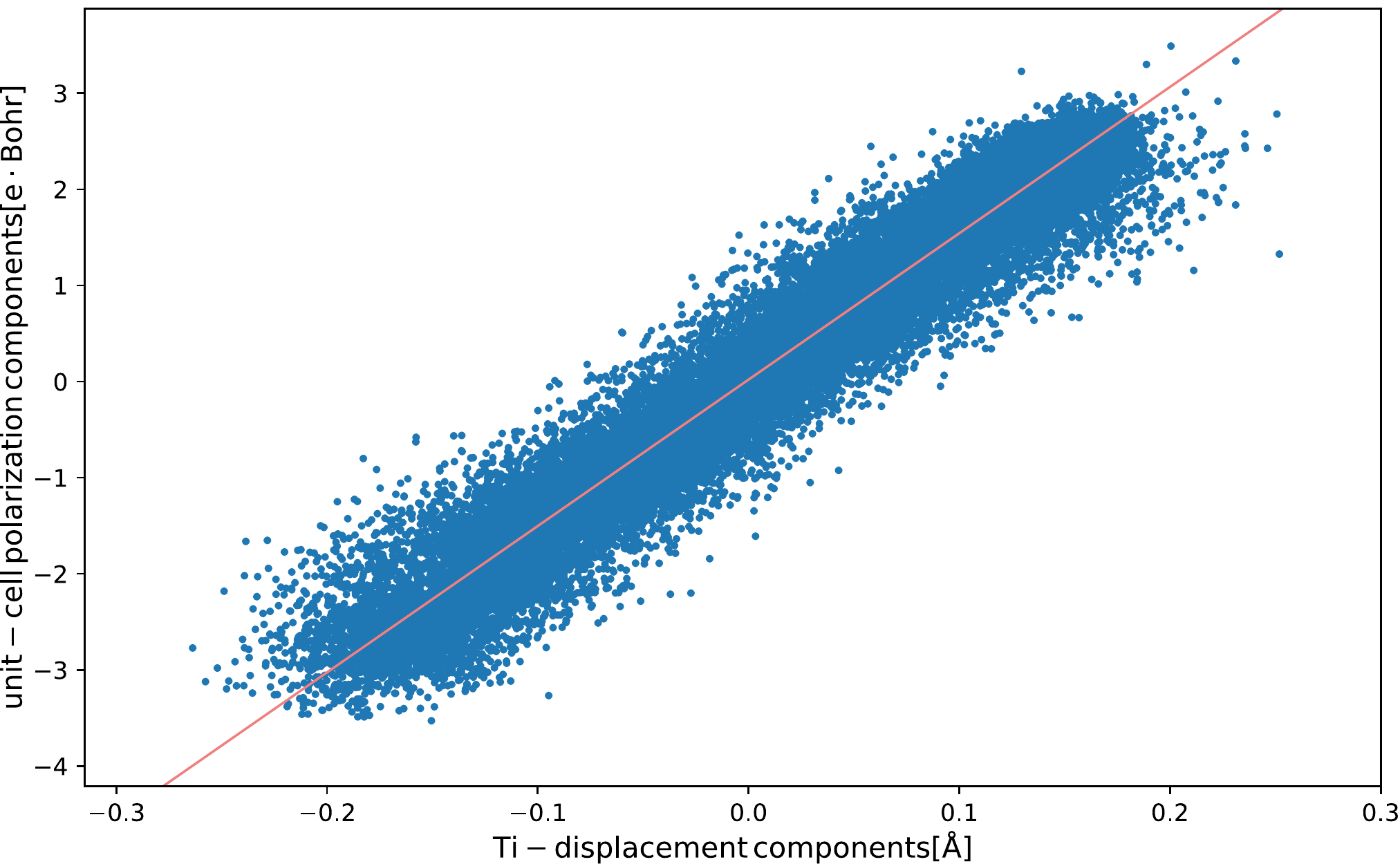}
     \captionsetup{labelformat=npj}
    \caption{Correlation plot between the unit-cell polarization components of ~\bto{} \npjcorr{as predicted by the ML polarization model} and the components of the Ti-displacement vector in the series of $5 \times 5 \times 5$ MD trajectories discussed in Sec.~II~B of the main paper. The two are strongly (linearly) correlated.}
   \label{fig:corr-disppol}
\end{figure}

\begin{figure}[tbhp]
    \centering
    \includegraphics[width=0.9\linewidth]{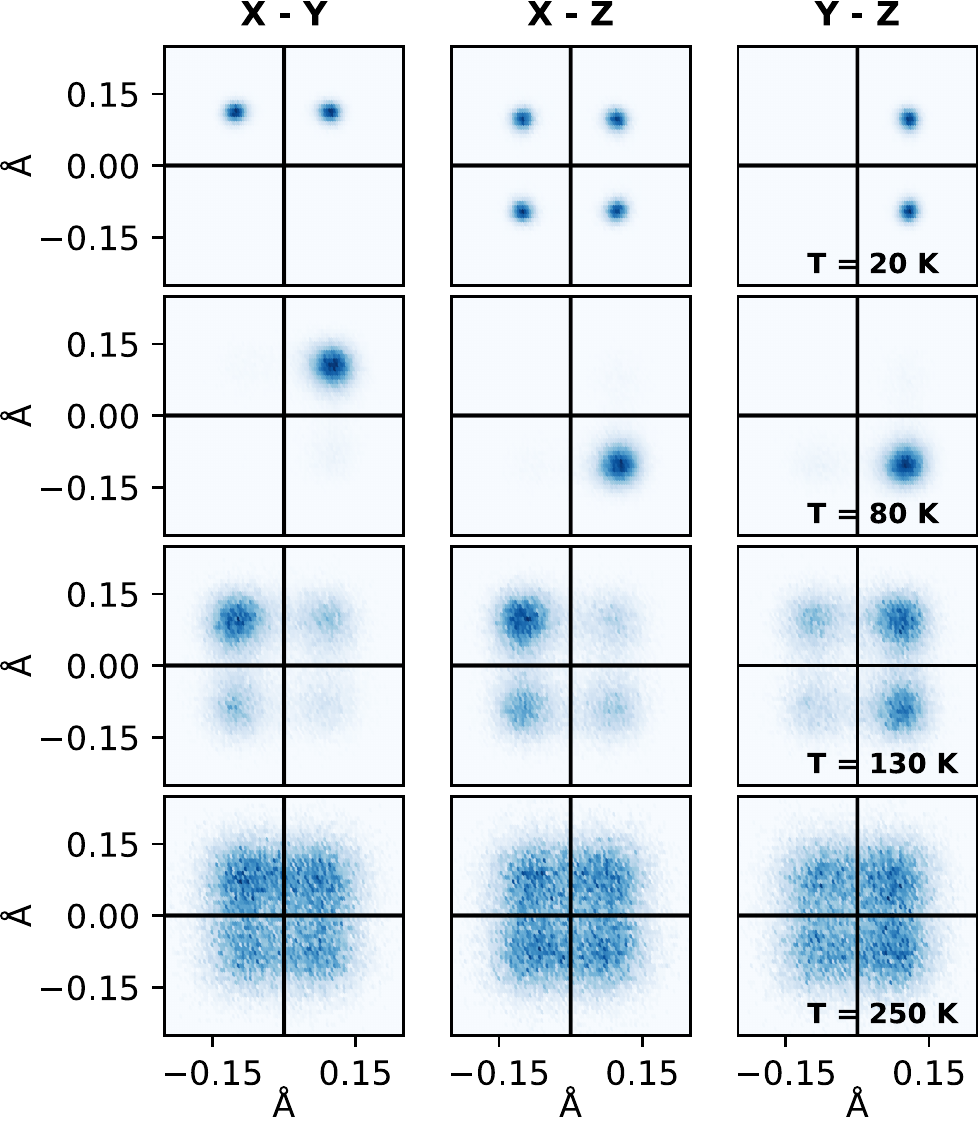}
     \captionsetup{labelformat=npj}
    \caption{2D displacement maps in a series of $NVT$ \sfour simulations with restricted cubic geometry, where the volume $V$ of the supercell is set to be equal to the equilibrium volume of an $NpT$ \sfour simulation at \SI{250}{\kelvin}. The phases at \SIlist{20;80;130;250}{\kelvin} are the same that we observe in Sec.~II~C of the main paper and show how the small volume fluctuations at finite pressure do not change the qualitative picture depicted in Fig.~3.}
   \label{fig:Ti-NVT}
\end{figure}

\begin{figure}[tbhp]
    \centering
    \includegraphics[width=0.8\linewidth]{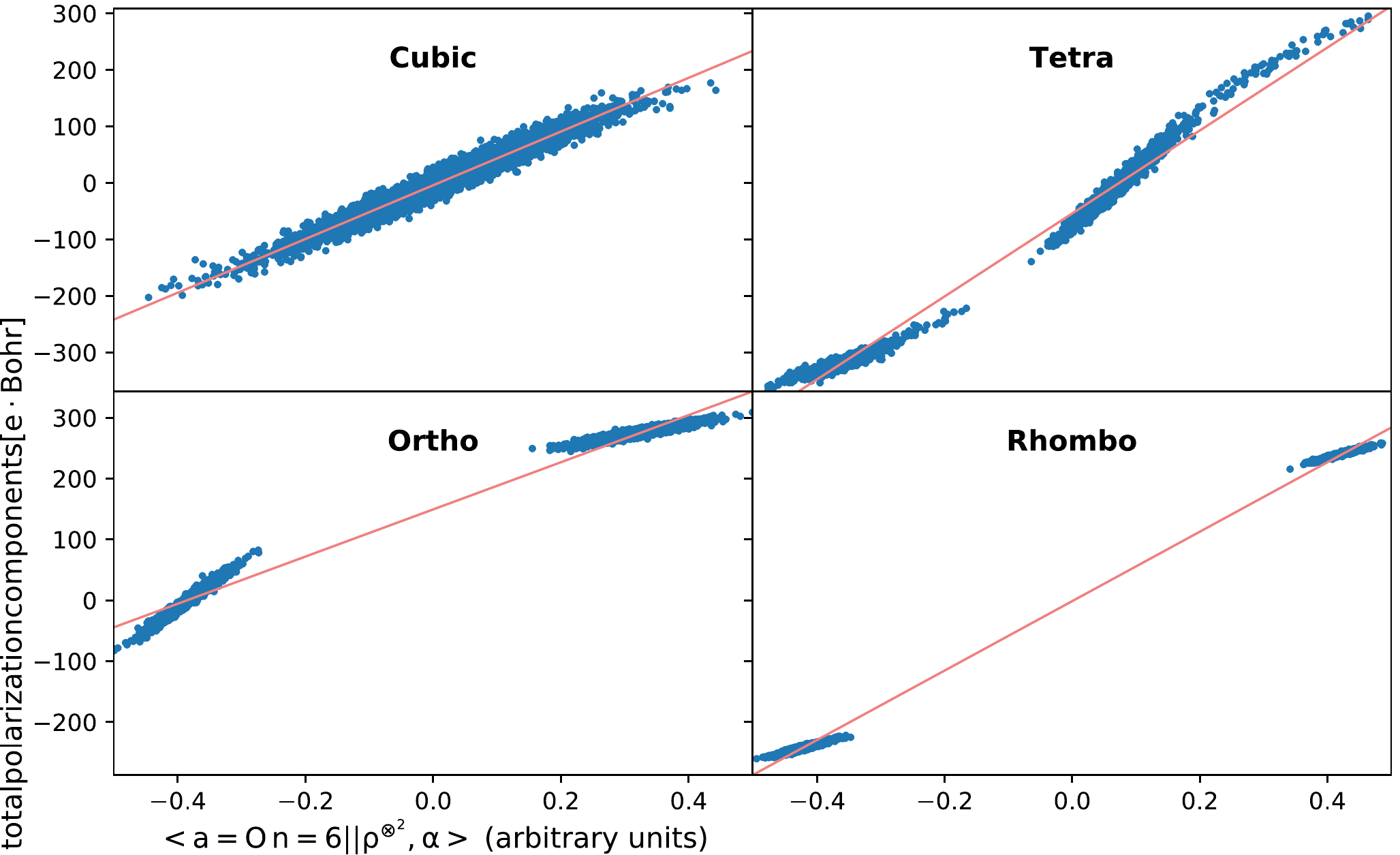}
     \captionsetup{labelformat=npj}
    \caption{Correlation plots between the total polarization components of ~\bto~and the components of the vector descriptor $\rep<a\!=\!\ce{O}\,n\!=\!6||\frho[\alpha]_i^1>$ which is observed to provide the best separation between the R-O-T-C clusters (\sfive{} MD trajectories). Red lines represent guides to the eye to observe the correlations. Given the high (albeit non-linear) correlations between $\mathbf{P}$ and this feature vector, the latter has been computed to reconstruct the phase diagram, as it provides a cheaper approach to compute the order parameter for the set of trajectories generated via MD.}
   \label{fig:corr-chipol}
\end{figure}

\begin{figure}[tbhp]
    \centering
    \includegraphics[width=\linewidth]{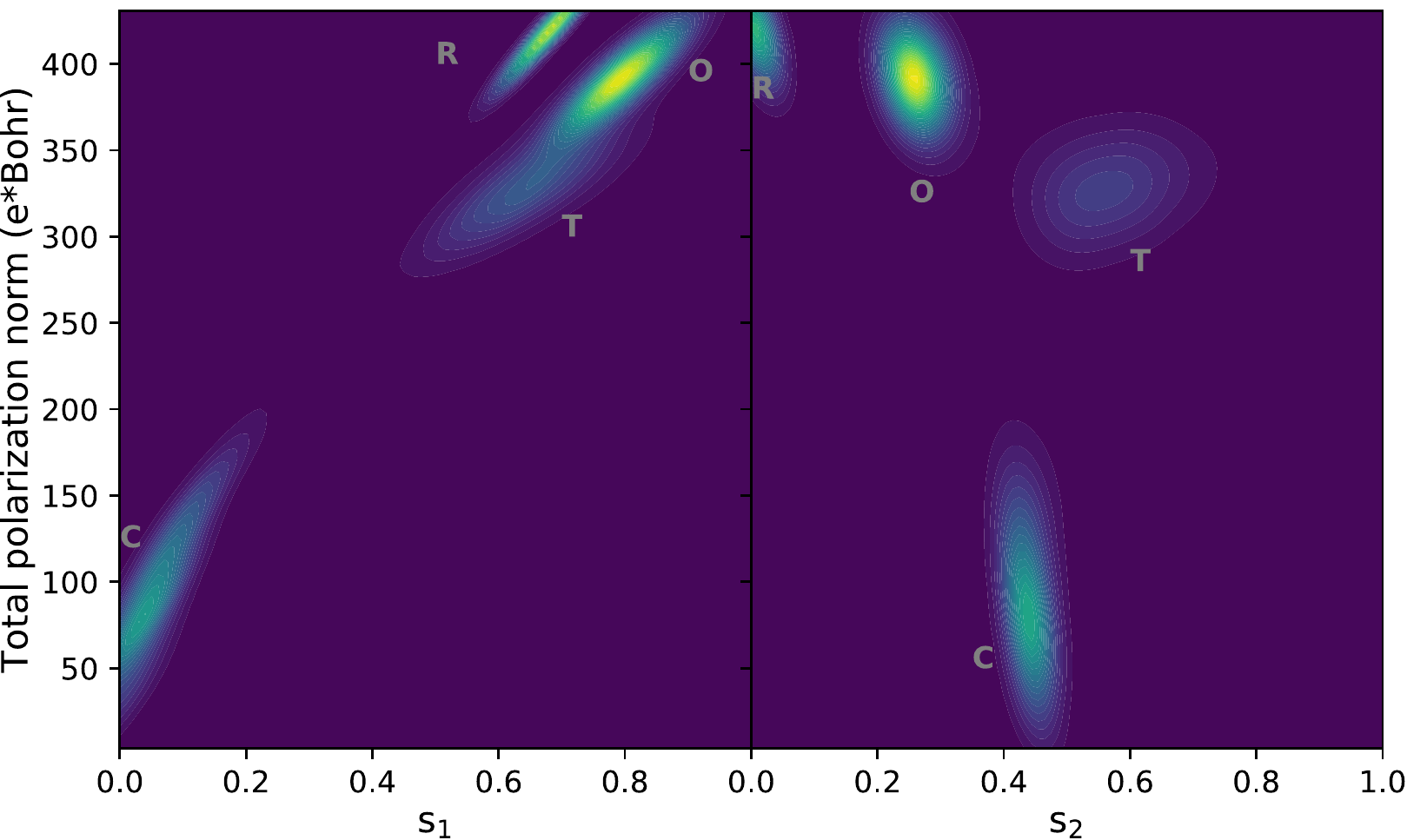}
     \captionsetup{labelformat=npj}
    \caption{2D contour plots showing the correlation between the total polarization norm and $s_1$ and $s_2$ in the series of \sfive{} MD trajectories analysed in Sec. II~B of the main paper. Yellow regions correspond to high-density peaks in the 2D map. The structure of the clusters reveals how $s_1$ is best suited to distinguish the cubic paraelectric from the ferroelectric phases and it is consequently strongly correlated with the polarization magnitude. $s_2$ instead separates the ferroelectric R-O-T phases and is correlated with the total polarization orientation.}
   \label{fig:qpolnorm}
\end{figure}

\begin{figure}[tbhp]
    \centering
    \includegraphics[width=\linewidth]{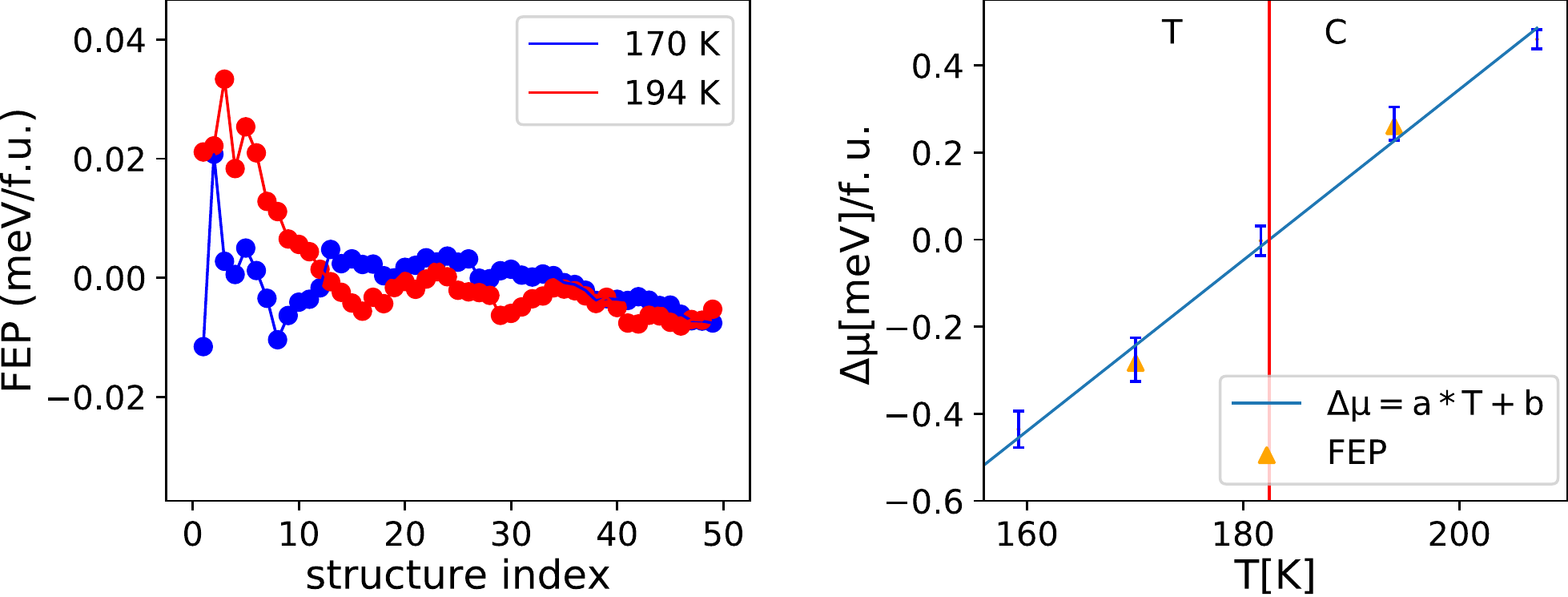}
     \captionsetup{labelformat=npj}
    \caption{FEP corrections to the chemical potentials across the T-C transition. The left panel shows the convergence plots of the FEP corrections, computed at \SIlist{170;194}{\kelvin}. These show the behavior of the free energy perturbation as a function of the number of cubic and tetragonal structures used for the calculation. The right panel shows a zoomed-in region of the $\Delta \mu$ profile about the transition point. Errobars represent the standard deviation computed over four independent MD trajectories at each temperature.}
   \label{fig:qpolnorm}
\end{figure}

\begin{figure}[tbhp]
    \centering
    \includegraphics[width=\linewidth]{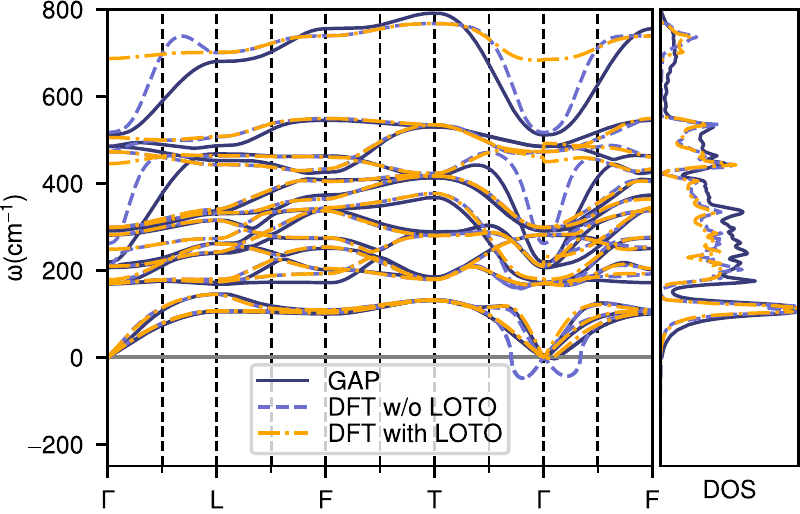}
     \captionsetup{labelformat=npj}
    \caption{Phonon spectra and corresponding DOS of the phonons with $\omega >0$ for the 5-atom rhombohedral ground state structure comparing GAP, DFT without and with LO-TO splitting, the latter being directly comparable to the experiments.  The vertical dashed lines indicate the explicitly calculated $q$-points.}
   \label{fig:qpolnorm}
\end{figure}

\begin{figure}[tbhp]
    \centering
    \includegraphics[width=\linewidth]{figures/thermal-expansion.pdf}
     \captionsetup{labelformat=npj}
    \caption{\npjcorr{Average lattice parameter (equal to $V^{1/3}$) as a function of the temperature in \sfour{} fully-flexible MD simulations, at $p = 1\,$atm and $p = -2\,$GPa. The shaded areas indicate the statistical uncertainty computed over four different simulations, while the vertical lines correspond to the critical transition temperatures predicted by the GAP model. In contrast with the effective Hamiltonian fitted in Tinte et \textit{al.} \cite{tinteQuantitativeAnalysis2003}, we do not observe any severe underestimation of the thermal expansion.}}
   \label{fig:thermal-exp}
\end{figure}

\begin{figure}[tbhp]
    \centering
    \includegraphics[width=\linewidth]{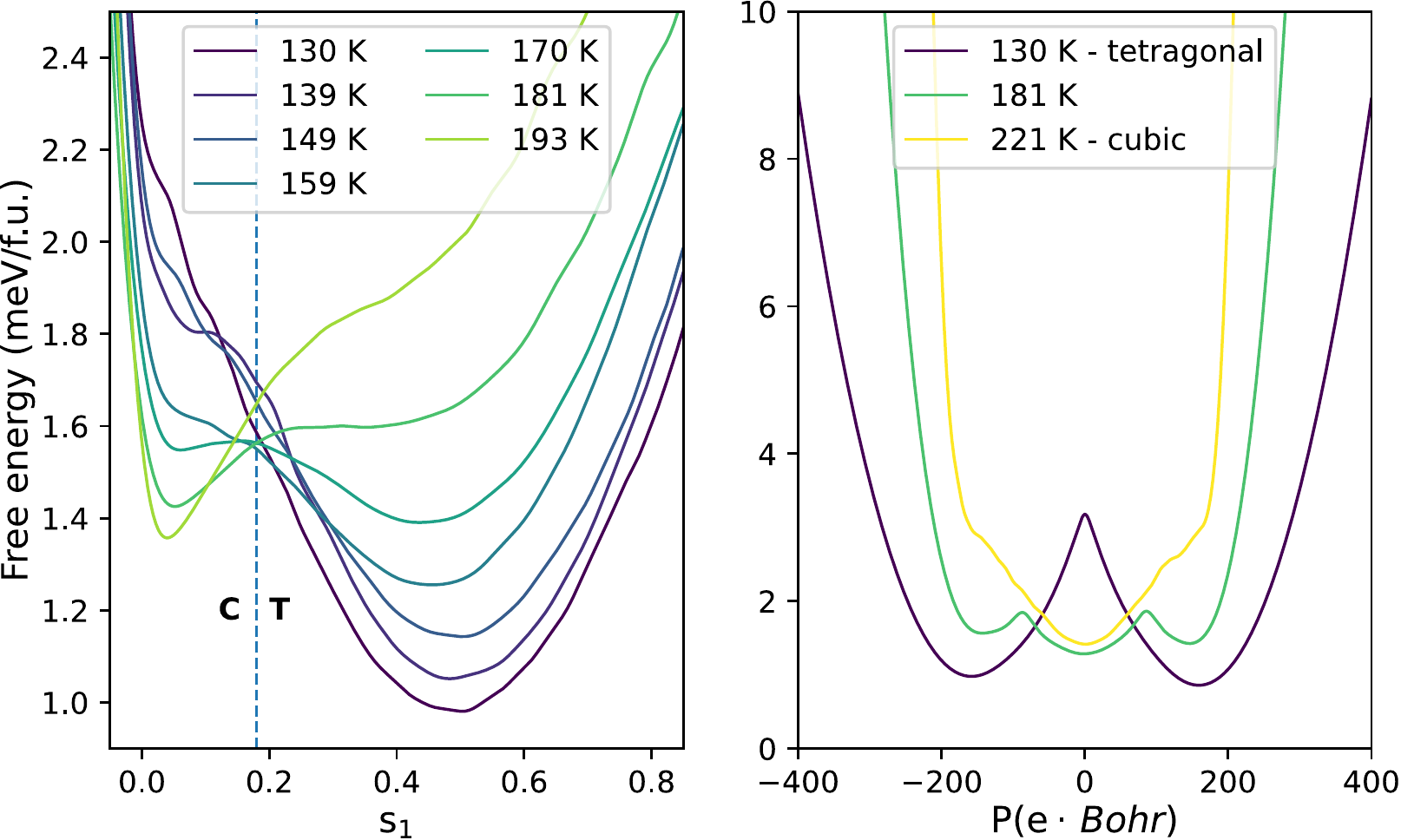}
     \captionsetup{labelformat=npj}
    \caption{\npjcorr{Free energy profiles as a function of $s_1$ (left panel) and the polarization $P$ (right panel) across the T-C transition, computed according to the relation $F(\xi) = -k_{B}T \log(\rho(\xi))$, where $\xi = (s_1, P)$. Here, $\rho(\xi)$ represents a probability distribution derived from a kernel-density estimate (kde) of the histograms of the variable $\xi$, across a few selected 4x4x4 fully flexible MD runs. The parameter $s_1$ is clearly a good CV to distinguish the T-C clusters and can thus capture their finite-temperature stability. The polarization free energy shows instead a Landau-like profile (as in Pitike et al. \cite{pitike_landaudevonshire_2019} for $\text{PbTiO}_3$), confirming the first-order character of the ferroelectric-to-paraelectric transition. These results also show how an extension of these calculations can potentially be used to fit the free energy predictions of the ML model with a Landau-Devonshire model (as in Supplementary Reference \cite{marton_first-principles-based_2017}).}}
   \label{fig:qpolnorm}
\end{figure}

\begin{figure}[tbp]
    \centering
    \includegraphics[width=\linewidth]{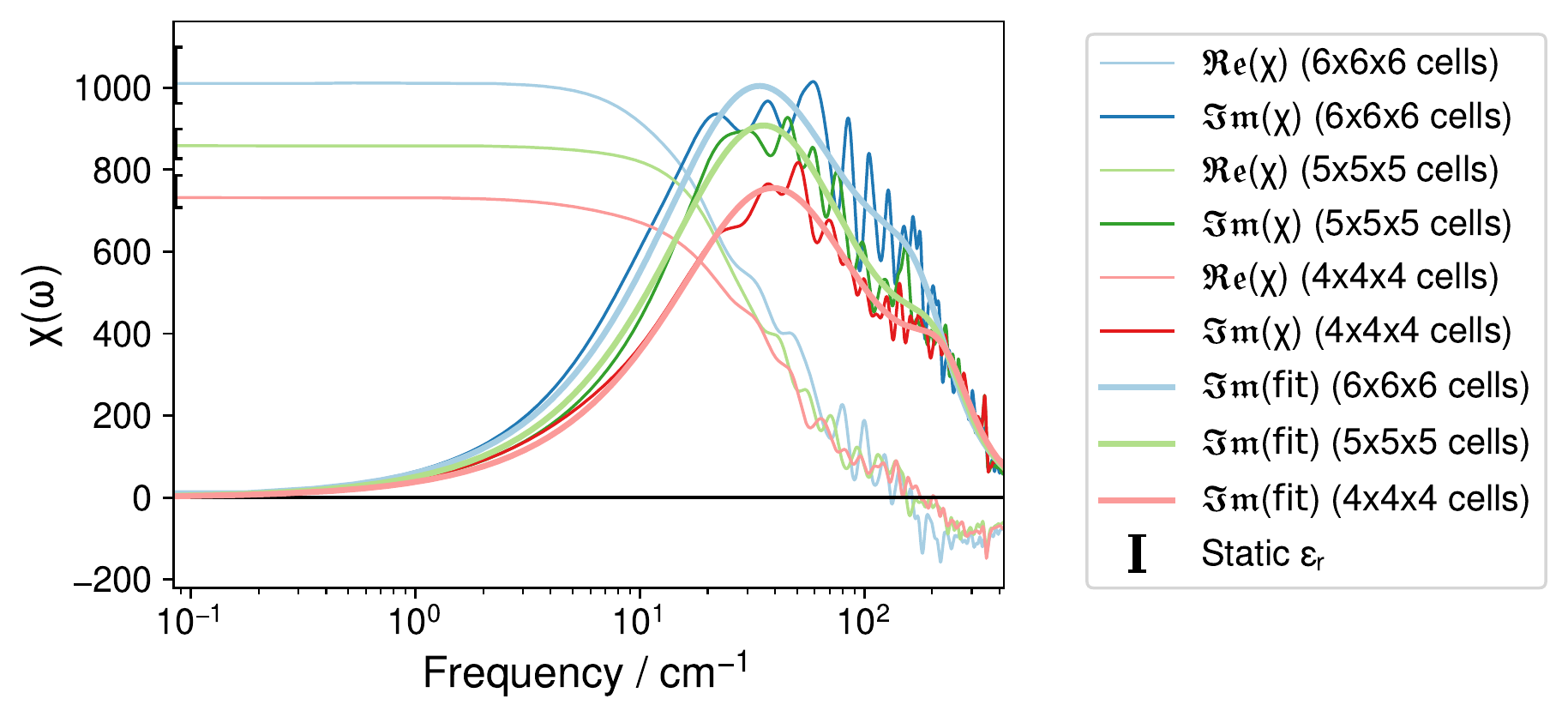}
     \captionsetup{labelformat=npj}
    \caption{\npjcorr{Behaviour of the frequency-dependent dielectric response spectrum as a function of supercell size used for the simulations.  All simulations run at \SI{250}{\kelvin}, except for the \sfour{} cell, which was run at \SI{252}{\kelvin}. The peak frequencies and damping constants do shift slightly, but not enough to explain the discrepancy from experimental results.  The \sfour{} and \sfive{} simulations were run for \SI{390}{\pico\second}, while the \ssix{} simulation was run for \SI{190}{\pico\second}.}}
    \label{fig:despectrum_size_convergence}
\end{figure}

\begin{figure}[tbp]
    \centering
    \includegraphics[width=0.6\linewidth]{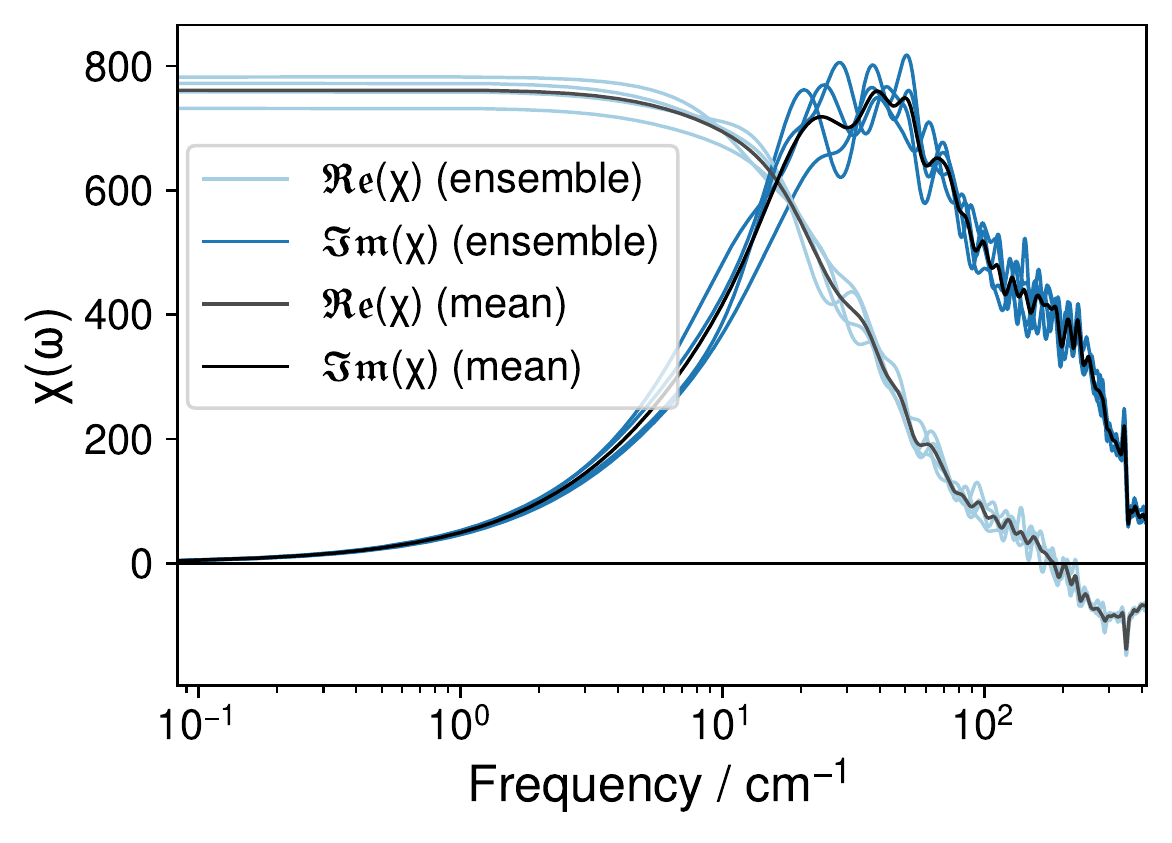}
     \captionsetup{labelformat=npj}
    \caption{\npjcorr{The dielectric response spectra calculated for four different MD simulations, on \sfour{} supercells at \SI{252}{\kelvin}, of \SI{390}{\pico\second} each.  The spectrum calculated using the average of the four dipole-dipole correlation functions $\left<\mathbf{M}(0)\mathbf{M(t)}\right>_\tau$ and the averaged static dielectric constants is shown on top in black, with the noisy oscillations much reduced.}}
    \label{fig:my_label}
\end{figure}

\clearpage

\section*{Supplementary references}
\begingroup
\renewcommand{\section}[2]{}%
\bibliography{mk.bib,mc.bib,BaTiO3_modeling.bib}
\endgroup

%% file: defs.tex
\input{dirac-rep}

\usepackage{bm,upgreek}

\newcommand{\mc}[1]{{\color{blue}{ #1}}}

\newcommand{\ipi}{\mbox{i-PI}}

\newcommand{\abinitio}{\textit{ab initio}}

\DeclareSIUnit{\electron}{\textit{e}}
\DeclareSIUnit{\electronvolt}{e\kern-0.08em V}
\DeclareSIUnit{\bohr}{Bohr}

\newcommand{\batiooo}{\ce{BaTiO3}}
\let\bto\batiooo
\newcommand{\btotitle}{$\text{B\lowercase{a}T\lowercase{i}O}_3$}
\newcommand{\pTi}{$\langle111\rangle$}
\newcommand{\mTi}{$\langle\bar{1}\bar{1}\bar{1}\rangle$}
\newcommand{\stwo}{2$\times$2$\times$2}
\newcommand{\sfour}{4$\times$4$\times$4}
\newcommand{\sfive}{5$\times$5$\times$5}
\newcommand{\ssix}{6$\times$6$\times$6}

%% file: dirac-rep.tex
\usepackage{physics}
\usepackage{xparse}

\newcommand{\mbf}[1]{\ensuremath{\mathbf{#1}}}

\NewDocumentCommand{\rep}{s d<| d|>}{%
\IfBooleanTF{#1}{
   \IfValueTF{#2}{
       \IfValueTF{#3}{\braket{#2}{#3}}{\bra{#2}}
       }{
       \IfValueTF{#3}{\ket{#3}}{}
       }
   }{
   \IfValueTF{#2}{
       \IfValueTF{#3}{\braket*{#2}{#3}}{\bra*{#2}}
       }{
       \IfValueTF{#3}{\ket*{#3}}{}
       }
   }
}

\NewDocumentCommand{\rbra}{sm}{\IfBooleanTF{#1}{\rep<#2|}{\rep*<#2|}}
\NewDocumentCommand{\rket}{sm}{\IfBooleanTF{#1}{\rep|#2>}{\rep*|#2>}}
\NewDocumentCommand{\rbraket}{smom}{
    \IfBooleanTF{#1}{
        \IfNoValueTF{#3}{\rep*<#2||#4>}{\matrixel{#2}{#3}{#4}}
    }{
        \IfNoValueTF{#3}{\rep<#2||#4>}{\matrixel*{#2}{#3}{#4}}
    }
}

\NewDocumentCommand{\field}{o m e{_} e{^} o e{_} e{^}}{
\IfValueTF{#5}{\overline{
  #2\IfValueT{#3}{_#3}\IfValueT{#4}{^{\otimes #4}} 
  \otimes
  #5\IfValueT{#6}{_#6}\IfValueT{#7}{^{\otimes #7}} 
  \IfValueT{#1}{;#1}
}}{
  \IfValueTF{#4}{\overline{
     #2\IfValueT{#3}{_#3}\IfValueT{#4}{^{\otimes #4}}
     \IfValueT{#1}{;#1}
  }}
  {#2\IfValueT{#3}{_#3}}
}
}

\NewDocumentCommand{\frho}{o e{_} e{^}}{
\field[#1]{\rho}_{#2}^{#3}
}
\NewDocumentCommand{\fdelta}{o e{_} e{^}}{
\field[#1]{\delta}_{#2}^{#3}
}

\newcommand{\e}{a}  

\newcommand{\bx}{\mbf{x}}

\NewDocumentCommand{\ex}{e_}{
\IfValueTF{#1}{\e_{#1}\bx_{#1}}{\e\bx}
}  

\NewDocumentCommand{\lm}{e_}{
\IfValueTF{#1}{l_{#1}m_{#1}}{lm}
}
\NewDocumentCommand{\nlm}{e_}{
\IfValueTF{#1}{n_{#1}\lm_{#1}}{n\lm}
}
\NewDocumentCommand{\enlm}{e_}{
\IfValueTF{#1}{\e_{#1}\nlm_{#1}}{\e\nlm}
}
\NewDocumentCommand{\en}{e_}{
\IfValueTF{#1}{\e_{#1}n_{#1}}{\e n}
}
\NewDocumentCommand{\nlk}{e_}{
\IfValueTF{#1}{n_{#1}l_{#1}k_{#1}}{nlk}
}
\NewDocumentCommand{\enlk}{e_}{
\IfValueTF{#1}{\e_{#1}\nlk_{#1}}{\e\nlk}
}
\NewDocumentCommand{\enl}{e_}{
\IfValueTF{#1}{\en_{#1}l_#1}{\en l}
}
\NewDocumentCommand{\nl}{e_}{
\IfValueTF{#1}{n_{#1}l_#1}{n l}
}

\NewDocumentCommand{\nnl}{s}{
\IfBooleanTF{#1}{n_1 n_2 l}{n_1; n_2; l}
}
\NewDocumentCommand{\ennl}{s}{
\IfBooleanTF{#1}{\en_1 \en_2 l}{\en_1; \en_2; l}
}

\NewDocumentCommand{\gslm}{s}{
\IfBooleanTF{#1}{\sigma\lambda\mu}{\sigma;\lambda\mu}
}

%% file: authors-finalsub.tex
\author[1,*]{Lorenzo Gigli}
\author[1]{Max Veit}
\author[2]{Michele Kotiuga}
\author[2]{Giovanni Pizzi}
\author[2]{Nicola Marzari}
\author[1]{Michele Ceriotti}

\affil[1]{Laboratory of Computational Science and Modeling (COSMO), Institute of Materials, \'Ecole Polytechnique F\'ed\'erale de Lausanne, CH-1015 Lausanne, Switzerland}
\affil[2]{Theory and Simulation of
  Materials (THEOS) and National Centre for Computational Design and
  Discovery of Novel Materials (MARVEL), \'Ecole Polytechnique
  F\'ed\'erale de Lausanne, CH-1015 Lausanne, Switzerland}
\affil[*]{lorenzo.gigli@epfl.ch}